%% file: matrixGeneral.tex
\documentclass[12pt]{article}
\usepackage{putex}
\usepackage{graphicx}
\usepackage{epstopdf}
\usepackage{enumerate}
\usepackage{cite}

\numberwithin{equation}{section}

\newcommand{\abs}[1]{\left\lvert #1 \right\rvert}

\newcommand {\be} {\begin {equation}}
\newcommand {\ee} {\end {equation}}

\newcommand {\bes} {\begin {equation*}}
\newcommand {\ees} {\end {equation*}}

\newcommand{\es}[2] {\begin{equation} \label{#1} \begin{split} #2 \end{split} \end{equation}}

\newcommand{\Z}{\mathbb{Z}}
\newcommand{\R}{\mathbb{R}}
\newcommand{\C}{\mathbb{C}}

\DeclareMathOperator {\Det} {det}

\definecolor{O1}{RGB}{255,178,128}
\definecolor{O2}{RGB}{255,103,1}
\definecolor{O3}{RGB}{212,85,0}
\definecolor{O4}{RGB}{128,51,0}

\begin{document}

\preprint{PUPT-2366}

\institution{IAS}{School of Natural Sciences, Institute for Advanced Study, Princeton, NJ 08540}
\institution{Harvard}{Center for the Fundamental Laws of Nature, Harvard University, Cambridge, MA 02138}
\institution{PU}{Joseph Henry Laboratories, Princeton University, Princeton, NJ 08544}
\institution{PCTS}{Center for Theoretical Science, Princeton University, Princeton, NJ 08544}

\title{Towards the $F$-Theorem: ${\cal N} = 2$ Field Theories on the Three-Sphere}

\authors{Daniel L.~Jafferis,\worksat{\IAS, \Harvard} Igor R.~Klebanov,\worksat{\PU, \PCTS}  Silviu S.~Pufu,\worksat{\PU} Benjamin R.~Safdi\worksat{\PU}}

\abstract{
For 3-dimensional field theories with ${\cal N}=2$ supersymmetry the Euclidean path integrals on the three-sphere can be calculated using the method of localization; they reduce to certain matrix integrals that depend on the R-charges of the matter fields. We solve a number of such large $N$ matrix models and calculate the free energy $F$ as a function of the trial R-charges consistent with the marginality of the superpotential. In all our ${\cal N}=2$ superconformal examples, the local maximization of $F$ yields answers that scale as $N^{3/2}$ and agree with the dual M-theory backgrounds $AdS_4\times Y$, where $Y$ are 7-dimensional Sasaki-Einstein spaces. We also find in toric examples that local $F$-maximization is equivalent to the minimization of the volume of $Y$ over the space of Sasakian metrics, a procedure also referred to as $Z$-minimization.  Moreover, we find that the functions $F$ and $Z$ are related for any trial R-charges.  In the models we study $F$ is positive and decreases along RG flows. We therefore propose the ``$F$-theorem'' that we hope applies to all 3-d field theories: the finite part of the free energy on the three-sphere decreases along RG trajectories and is stationary at RG fixed points.  We also show that in an infinite class of Chern-Simons-matter gauge theories where the Chern-Simons levels do not sum to zero, the free energy grows as $N^{5/3}$ at large $N$.  This non-trivial scaling matches that of the free energy of the gravity duals in type IIA string theory with Romans mass. }

\date{March 2011}

\maketitle

\tableofcontents

\section{Introduction}

Among the earliest tests of the $AdS_5$/CFT$_4$ correspondence \cite{Maldacena:1997re,Gubser:1998bc,Witten:1998qj} were comparisons of the Weyl anomaly coefficients $a$ and $c$. On the gravity side these coefficients were calculated in \cite{Henningson:1998gx} and were found to be equal; their values match the corresponding results in a variety of large $N$ superconformal 4-d gauge theories.

Thanks to the important progress during the past several years, there now also exists a large set of precisely formulated $AdS_4$/CFT$_3$ conjectures. In particular, M-theory in the $AdS_4\times S^7/\Z_k$ background is now thought to be dual to the ABJM theory---the ${\cal N}=6$ superconformal $U(N)_k\times U(N)_{-k}$ Chern-Simons-matter gauge theory constructed in \cite{Aharony:2008ug} (see also \cite{Hosomichi:2008jd,Benna:2008zy}). Many similar duality conjectures with lower amounts of supersymmetry are also available. While various successful tests of some of these $AdS_4$/CFT$_3$ conjectures have been made, it is interesting to ask whether there exists an analog in this dimensionality of the Weyl anomaly matching. At first this question seems silly: of course, there are no anomalies in 3-d field theories. Nevertheless, an idea has emerged in recent research \cite{Kapustin:2009kz, Drukker:2010nc, Herzog:2010hf, Jafferis:2010un } that the 3-d quantity that plays a special role, and may be analogous to the anomaly $a$-coefficient in 4 dimensions, is the free energy of the Euclidean CFT on a three-dimensional sphere:\footnote{Similarly, in a 4-d CFT the anomaly $a$-coefficient may be extracted from the free energy on the four-sphere after removing the power-law divergences and differentiating with respect to $\ln R$.}
\es{basicF}{
  F = -\ln \abs{Z_{S^3}} \,,
 }
where $Z$ is the Euclidean path integral.\footnote{A seemingly different measure of the number of degrees of freedom in a 3-d CFT was proposed in \cite{Myers:2010xs, Myers:2010tj}; it is the entanglement entropy between the two hemispheres in the CFT on $\R\times S^2$. In \cite{Casini:2011kv} it was shown that this quantity, which is the same as the entanglement entropy between a circle and its complement on a plane, is also equal to minus the free energy of the Euclidean theory on $S^3$. We thank Rob Myers for pointing this out to us.}   In a general 3-d CFT the free energy has power law divergences. After they are subtracted, the finite part is independent of the radius of the three-sphere and appears to be an unambiguous, invariant quantity that provides a good measure of the number of degrees of freedom.  Explicit calculations in unitary 3-d CFTs give {\it positive} values for $F$ \cite{Drukker:2010nc,Herzog:2010hf, Santamaria:2010dm, Jafferis:2010un}, in contrast with the thermal free energy on $\R^2 \times S^1$, which is negative.
The corresponding calculations in Euclidean $AdS_4$ also give a positive free energy \cite{Emparan:1999pm}
 \es{FHolography}{
  F = \frac{\pi L^2}{2 G_N}
 }
after the counterterms that remove the power law divergences are included \cite{Emparan:1999pm, Henningson:1998gx, Balasubramanian:1999re}.   Here, $G_N$ is the effective four-dimensional Newton constant and $L$ is the radius of $AdS_4$.  It has been shown that $L$ decreases along holographic RG flow in the leading supergravity approximation \cite{Freedman:1999gp, Myers:2010tj}.  For M-theory on $AdS_4 \times Y$, where $Y$ is a seven-dimensional Einstein space threaded by $N$ units of flux, the gravitational free energy is \cite{Herzog:2010hf}
\es{MtheoryExpectation}{
  F = N^{3/2} \sqrt{\frac{2 \pi^6}{27 \Vol(Y)}} + o(N^{3/2}) \,,
 }
where the metric on $Y$ is normalized so that $R_{ij} = 6 g_{ij}$.   This formula exhibits the characteristic $N^{3/2}$ scaling of the number of degrees of freedom on coincident M2-branes \cite{Klebanov:1996un}, and it comes with a specific normalization that can be compared with the dual field theory calculations.

For field theories with extended supersymmetry, the free energy on the three-sphere can be calculated using the method of localization that reduces it to certain matrix integrals. For all ${\cal N}\geq 3$ supersymmetric theories, where all field dimensions are determined by supersymmetry, the necessary integrals were written down in \cite{Kapustin:2009kz}. Large $N$ calculations of these matrix integrals for theories with known M-theory duals \cite{Drukker:2010nc, Herzog:2010hf, Santamaria:2010dm} produce perfect agreement with (\ref{MtheoryExpectation}) and thus provide impressive tests of the $AdS_4$/CFT$_3$ conjectures. In the present paper we extend these successes to large $N$ theories with ${\cal N}=2$ supersymmetry. For such theories the modification of the localization procedure that takes into account anomalous dimensions was derived in \cite{Jafferis:2010un,Hama:2010av}. We will solve a variety of corresponding large $N$ matrix models and provide many new tests of $AdS_4$/CFT$_3$ conjectures.

These solvable ${\cal N}=2$ theories give rise to some new phenomena that could not be seen in models with higher supersymmetry.
 In ${\cal N}=2$ theories the constraints of conformal invariance are in general not sufficient to fix all the R-charges of gauge-invariant operators. In such cases it was proposed \cite{Jafferis:2010un} that the remaining freedom in the R-charges should be fixed by extremizing the free energy on $S^3$. We apply this idea to various large $N$ models and show that the R-charges determined this way are in agreement with the AdS/CFT correspondence.
In fact, in all cases we find that the R-charges locally {\it maximize} $F$. This is analogous to the well-known statement that R-charges in four-dimensional ${\cal N}=1$ theories locally maximize the anomaly coefficient $a$ \cite{Intriligator:2003jj}.

If the cone over the seven-dimensional internal space $Y$ is a toric Calabi-Yau four-fold, then one can go a little further.  In toric geometry the volume of $Y$ can be found by performing a similar extremization procedure that is usually referred to as $Z$-minimization \cite{Martelli:2005tp}.  The function $Z$ that one is supposed to minimize is nothing but the Einstein-Hilbert action with positive cosmological constant evaluated on the set of Sasakian metrics on $Y$.  It turns out that $Z$ depends only on how the R-symmetry (given by the Reeb vector) is embedded within the $U(1)^4$ isometry of the toric Calabi-Yau, which can be identified with a $U(1)^4$ symmetry of the gauge theory.  In particular, using this identification one can express $Z$ in terms of the trial R-charges consistent with the marginality of the superpotential.  In the cases we examine we show that $Z(\Delta) \sim 1/F(\Delta)^2$, where $F$ is the three-sphere free energy computed using the trial R-charges $\Delta$.  In other words, not only do $Z$-minimization and $F$-maximization yield the same answer, but also the functions of $\Delta$ that one is supposed to extremize are related.  A similar relation in the case of $(3+1)$-dimensional gauge theories between $Z$ and the anomaly coefficient $a$ was found in \cite{Butti:2005vn}.

In this paper we also study some pairs of fixed points connected by RG flow and find that $F$ decreases along the flow, just like $a$ does in 4 dimensions (there is growing evidence for the $a$-theorem in 4-d that states that $a$ decreases along RG trajectories and is stationary at RG fixed points \cite{Cardy:1988cwa}). We also find that, just like $a$, the free energy $F$ stays constant under exactly marginal deformations.  It is therefore tempting to conjecture that there exists a similar $F$-theorem in 3-d, stating that the free energy on the three-sphere decreases along RG trajectories and is stationary at RG fixed points.

The rest of this paper is organized as follows.  In section~\ref{MATRIX} we review the rules by which one can construct the matrix model associated with a particular ${\cal N} = 2$ quiver.  We show that in gauge theories where the bifundamentals are non-chiral, the total number of fundamentals equals the total number of anti-fundamentals, and the Chern-Simons levels sum to zero, the free energy scales as $N^{3/2}$.  In section~\ref{ADJOINTS} we discuss an infinite class of the necklace quiver gauge theories with ${\cal N} = 2$ supersymmetry where the ${\cal N} = 3$ models proposed in  \cite{Jafferis:2008qz, Imamura:2008nn}, and studied in \cite{Herzog:2010hf}, are deformed by adding a cubic superpotential for the adjoints \cite{Martelli:2009ga}.  In sections~\ref{FLAVORC3} and~\ref{FLAVORABJM} we display many examples of flavored quivers whose quantum corrected moduli space of vacua was constructed in \cite{Benini:2009qs, Jafferis:2009th} and perform $F$-maximization to find the R-symmetry in the IR\@.   In section~\ref{ABJMNOFLAVORS} we discuss deformations of ABJM theory and RG flows.  In section~\ref{TORIC} we show that for the examples studies in sections~\ref{FLAVORC3} and~\ref{FLAVORABJM} that $Z$-minimization is equivalent to $F$-maximization.  In particular, this equivalence represents a test of the proposed $AdS_4$/CFT$_3$ dualities.  In section~\ref{FIVETHIRDS} we study the ${\cal N} = 3$ necklace quiver theories when the Chern-Simons levels don't add up to zero.  We find that in this case the free energy on $S^3$ as computed from the matrix model behaves as $N^{5/3}$ at large $N$, in agreement with a dual massive type IIA construction \cite{Aharony:2010af}.  We end with a discussion in section~\ref{DISCUSSION}.  Various additional technical details are delegated to the appendices.

\section{Matrix models for ${\cal N} = 2$ quiver gauge theories}
\label{MATRIX}

Using localization it was shown that the $S^3$ partition function of ${\cal N}=2$ Chern-Simons-matter theories is given by a matrix integral over the Cartan subalgebra of the gauge groups \cite{Jafferis:2010un, Hama:2010av}.  The integrand involves both gaussians determined by the Chern-Simons levels as well as factors appearing from one-loop determinants. The latter depend on the curvature couplings on the sphere, parameterized by trial R-charges $\Delta$:
 \es{PartitionFunction}{
  F(\Delta) = -\ln \int \left( \prod_{\rm Cartan} \frac{d \sigma}{2 \pi} \right)
   \exp \left[ \frac{i}{4 \pi} \tr_k \sigma^2 - \tr_m \sigma\right] \Det_{\rm Ad} \left(2 \sinh \frac{\sigma}{2} \right) \\
    \times \prod_{\substack{\text{chirals}\\\text{in rep $R_i$}}} \Det_{R_i} \left(e^{\ell\left(1 - \Delta_i + i \frac{\sigma}{2 \pi}\right)} \right) \,,
 }
where the function
 \es{ellDef}{
  \ell(z) = -z \ln \left(1 - e^{2 \pi i z} \right) + \frac i2 \left(\pi z^2 + \frac{1}{\pi} \text{Li}_2
   \left(e^{2 \pi i z} \right) \right) - \frac{i \pi}{12}
 }
satisfies the differential equation $d \ell / d z = - \pi z \cot (\pi z)$.  The trace $\tr_k$ is normalized so that for each gauge group $a$ it equals the Chern-Simons level $k_a$  times the trace in the fundamental representation, while the trace $\tr_m$ is normalized so that for each gauge group it equals the bare monopole R-charge $\Delta_m^{(a)}$ that we define below times the trace in the fundamental representation.  In eq.~\eqref{PartitionFunction} the integration variables $\sigma$ are the scalars in the vector multiplets.  Since these scalars transform in the adjoint representation of the gauge group, the integration contour should be taken to be the real axis for each integration variable.

Some of the important ingredients of the $U(N)^p$ CS gauge theories we study are the topological conserved currents $j_{{\rm top}, a} = *\tr F_a$ and monopole operators $T_{\vec{q}}$ that create $q_a$ units of $\tr F_a$ flux through a two-sphere surrounding the insertion point.  In general, the R-symmetry can mix with these topological global symmetries, and the monopole operators $T_{\vec{q}}$ acquire R-charges $R[T_{\vec{q}}] = \gamma_{\vec{q}} + \sum_a \Delta_m^{(a)} q_a$, where $\gamma_{\vec{q}}$ is an anomalous dimension invariant under sending $\vec{q} \to -\vec{q}$, and the $\Delta_m^{(a)}$ are what we call bare monopole R-charges.  The anomalous dimensions $\gamma_{\vec{q}}$ can be computed exactly at one-loop in perturbation theory from the matter R-charges, as in refs.~\cite{Jafferis:2009th, Benini:2009qs} based on the work of \cite{Borokhov:2002cg}.   Of special interest will be the ``diagonal'' monopole operators $T$ corresponding to $\vec q=(1,1,1, \ldots)$ and $\tilde T$ corresponding to $\vec{q} = (-1, -1, -1, \ldots)$, because they play a crucial role in the construction of the quantum-corrected moduli space of vacua in these theories \cite{Jafferis:2009th, Benini:2009qs}.  Their R-charges satisfy the relation
 \es{TTtilde}{
  R[T] - R[\tilde T] = 2 \Delta_m \,, \qquad
   \Delta_m \equiv \sum_a \Delta_m^{(a)} \,.
 }
One may worry already that the bare R-charges of the diagonal monopole operators are not gauge-invariant observables because the Chern-Simons coupling makes the monopole operators not gauge-invariant.  As we will explain in more detail in section~\ref{SYMMETRY}, with an appropriate choice of gauge group one can construct gauge-invariant operators out of $T$ or out of $\tilde T$, and from the R-charges of these gauge-invariant operators one can calculate $\Delta_m$.  (In passing, note that the same concern can be raised about the R-charges of the bifundamental fields, and the same resolution holds.)  In theories with charge conjugation symmetry the R-charge of $T$ should equal that of $\tilde T$, which implies $\Delta_m = 0$.  Indeed, $F$-maximization in non-chiral theories is consistent with this observation.

Since the R-symmetry can mix with any other abelian global symmetry, it would be interesting to ask how many such global symmetries there are for a given quiver.  We will be interested in quivers with gauge group $U(N)^p$ as well as quivers with gauge group $SU(N)^p \times U(1)$, where the second factor is the diagonal $U(1)$ in $U(N)^p$.  If there is no superpotential, we can show that the number of abelian flavor symmetries equals the number of matter representations regardless of which choice of gauge group.  Indeed, if all the gauge groups are $SU(N)$, then for each matter field there is a $U(1)$ global symmetry that acts by multiplying that field by a phase.   Replacing some of the $SU(N)$ gauge groups by $U(N)$ gauges some of these $U(1)$ symmetries.  However, for each new $U(1)$ gauge symmetry there is now an additional topological conserved current $j_{\rm top} = *F$ in addition to the old conserved current $j_{\rm matter}$. For Chern-Simons level $k$ the $U(1)$ gauge field now couples to $j_{\rm matter} + k j_{\rm top}$.  Going from $SU(N)$ to $U(N)$ gauge theory therefore introduces a new topological $U(1)$ symmetry and gauges a linear combination of this $U(1)$ and the diagonal $U(1)$ in $U(N)$.  Thus, the total number of global symmetries does not change and stays equal to the number of matter representations for any of choice of gauge group.  A non-trivial superpotential will generically break some of these flavor symmetries.

%The number of abelian flavor symmetries is always simply given by the number of matter representations. This is obvious if the gauge group is $SU(N)$, since then the $U(1)$ part of the flavor symmetry is not gauged.   If the gauge group contains some of the $U(1)$ factors, then the monopole flavor symmetry, associated to the topological current $*\tr F$, replaces the gauged symmetry. In the presence of a non-zero Chern-Simons term, the abelian current that is gauged is in fact $j_{\rm matter} + k j_{\rm top}$. In any case, the total number of abelian symmetries is always the number of matter representations.

\subsection{The forces on the eigenvalues}

Suppose we have a quiver with nodes $1, 2, \ldots, p$ with $U(N)$ gauge groups and CS levels $k_a$.  Let's denote the eigenvalues corresponding to the $a$th node by $\lambda^{(a)}_i$, with $i = 1, 2, \ldots, N$.  In the saddle point approximation the force acting on $\lambda^{(a)}_i$ can be split into several pieces:
 \es{SplitForce}{
  F_i^{(a)} =  F_{i, \text{ext}}^{(a)} + F_{i, \text{self}}^{(a)}  + \sum_b F_{i, \text{inter}}^{(a, b)}  + \sum_b F_{i, \text{inter}}^{(b, a)} \,.
 }
The first term is the external force
 \es{External}{
  F_{i, \text{ext}}^{(a)} =  \frac{i k_a}{2 \pi}\lambda_i^{(a)}  - \Delta_m^{(a)} \,,
 }
where $\Delta_m^{(a)}$ is the corresponding bare monopole R-charge.  The second term is due to interactions with eigenvalues belonging to the same node:
 \es{Fself}{
  F_{i, \text{self}}^{(a)} = \sum_{j \neq i} \coth \frac{\lambda^{(a)}_i - \lambda^{(a)}_j}{2} \,.
 }
Finally, the last two terms in eq.~\eqref{SplitForce} correspond to contributions of bifundamental fields $(a, b)$ that transform in the fundamental representation of node $a$ and the anti-fundamental representation of node $b$.  We have
 \es{FinterFundamental}{
 F_{i, \text{inter}}^{(a, b)} =\sum_j \left[ \frac {\Delta_{(a, b)} - 1}{2} - i \frac{\lambda^{(a)}_i - \lambda^{(b)}_j}{4 \pi} \right]
   \coth\left[\frac{\lambda^{(a)}_i - \lambda^{(b)}_j}{2} - i \pi \left(1 - \Delta_{(a, b)}\right) \right] \,,
 }
 \es{Finteranti-fundamental}{
 F_{i, \text{inter}}^{(b, a)} =\sum_j \left[ \frac {\Delta_{(b, a)} - 1}{2} + i \frac{\lambda^{(a)}_i - \lambda^{(b)}_j}{4 \pi} \right]
   \coth\left[\frac{\lambda^{(a)}_i - \lambda^{(b)}_j}{2} + i \pi \left(1 - \Delta_{(b, a)}\right) \right] \,.
 }

We can split the interaction forces between the eigenvalues into long-range forces and short-range forces.  We define the long-range forces to be those forces obtained by replacing $\coth(u)$ with its large $u$ approximation, $\sgn \Re(u)$.  Since $\sgn \Re (\alpha u) = \sgn \Re (u)$ if $\alpha>0$, we have
 \es{LongRangeSelf}{
  F_{i, \text{self}}^{(a)} &\approx \hat F_{i, \text{self}}^{(a)}
    = \sum_{j \neq i} \sgn \Re \left(\lambda^{(a)}_i - \lambda^{(a)}_j \right) \,, \\
  F_{i, \text{inter}}^{(a, b)} &\approx \hat F_{i, \text{inter}}^{(a, b)}
   = \sum_j \left[ \frac {\Delta_{(a, b)} - 1}{2} - i \frac{\lambda^{(a)}_i - \lambda^{(b)}_j}{4 \pi} \right]
   \sgn \Re \left(\lambda^{(a)}_i - \lambda^{(b)}_j\right) \,, \\
  F_{i, \text{inter}}^{(b, a)} &\approx
   \hat F_{i, \text{inter}}^{(b, a)}
    = \sum_j \left[ \frac {\Delta_{(b, a)} - 1}{2} + i \frac{\lambda^{(a)}_i - \lambda^{(b)}_j}{4 \pi} \right]
   \sgn \Re \left(\lambda^{(a)}_i - \lambda^{(b)}_j\right)  \,.
 }

\subsection{General rules for matrix models with no long-range forces}
\label{RULES}

We want to study quiver gauge theories with free energies that scale as $N^{3/2}$ in the large $N$ limit, because these theories are thought to have M-theory duals.  One way of achieving this is for the real part of the eigenvalues to scale as $N^{1/2}$ and the imaginary parts to stay order $N^0$ in the large $N$ limit (see appendix~\ref{DERIVATION} for more details).  A necessary condition for this scaling is that the long range forces must vanish at the saddle point of the matrix integral.

A large class of such theories are quiver gauge theories with non-chiral bifundamental superfields, meaning that for each ${\cal N} =2$ chiral superfield $X_{(a, b)}$ transforming in  $(\bf N, \bf \overline N)$ of the gauge groups $U(N)_a \times U(N)_b$ there exists another chiral superfield $X_{(b, a)}$ transforming in $(\bf \overline N, \bf N)$.  The two fields $X_{(a, b)}$ and $X_{(b, a)}$ need not be related by supersymmetry, and thus their R-charges $\Delta_{(a, b)}$ and $\Delta_{(b, a)}$ need not be equal.  In addition to these bifundamental fields we will also allow for equal numbers of fundamental and anti-fundamental fields.\footnote{Equal in total number; the number of fundamental and anti-fundamental fields charged under a given gauge group are allowed to differ.} The kinetic terms for the vector multiplets could be either Chern-Simons with level $k_a$ or Yang-Mills.  Additionally we require
 \es{ChernSimonsSum}{
  \sum_a k_a = 0 \,.
 }

For such theories the condition that the long-range forces \eqref{LongRangeSelf} vanish is equivalent to
 \es{DeltaSum}{
  \sum \Delta_{(a, b)}  + \sum \Delta_{(b, a)} = n_{(a)} - 2
 }
for each node $a$, where the sum is taken over all the bifundamental fields transforming non-trivially under $U(N)_a$, and $n_a$ denotes the number of such fields (adjoint fields are supposed to be counted twice:  once as part of the first sum and once as part of the second sum).

With these assumptions, it is consistent to assume that in the large $N$ limit the eigenvalues $\lambda_i^{(a)}$ behave as \cite{Herzog:2010hf} (see also \cite{Suyama:2009pd})
 \es{LargeNScalingHalf}{
  \lambda_i^{(a)} = N^{1/2} x_i + i y_{a, i} + o(N^0).
 }
As we take $N$ to infinity, we can replace $x_i$ and $y_{a, i}$ by continuous functions $x(s)$ and $y_a(s)$ such that $x_i = x(i/N)$ and $y_{a, i} = y_a (i/N)$.  In the following discussion, it will be useful to consider the density
 \es{density}{
  \rho(x) = \frac{ds}{dx}
 }
and express the imaginary parts of the eigenvalues as functions $y_a(x)$.

That the long-range forces \eqref{LongRangeSelf} vanish implies that the free energy functional is local.  Here are the rules for constructing the free energy functional for any ${\cal N} = 2$ quiver theory that satisfies the conditions described above:
 \begin{enumerate}
  \item For each gauge group $a$ with CS level $k_a$ and bare monopole R-charge $\Delta_m^{(a)}$ one should add the term
   \es{CSContribution}{
    \frac{k_a}{2 \pi} N^{3/2} \int dx\, \rho(x) x y_a(x)
     + \Delta_m^{(a)} N^{3/2} \int dx\, \rho(x) x \,.
   }
  \item
     For a pair of bifundamental fields, one of R-charge $\Delta_{(a, b)}$ transforming in the $(\bf N, \bf \overline N)$ of $U(N)_a \times U(N)_b$ and one of R-charge $\Delta_{(b, a)}$ transforming in the $(\bf \overline N, \bf N)$ of $U(N)_a \times U(N)_b$, one should add
 \es{Pair}{
    -N^{3/2}\frac{ 2- \Delta_{(a, b)}^+}{2} \int dx\, \rho(x)^2
      \left[ \left(y_a-y_b + \pi \Delta_{(a, b)}^- \right)^2
      - \frac 13 {\pi^2 \Delta_{(a, b)}^+  \left(4-\Delta_{(a, b)}^+\right)} \right] \,,
 }
where $\Delta_{(a, b)}^{\pm} \equiv \Delta_{(a, b)} \pm \Delta_{(b, a)}$ satisfies $\Delta_{(a, b)}^+ < 2$, and $y_a - y_b$ is in the range
 \es{Range}{
   \abs{y_a - y_b + \pi \Delta_{(a, b)}^-} \leq \pi \Delta_{(a, b)}^+ \,.
 }
Outside this range the formula \eqref{Pair} is no longer valid, and in fact for arbitrary $y_a - y_b$ the integrand is a non-smooth function.  The boundaries of the range \eqref{Range} are points where the integrand should be considered to be non-differentiable.  In practice, this means that the equations obtained from varying the free energy functional with respect to $y_a - y_b$ need not hold whenever $\abs{y_a - y_b + \pi \Delta_{(a, b)}^-} = \pm \pi \Delta_{(a, b)}^+$.

 \item For an adjoint field of R-charge $\Delta_{(a, a)}$, one should add
 \es{Adjoint}{
    \frac {2\pi^2}3 N^{3/2} \Delta_{(a, a)}  \left( 1- \Delta_{(a, a)}  \right) 
        \left(2-\Delta_{(a, a)}\right)  \int dx\, \rho(x)^2 \,.
 }

   \item For a field $X_a$ with R-charge $\Delta_a$ transforming in the fundamental of $U(N)_a$, one should add
    \es{FundContribution}{
     N^{3/2} \int dx\, \rho(x) \abs{x} \left( \frac{1- \Delta_a}{2} -\frac{1}{4 \pi}  y_a(x) \right) \,,
    }
  while for an anti-fundamental field of R-charge $ \tilde \Delta_a$ one should add
    \es{AntifundContribution}{
     N^{3/2} \int dx\, \rho(x) \abs{x} \left( \frac{1- \tilde \Delta_a}{2} +\frac{1}{4 \pi}  y_a(x) \right) \,.
    }

  \end{enumerate}

\subsection{Flat directions and $U(N)$ vs.~$SU(N)$}
\label{SYMMETRY}

In a theory with $p$ $U(N)$ gauge groups, the matrix integral \eqref{PartitionFunction}, seen as a function of the R-charges of the matter fields as well as the bare monopole R-charges $\Delta_m^{(a)}$, has the following symmetries parameterized by $p$ real numbers $\delta^{(a)}$:
 \es{Symmetries}{
  \text{eigenvalues $\lambda_i^{(a)}$ for $a$th gauge group:} \qquad &
   \lambda_i^{(a)} \to \lambda_i^{(a)} - 2 \pi i \delta^{(a)} \,, \\
  \text{$U(N)_a \times U(N)_b$ bifundamental of R-charge $\Delta_{(a, b)}$:} \qquad &
   \Delta_{(a, b)} \to \Delta_{(a, b)} + \delta^{(a)} - \delta^{(b)} \,,\\
  \text{$U(N)_a$ fundamental of R-charge $\Delta_a$:} \qquad &
   \Delta_a \to \Delta_a + \delta^{(a)} \,, \\
  \text{$U(N)_a$ anti-fundamental of R-charge $\tilde \Delta_a$:} \qquad &
   \tilde \Delta_a \to \tilde \Delta_a - \delta^{(a)} \,, \\
  \text{bare monopole R-charge $\Delta_m^{(a)}$ for $a$th gauge group:} \qquad &
   \Delta_m^{(a)} \to \Delta_m^{(a)} + k_a \delta^{(a)} \,.
 }
The transformations \eqref{Symmetries} leave the matrix integral \eqref{PartitionFunction} invariant (up to a phase) because they are equivalent to a change of variables where the integration contour for each set of eigenvalues is shifted by a constant amount.  

A consequence of this symmetry is that at finite $N$ the free energy $F(\Delta)$ (namely the extremum of the free energy functional for fixed R-charges $\Delta$) has $p$ flat directions parameterized by $\delta^{(a)}$.  In the $U(N)^p$ theory, these flat directions are to be expected because, for example, the bifundamental fields in the theory are not gauge-invariant operators.
 %and assigning them R-charges would be meaningless.
 The free energy should depend only on the R-charges of gauge-invariant operators.  One then has two options:  work with the $U(N)^p$ theory where by maximizing $F$ one can only determine the R-charges of composite gauge-invariant operators (for example, $\Delta_{(a, b)} + \Delta_{(b, a)}$ would be a well-defined number since it is the R-charge of the gauge-invariant operator $\tr X_{(a, b)} X_{(b, a)}$), or ungauge some of the diagonal $U(1)_a$'s in the $U(N)_a$ gauge groups.  If the $U(N)_a$ gauge group is replaced by $SU(N)_a$ then the corresponding eigenvalues $\lambda_i^{(a)}$ should satisfy the tracelessness condition
 \es{Tracelessness}{
  \sum_i \lambda_i^{(a)} = 0 \,.
 }
This condition fixes $\delta^{(a)}$ and removes a flat direction from $F$.

In the large $N$ limit the free energy will generically have more flat directions than at finite $N$.  For example, at large $N$ there are an additional $n_f$ flat directions coming from the flavors for the following reason.  In the theories we consider each fundamental field is paired with an anti-fundamental field.  Let the R-charge of one of the fundamental fields be $\Delta_f$ and the R-charge of the corresponding anti-fundamental field be $\tilde \Delta_f$.  At finite $N$ the sum $\Delta_f + \tilde \Delta_f$ is fixed by the marginality of the superpotential, leaving one free R-charge.  However the finite $N$ free energy will typically be a non-trivial function of both $\Delta_f$ and $\tilde \Delta_f$.  At large $N$, on the other hand, the free energy really only depends on the sum $\Delta_f + \tilde \Delta_f$, as can be seen from eqs.~\eqref{FundContribution} and \eqref{AntifundContribution}.  This gives us an additional $n_f$ ``accidental" flat directions at large $N$.   

Looking at equation~\eqref{CSContribution}, one can see that at large $N$ the free energy only depends on the sum 
 \es{DeltamDef}{
  \Delta_m = \sum_{a=1}^p \Delta_m^{(a)}.
 }
 Naively one would think this gives us $p-1$ additional flat directions corresponding to shifts in the individual $\Delta_m^{(a)}$, which leave the sum in equation~\eqref{DeltamDef} invariant.  However, in theories where  $\sum_a k_a = 0$, which are all the theories presented in this paper except for the theories in section~\ref{FIVETHIRDS}, the story is slightly more subtle.  In these theories we actually only gain $p-2$ additional flat directions in the large $N$ limit.  This is because at order  ${\cal O}(N^{3/2})$ the symmetry corresponding to $\delta^{(a)} = \delta$ is equivalent to one of the ``new" flat directions, which correspond to symmetries of the sum~\eqref{DeltamDef}.  
% The reasons why shifting all $\delta^{(a)}$ by the same amount is equivalent to one of the ``new" large $N$ symmetries of equation~\ref{DeltamDef}  come from the following facts:
% \begin{enumerate}
%  \item  The R-charges of the bifundamental fields are invariant under this shift.
%  \item  While the individual bare monopole R-charges $\Delta_m^{(a)}$ transform under this shift, the sum in equation~\eqref{DeltamDef} is invariant provided that $\sum_a k_a = 0$. From \eqref{CSContribution} it follows that to order ${\cal O}(N^{3/2})$ the free energy depends only on $\Delta_m$ and not on the individual $\Delta_m^{(a)}$.
% \item While the R-charges of the various fundamental or anti-fundamental fields get shifted by $\pm \delta$, the sum of these shifts vanishes provided we have an equal total number of fundamental and anti-fundamental fields.  From \eqref{FundContribution}--\eqref{AntifundContribution} one can see that it is only the sum of the R-charges of the flavor fields that enters the ${\cal O}(N^{3/2})$ expression for the free energy.
% \end{enumerate}   
To summarize, at large $N$ we have a total of $2 (p-1) + n_f$ flat directions of the free energy.  However, only $p$ of these flat directions correspond to gauge symmetries.  The other $p - 2 + n_f$ flat directions are only there at infinite $N$.   
 
 One could choose to eliminate some of the flat directions in the free energy by changing the gauge groups from $U(N)$ to $SU(N)$.  Since the diagonal monopole operators $T$ and $\tilde T$ are essential in obtaining the quantum-corrected moduli space in these theories, we would like to keep the diagonal $U(1)$ in $U(N)^p$ as a gauge symmetry.  So, let's choose to eliminate all the flat directions in the free energy coming from the abelian gauge symmetries, except for the flat direction corresponding to this diagonal $U(1)$.  The R-charges of the (bi)fundamental fields are then gauge invariant quantities, as we will explain below.  The residual abelian gauge symmetry gives us $p-1$ gauge invariant combinations of the $p$ bare monopole R-charges $\Delta_m^{(a)}$.  However, at large $N$ we will only be able to compute the sums $\Delta_m$ and $\Delta_f + \tilde \Delta_f$ because of the accidental flat directions.
 
   In going from $U(N)^p$ to $SU(N)^p \times U(1)$ we should regard
 \es{ApludDef}{
  A_+ = \sum_{a=1}^p \tr A_{a}
 }
as a dynamical gauge field, while the other gauge fields $A^{(b)} = \sum_a \alpha_{a}^b \tr A_a$, where $\alpha_a^b$ is a basis of solutions to $\sum_{a=1}^p \alpha_a^b = 0$, should be treated as background fields that we set to zero.  The ungauging procedure \cite{Witten:2003ya} can be done rigorously by adding $p-1$ vector multiplets whose vector components are $B_b$, $b = 1, 2, \ldots, p-1$, and that couple to the topological currents $*F^{(b)} = *\sum_{a=1}^p \alpha_a^b \tr F_a$ through
 \es{BCoupling}{
  \delta S = \sum_{b=1}^{p-1} \int B_b \wedge F^{(b)}  \,,
 }
with an appropriate supersymmetric completion.  Making the fields $B_b$ dynamical, the integration over them in the path integral essentially ungauges $A^{(b)}$.  For related discussions, see \cite{Imamura:2008ji, Imamura:2009ur, Klebanov:2010tj, Benishti:2010jn}.  

To summarize so far, in the $U(N)^p$ gauge theory the large $N$ free energy $F(\Delta)$ generically has $2 p - 2 +n_f$ flat directions.  However, only $p$ of these symmetries correspond to gauge symmetries with the rest being accidental flat direction appearing only at large $N$.  If we want to remove $p-1$ of the flat directions corresponding to gauge symmetries, we should consider the $SU(N)^p \times U(1)$ gauge theory, where the $U(1)$ gauge field is $A_+$.  In this theory one can construct the baryonic operator
\es{baryonGeneral}{
\mathcal{B} \left(X_{(a, b)}\right) = \epsilon_{i_1 \cdots i_N} \epsilon^{j_1 \cdots j_N} \left( X_{(a, b)} \right)^{i_1}_{j_1}
 \cdots \left( X_{(a, b)} \right)^{i_N}_{j_N} \, ,
}
which is a gauge-invariant chiral primary with R-charge $N \Delta_{(a, b)}$.  In other words, the operator $X_{(a, b)}$ can be assigned a unique R-charge $\Delta_{(a, b)}$ because the baryon ${\cal B}  \left(X_{(a, b)}\right)$ has the well-defined R-charge $N \Delta_{(a, b)}$.  Minimizing $F(\Delta)$ in this theory one can then determine the R-charges of the bifundamenal fields.

Ungauging the $p-1$ off-diagonal $U(1)$ gauge fields makes it possible to define gauge-invariant baryonic operators at the expense of removing from the chiral ring the off-diagonal monopole operators that generate non-zero numbers of $F^{(b)}$ flux units that exist in the $U(N)^p$ theory.  This ungauging doesn't remove, however, the diagonal monopole operators $T$ and $\tilde T$, because these operators generate equal numbers of $\tr F_a$ flux units and thus no $F^{(b)}$ flux units.  Moreover, the bare monopole R-charges $\Delta_m$ of $T$  and $-\Delta_m$ of $\tilde T$ are well-defined quantities because one can construct a baryonic-like operator out of $T$ or $\tilde T$.

From an AdS/CFT point of view,  ungauging $U(1)$ symmetries in the boundary theory is equivalent to changing boundary conditions in the bulk for the bulk gauge fields dual to those $U(1)$ symmetries.  In M-theory, the boundary conditions corresponding to the $U(N)^p$ gauge theory allow the existence of M2-branes wrapping topologically non-trivial two-cycles, but disallow the existence of the magnetic dual objects, which would be the M5-branes wrapping the dual five-cycles.  The boundary conditions for the $SU(N)^p \times U(1)$ gauge theory allow the existence of M5-branes but disallow M2-branes wrapped on topologically non-trivial cycles.  Since these wrapped M2-branes are dual to off-diagonal monopole operators and the M5-branes wrapping topologically non-trivial cycles are dual to baryonic operators, the general picture on the gravity side is consistent with the field theory analysis.  See \cite{Klebanov:2010tj, Benishti:2010jn} for a more detailed discussion.

In addition to M5-branes wrapping topologically non-trivial cycles that are allowed only in the $SU(N)^p \times U(1)$ gauge theory, on the gravity side one can also consider giant gravitons, which are BPS configurations of M5-branes wrapping topologically trivial five-cycles and rotating within the 7-d space $Y$ \cite{McGreevy:2000cw}.  On the field theory side, these objects are thought to be dual to determinants of operators that transform in the adjoint representation of one of the gauge groups (such as determinants of products of bifundamental fields).  These determinant operators are gauge invariant in both the $U(N)^p$ and $SU(N)^p \times U(1)$ gauge theory.

In general, the relation between the volume of a five-cycle wrapped by an M5-brane and the dimension of the corresponding gauge theory operator is \cite{Fabbri:1999hw}
\es{DeltaBaryon}{
  \Delta = \frac{\pi N}{6} \frac{\Vol(\Sigma_5)}{\Vol(Y)} \,,
 }
regardless of whether the five-cycle the brane is wrapping is topologically trivial or not.  We will make extensive use of this formula, as it provides a way of extracting the expected R-charge of the bifundamental fields (or of certain products of bifundamental fields) from the gravity side.  Indeed, after performing $F$-maximization, we check not only that the extremum of $F$ matches the supergravity prediction \eqref{MtheoryExpectation} computed using the volume of $Y$, but also that the dimensions of the operators dual to wrapped M5-branes agree with eq.~\eqref{DeltaBaryon}, which involves the volumes of the various five-cycles computed from the gravity side.

\section{A class of ${\cal N} = 2$ necklace quivers}
\label{ADJOINTS}

The first class of quiver gauge theories where we apply the formalism developed in the previous section involves a modification of the ``necklace'' ${\cal N} = 3$ Chern-Simons theories proposed in \cite{Jafferis:2008qz, Imamura:2008nn} and examined in a similar context in \cite{Herzog:2010hf}.  The ${\cal N} = 3$ quivers involve $p$ gauge groups with CS levels $k_a$ that satisfy $\sum_{a=1}^p k_a = 0$ as well as bifundamental chiral superfields $A_{a, a+1}$ and $B_{a+1, a}$.
\begin{figure}[htb]
\begin{center}
\leavevmode
\newcommand{\svgwidth}{.7\textwidth}
\definecolor{orange}{RGB}{255,165,0}
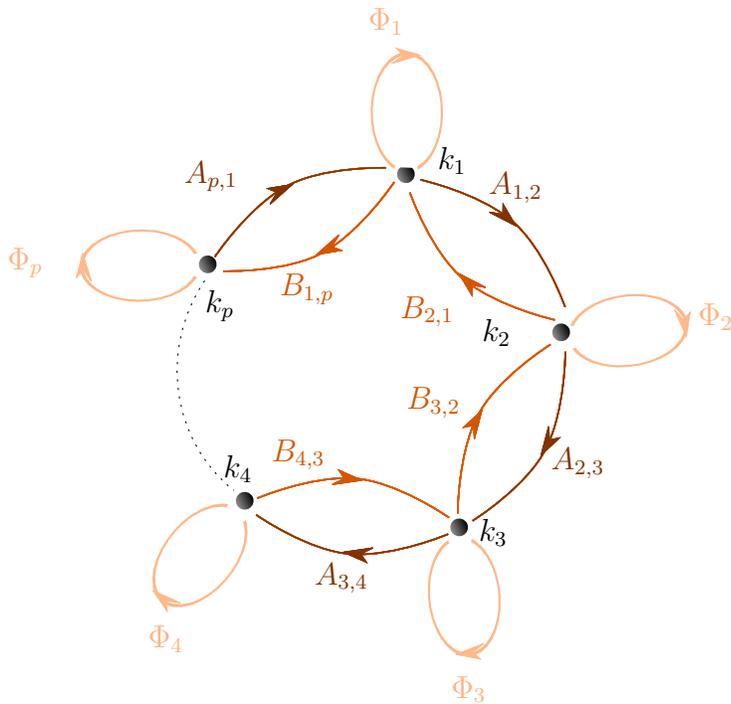
\end{center}
\caption{A ``necklace'' quiver diagram for the ${\cal N} = 3$ Chern-Simons-matter gauge theories with superpotential \eqref{WN2} or the ${\cal N} = 2$ CS-matter gauge theories with superpotential \eqref{WN3}.  We impose the condition that the CS levels $k_a$ should sum to zero.}
\label{NAQUIV}
\end{figure}
These theories are natural generalizations of the ABJM model, which corresponds to $p=2$. They have quartic superpotentials $W \sim \sum_a {1\over k_a} \tr (A_{a, a+1} B_{a+1, a} - B_{a, a-1} A_{a-1, a})$.  An equivalent description of these theories involves extra adjoint chiral multiplets $\Phi_a$ and the superpotential
 \es{WN2}{
   W_{{\cal N} = 3} \sim \sum_a \tr \left(k_a \Phi_a^2 + \Phi_a (A_{a, a+1} B_{a+1, a} - B_{a, a-1} A_{a-1, a}) \right)
  }
(see figure~\ref{NAQUIV}).  That the two descriptions are equivalent can be seen by simply integrating out the fields $\Phi_a$.  If one now changes the superpotential to\footnote{I.R.K. thanks M. Kiermaier and T. Klose for earlier discussions of these models.}
 \es{WN3}{
   W_{{\cal N} = 2} \sim \sum_a \tr \left(\mu_a \Phi_a^3 + \Phi_a (A_{a, a+1} B_{a+1, a} - B_{a, a-1} A_{a-1, a}) \right) \,,
  }
for some set of parameters $\mu_a$, the resulting theories have only ${\cal N} = 2$ supersymmetry.\footnote{In the two-node case this model is equivalent to the Martelli-Sparks proposal for the dual of $AdS_4\times V_{5,2}$ \cite{Martelli:2009ga}.}  If we perturb such an ${\cal N}=2$ fixed point by the relevant superpotential deformation $\delta W= \sum_a \tr \left(k_a \Phi_a^2 \right)$ then it should flow to the corresponding ${\cal N}=3$ theory.

To keep the discussion as general as possible, let us consider the class of superpotentials
 \es{SuperpotNecklace}{
  W \sim \sum_a\tr \left[ \mu_a \Phi_a^{n+1} +  \Phi_a (A_{a, a+1} B_{a+1, a} - B_{a, a-1} A_{a-1, a}) \right) \,,
 }
where we assume that all the parameters $\mu_a$ are non-vanishing.  If $n=1$ or $2$, this theory is dual to $AdS_4 \times Y_n(\vec{k})$. The spaces $Y_n(\vec{k})$ probably have a Sasaki-Einstein metric only when $n \leq 2$, though, because of the Lichnerowicz obstruction of \cite{Gauntlett:2006vf,Martelli:2009ga}.  Let us denote by $\Delta_A$ and $\Delta_B$ the conformal dimensions of the bifundamental fields $A_a$ and $B_a$, respectively, and by $\delta$ the conformal dimensions of the adjoints $\Phi_a$.  The condition that the superpotential is marginal implies
 \es{MarginalityNecklace}{
  \delta = 2/(n+1) \,, \qquad \Delta_+ \equiv \Delta_A + \Delta_B = 2n/(n+1) \,.
 }
 Setting the bare monopole R-charge $\Delta_m = 0$,\footnote{ If one includes a non-zero $\Delta_m$ in the free energy, $F$-maximization requires the bare monopole R-charge to vanish.} eqs.~\eqref{CSContribution}--\eqref{Pair} then imply that the free energy functional is
 \es{FreeNecklace}{
    F_n[\rho, y_a] &= \sum_{a=1}^p \frac{k_a}{2 \pi} N^{3/2} \int dx\, \rho x y_a
      + \frac {2\pi^2 p}{3} N^{3/2}\delta (\delta - 1) (\delta-2) \int dx\, \rho^2  \\
       {}&-N^{3/2}\frac{ 2- \Delta_+}{2} \sum_{a=1}^p \int dx\, \rho^2
      \left[ (y_a-y_{a-1} + \pi \Delta_-)^2
      - \frac{\pi^2 \Delta_+ (4-\Delta_+)}{3} \right] \,,
 }
with $\Delta_- \equiv \Delta_A - \Delta_B$.  Using \eqref{MarginalityNecklace}, this equation can be simplified to
 \es{FreeNecklaceSimp}{
  F_n[\rho, y_a] &=  \sum_{a=1}^p \frac{k_a}{2 \pi} N^{3/2} \int dx\, \rho x y_a
      - N^{3/2}  \sum_{a=1}^p \int dx\, \rho^2 \left[\frac{ (y_a-y_{a-1} + \pi \Delta_-)^2}{n+1} - \frac{4 \pi^2 n^2}{(n+1)^3} \right] \,.
 }
As discussed after eq.~\eqref{Pair}, this expression holds as long as $\abs{y_a-y_{a-1} + \pi \Delta_-} \leq \pi \Delta_+ = 2 \pi n / (n+1)$.

Since the quiver is symmetric under interchanging the $A$ fields with the $B$ fields, we expect that the saddle point has $\Delta_A = \Delta_B$, so $\Delta_- = 0$.  In this case, we can absorb the dependence on $n$ into a redefinition of $y_a$ and $\rho$.  By writing
 \es{Scalings}{
  y_a = \frac{2n}{n+1} \hat y_a \,, \qquad \rho \to \frac{n+1}{2 \sqrt{n}} \hat \rho \,,
   \qquad x \to \frac{2 \sqrt{n}}{n+1} \hat x \,,
 }
one can easily show that
 \es{FScaled}{
  F_n[\rho, y_a] = \frac{4 n^{3/2}}{(n+1)^2} F_1[\hat \rho, \hat y_a] \,.
 }
Clearly, this relation is also satisfied by the extrema $F_n$ and $F_1$ of the functionals $F_n[\rho, y_a]$ and $F_1[\hat \rho, \hat y_a]$, respectively, which given \eqref{MtheoryExpectation} implies
 \es{VolRelation}{
  \Vol(Y_n(\vec{k})) = \frac{(n+1)^4}{16 n^3} \Vol(Y_1(\vec{k})) \,.
 }
In particular, we have
 \es{VolRelationPart}{
  \Vol(Y_2(\vec{k})) = \frac{81}{128} \Vol(Y_1(\vec{k})) \,.
 }
When $\vec{k} = (1, -1)$ then $Y_1(\vec{k}) = S^7$ with volume $\Vol(S^7) = \pi^4/3$ and $Y_2(\vec{k}) = V_{5, 2}$ \cite{Martelli:2009ga} with volume $\Vol(V_{5, 2}) = 27 \pi^4/128$ \cite{Bergman:2001qi}, in agreement with eq.~\eqref{VolRelationPart}.

We have just shown that for the RG flow between the ${\cal N} = 2$ theory \eqref{WN2} in the UV deformed by the relevant superpotential deformation $\delta W= \sum_a \tr \left(k_a \Phi_a^2 \right)$ and the ${\cal N} = 3$ theory \eqref{WN3} in the IR, we have $(F_{\rm IR}/F_{\rm UV})^2= 81/128$.  The universal ratio $81/128$ is reminiscent of the $a_{\rm IR} / a_{\rm UV}= 27/32$ that often arises in $(3+1)$-dimensional RG flows; see \cite{Tachikawa:2009tt} for a general argument.

\section{Flavored gauge theories with one gauge group}
\label{FLAVORC3}

The first examples we consider are flavored variations of the 3-d ${\cal N} = 8$ Yang-Mills theory, which can be obtained as the dimensional reduction of the ${\cal N} = 4$ gauge theory in four dimensions.  In ${\cal N} = 2$ notation, the 3-d ${\cal N} = 8$ vector multiplet consists of an ${\cal N} = 2$ vector multiplet with gauge group $U(N)$ or $SU(N)$ as well as three adjoint chiral superfields $X_i$, $1 \leq i \leq 3$.  The superpotential
 \es{W0}{
  W_0 = \tr X_1 [X_2, X_3]
 }
ensures that the long-range forces between the eigenvalues vanish, because the requirement that the superpotential is marginal is equivalent to eq.~\eqref{DeltaSum}.  The flavoring of this model consists of adding fields $q_\alpha$ and $\tilde q_\alpha$ transforming in the anti-fundamental and fundamental representations of the gauge group, respectively, coupled to the adjoints $X_i$ through the superpotential coupling
 \es{Superpot}{
  \sum_\alpha q_\alpha {\cal O}_\alpha(X_i) \tilde q_\alpha \,.
 }
Here, ${\cal O}_\alpha(X_i)$ are polynomials in the $X_i$ with no constant term, which, as operators, also transform in the adjoint representation of the gauge group.  It was conjectured in \cite{Jafferis:2009th, Benini:2009qs} that the $U(1)$ quantum corrected moduli space in this case can be described as the embedded codimension one surface
 \es{MSOneGeneral}{
  T \tilde T = \prod_\alpha {\cal O}_\alpha(X_i)
 }
in $\C^5$, where the monopole operators $T$ and $\tilde T$ as well as the three fields $X_i$ should be regarded as the five complex coordinates in $\C^5$.  This moduli space is a Calabi-Yau space with a conical singularity at $T = \tilde T = X_i = 0$.  The field theory we just described is then conjectured to be the theory on M2-branes placed at the tip of the Calabi-Yau cone \eqref{MSOneGeneral} and is therefore dual to $AdS_4 \times Y$, where $Y$ is the Sasaki-Einstein base of the cone \eqref{MSOneGeneral}.

\subsection{An infinite family of $AdS_4$/CFT$_3$ duals}
\label{TORICONE}

\begin{figure}[htb]
\begin{center}
\leavevmode
\newcommand{\svgwidth}{.3\textwidth}
\definecolor{orange}{RGB}{255,165,0}
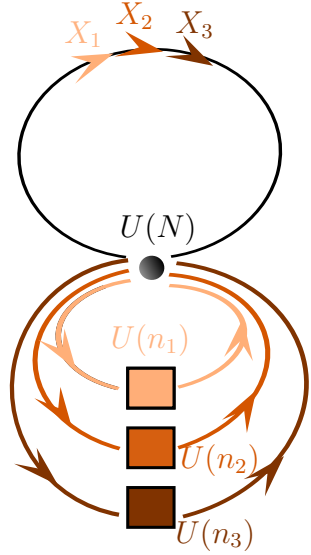
\end{center}
\caption{The quiver diagram for the flavored theories corresponding to the superpotential in equation~\eqref{FlavoredSuperpot}.   }
\label{C3QUIV}
\end{figure}

Let's first couple the basic model with superpotential \eqref{W0} to three sets of pairs of chiral superfields $\left(q_j^{(a)}, \tilde q_j^{(a)}\right)$, where $i = 1, 2, 3$ and $j = 1, 2, \ldots, n_i$ for some integers $n_i \geq 0$ with at least one of the $n_i$ being strictly positive.  The quiver diagram for this theory is shown in figure~\ref{C3QUIV}.
The superpotential of the flavored theory is
\es{FlavoredSuperpot}{
  W \sim W_0 + \tr \left[ \sum_{j=1}^{n_1} q_j^{(1)} X_1 \tilde q_j^{(1)}
    + \sum_{j=1}^{n_2} q_j^{(2)} X_2 \tilde q_j^{(2)}
    + \sum_{j=1}^{n_3} q_j^{(3)} X_3 \tilde q_j^{(3)}\right] \,.
 }
These theories were considered in detail in \cite{Benini:2009qs} where it was shown that for each such theory the quantum corrected moduli space of vacua is a toric Calabi-Yau cone.  This cone can be parameterized by the complex coordinates $X_i$ as well as the monopole operators $T$ and $\tilde T$ subject to the constraint
 \es{TTOPE}{
  T \tilde T = X_1^{n_1} X_2^{n_2} X_3^{n_3} \,.
 }
The fact that the superpotential should have R-charge $R[W] = 2$ as well as the constraint \eqref{TTOPE} imposes a number of constraints on the R-charges of the various fields:
 \es{RChargeConstraints}{
  \sum_{i = 1}^3 R[X_i] = 2 \,, \qquad R[T] + R[\tilde T] = \sum_{i = 1}^3 n_i R[X_i] \,, \qquad
   R[q_j^{(i)}] + R[\tilde q_j^{(i)}] + R[X_i]  = 2 \,.
 }
With these assumptions, the rules of section \ref{RULES} imply that the free energy functional is\footnote{In these non-chiral theories, $F$-maximization will give $\Delta_m = 0$ due to charge conjugation symmetry, but we will nevertheless keep $\Delta_m$ explicitly in the intermediate steps.}
 \es{FreeOneNode}{
  F[\rho] = 2 \pi^2 N^{3/2} \Delta_1 \Delta_2 \Delta_3  \int dx\, \rho^2
    + \frac{N^{3/2}}{2} \left( \sum_{i=1}^3 n_i \Delta_i \right) \int dx\, \rho \abs{x}
    + N^{3/2} \Delta_m \int dx\, \rho x \,,
 }
where we denoted $R[X_i] = \Delta_i$.  We also have $R[T] - R[\tilde T] = 2 \Delta_m$ (see eq.~\eqref{TTtilde}).

The eigenvalue density $\rho(x)$ that maximizes $F$ is supported on $[x_-, x_+]$ with $x_- < 0 < x_+$:
 \es{DensityOneNode}{
  \rho = \begin{cases}
    \displaystyle{\frac{\left( \sum_{i = 1}^3 n_i \Delta_i\right)  - 2 \Delta_m}{8 \pi^2 \Delta_1 \Delta_2 \Delta_3}}
      (x-x_-) & \text{if $x<0$} \,, \\[20pt]
    \displaystyle{\frac{\left( \sum_{i = 1}^3 n_i \Delta_i\right) + 2 \Delta_m}{8 \pi^2 \Delta_1 \Delta_2 \Delta_3}}
      (x_+ - x ) & \text{if $x \geq 0$} \,,
  \end{cases}
 }
where the endpoints of the distribution are such that $\rho$ is continuous at $x = 0$,
 \es{xpmDef}{
  x_\pm \equiv \pm \sqrt{\frac{8 \pi^2 \Delta_1 \Delta_2 \Delta_3 \left[\left( \sum_{i = 1}^3 n_i \Delta_i\right) \mp 2 \Delta_m \right]}
  {\left( \sum_{i = 1}^3 n_i \Delta_i\right) \left[\left( \sum_{i = 1}^3 n_i \Delta_i \right) \pm 2 \Delta_m \right]}
  } \,.
 }
Plugging these expressions into eq.~\eqref{FreeOneNode}, we find that the extremum of $F[\rho]$ at given $\Delta_i$ and $\Delta_m$ is given by
 \es{FExtOneNode}{
  F = \frac{2 \sqrt{2} \pi N^{3/2}}{3} \sqrt{\Delta_1 \Delta_2 \Delta_3 \left[ \left( \sum_{i = 1}^3 n_i \Delta_i \right)
    -  \frac{4 \Delta_m^2 }
    {\left( \sum_{i = 1}^3 n_i \Delta_i \right)} \right]} \,.
 }

In order to find $\Delta_i$ and $\Delta_m$, one just has to maximize $F$ under the constraint that $\sum_{i=1}^3 \Delta_i = 2$.  The maximization problem clearly implies that $\Delta_m = 0$, so
 \es{FExtFurther}{
  F = \frac{2 \sqrt{2} \pi N^{3/2}}{3} \sqrt{\Delta_1 \Delta_2 \Delta_3 \left( \sum_{i = 1}^3 n_i \Delta_i \right)} \,.
 }
Finding $\Delta_i$ requires solving a system of algebraic equations with no simple closed-form solutions.  However, in section \ref{pcases} we will examine a variety of special cases where closed-form solutions are available.

In section~\ref{TORIC} we will show using toric geometry techniques that the extremum of the free energy \eqref{FExtOneNode} matches with the gravity prediction based on the volume of the internal space $Y$ and eq.~\eqref{MtheoryExpectation}.  As we will explain, in toric geometry one finds the volume of a toric Sasaki-Einstein space $Y$ by extremizing a certain function of three variables.  By matching determinant operators in field theory to giant gravitons in gravity, we may write those three variables in terms of the three free R-charge parameters.  We show that combining this function with equation \eqref{MtheoryExpectation} exactly gives equation \eqref{FExtOneNode}, even for R-charges that don't extremize these functions.

\subsection{Particular cases}
\label{pcases}

\subsubsection{$\C^2 \times (\C^2 / \Z_{n_1})$}
\label{BMODEL}

It is instructive to examine particular cases of our general formula \eqref{FExtFurther}.  The first particular case we study is $n_2 = n_3 = 0$ with $n_1$ arbitrary.  The moduli space \eqref{TTOPE} is in this case $\C^2 \times (\C^2 / \Z_{n_1})$, where the $\Z_{n_1}$ is generated by $(z_3, z_4)  \sim \left(z_3 e^{2 \pi i/n_1}, z_4 e^{-2 \pi i/n_1}\right)$.  This theory should therefore be dual to $AdS_4 \times S^7 / \Z_{n_1}$, where the $\Z_{n_1}$ action on $S^7$ is that induced by the corresponding $\Z_{n_1}$ action on $\C^4$ \cite{Bashkirov:2010kz}.  Eq.~\eqref{FExtFurther} is extremized for $\Delta_1 = 1$ and $\Delta_2 = \Delta_3 = 1/2$, which, when combined with \eqref{MtheoryExpectation} gives
 \es{volOneField}{
  \Vol(Y) = \frac{\pi^4}{3 n_1} \,.
 }
Since the volume of the round seven-sphere is $\Vol(S^7) = \pi^4/3$, this formula is consistent with the expectation that the internal space $Y$ is a $\Z_{n_1}$ orbifold of $S^7$.  Indeed, it was argued in \cite{Bashkirov:2010kz} that there is a supersymmetry enhancement to maximal ${\cal N} = 8$ supersymmetry when $n_1 = 1$.

\subsubsection{$\text{CY}_3 \times \C$ theories}
\label{CY3THEORIES}

Consider $n_3 = 0$ with arbitrary $n_1$ and $n_2$.  The equation describing the moduli space reduces to $T\tilde T = X_1^{n_1} X_2^{n_2}$, which describes a toric $\text{CY}_3$ cone times $\C$, the complex coordinate in $\C$ being $X_3$.  Since the $\text{CY}_3$ is singular at $X_1 = X_2 = 0$, the space $\text{CY}_3 \times \C$ has non-isolated singularities and so does the base of this cone, the Sasaki-Einstein space $Y$.  These non-isolated singularities might be a reason to worry to what extent AdS/CFT results are applicable in this case, as additional states in M-theory might appear from these singularities.  As we will explain, the matrix model computation of the free energy matches the M-theory expectation \eqref{MtheoryExpectation} in spite of these potential problems.  The free energy \eqref{FExtFurther} is extremized by
 \es{DeltaCY3}{
  \Delta_1 = \frac{n_1 - 2 n_2 + \sqrt{n_1^2 + n_2^2 - n_1 n_2}}{2 (n_1 - n_2)} \,,
  \qquad
  \Delta_2 = \frac{n_2 - 2 n_1 +  \sqrt{n_1^2 + n_2^2 - n_1 n_2}}{2(n_2 - n_1)} \,,
   \qquad
   \Delta_3 = \frac 12\,,
 }
giving
 \es{FreeExtCY3}{
  F_{n_1, n_2} &= \frac{\pi N^{3/2}}{3 \sqrt{2}\abs{n_1-n_2}} \Bigg[
  \left(n_1 + n_2 + \sqrt{n_1^2 + n_2^2 - n_1 n_2} \right)\\
    &\times \left(n_1 - 2 n_2 + \sqrt{n_1^2 + n_2^2 - n_1 n_2} \right)
    \left(-2 n_1 + n_2 + \sqrt{n_1^2 + n_2^2 - n_1 n_2} \right) \Bigg]^{1/2} \,.
 }
Note that the field $X_3$ corresponding to the $\C$ factor in $\text{CY}_3\times \C$ has the canonical R-charge $\Delta_3 = 1/2$.

When $n_1 = n_2 = 1$ the Calabi-Yau three-fold is the well-known conifold ${\cal C}$.  In this case $\Delta_1 = \Delta_2 = 3/4$, and from eqs.~\eqref{FreeExtCY3} and \eqref{MtheoryExpectation} one obtains $\Vol(Y) = 16 \pi^4/81$.  In appendix~\ref{CCY5} we confirm this number from a direct computation of the volume of $Y$ using an explicit metric.  In appendix~\ref{CCY5} we also show that the space $Y$ has topologically non-trivial five-cycles that using \eqref{DeltaBaryon} would yield M5-branes dual to operators of dimension $3N/8$.  From $T\tilde T = X_1 X_2$ it follows that $\det X_1$ and $\det X_2$ correspond to giant gravitons wrapping topologically trivial cycles constructed from two topologically non-trivial cycles, so one expects the dimensions of $\det X_1$ and $\det X_2$ to be $3N/4$, in agreement with the value obtained by $F$-maximization.

\subsubsection{The $D_3$ theory}
\label{D3ONE}

Another fairly simple particular case is $n_1 = n_2 = n_3 = 1$.  The associated $\text{CY}_4$ is described by the equation $T\tilde T = X_1 X_2 X_3$ and is therefore a complete intersection.  While the volume of the Sasaki-Einstein base $Y$ can of course be obtained as a particular case from the toric geometry computation in section~\ref{TORIC}, there is actually a simpler way of computing this volume using the results of \cite{Bergman:2001qi}.  Indeed, eq.~(16) of that paper with $n = 4$, $d = 6$, $\vec{w} = (3, 3, 2, 2, 2)$ (so $w = 72$ and $\abs{w} = 12$) gives $\Vol(Y) = 9 \pi^4/64$.  From the matrix model, the extremum of the free energy \eqref{FExtFurther} can be found to be
 \es{D3C3Free}{
  F = \frac{8 \pi}{9} \sqrt{ \frac{2}{3}} N^{3/2} \,,
}
in agreement with the value we found for $\Vol(Y)$.

\subsection{Universal RG flows}
\label{STRETCHEDSQUASHED}

The theories discussed in section~\ref{CY3THEORIES} dual to $\text{CY}_3 \times \C$ have two obvious relevant superpotential deformations:  $\tr (X_3)^2$ of R-charge $1$ and $\tr (X_3)^3$ of R-charge $3/2$.  Adding either of these operators to the superpotential causes an RG flow to a new IR fixed point.  The RG flows obtained this way are universal in the sense that, as we will now show, the ratio of the IR and UV free energies is independent of the details of the three-fold $\text{CY}_3$.  We will only compute this ratio for the toric $\text{CY}_3$ examples of section~\ref{CY3THEORIES}, but we believe that the same ratio can be obtained for non-toric examples.

To give a unified treatment of the $\tr (X_3)^2$ and $\tr (X_3)^3$ deformations, let's examine the theory obtained by adding $\tr (X_3)^p$ to the superpotentials \eqref{FlavoredSuperpot} with $n_3 = 0$ but otherwise arbitrary $n_1$ and $n_2$.  This extra term in the superpotential fixes the R-charge of $X_3$ to be $\Delta_3 = 2/p$.  Fixing $\Delta_3$ to this value and writing for example $\Delta_2 = 2 - \Delta_1 - \Delta_3$ one can find the R-charges of the new IR fixed point by maximizing \eqref{FExtFurther}.  A simple computation shows that the IR R-charges are related to the UV R-charges through
 \es{IRUVRelation}{
  \Delta_1^{\rm IR} = \frac{4(p-1)}{3p} \Delta_1^{\rm UV} \,, \qquad
   \Delta_2^{\rm IR} = \frac{4(p-1)}{3p} \Delta_2^{\rm UV} \,, \qquad
   \Delta_3^{\rm IR} = \frac 2p \,,
 }
where $\Delta_1^{\rm UV}$ and $\Delta_2^{\rm UV}$ have the values given in \eqref{DeltaCY3}.  Consequently, the IR free energy is also related to the UV free energy in a way independent of which $\text{CY}_3$ space one may want to consider:
 \es{IRUVFree}{
  F^{\rm IR} = \frac{16 (p-1)^{3/2}}{3 \sqrt{3} p^2} F^{\rm UV} \,.
 }
In particular, for $p= 2$ one obtains $F^{\rm IR} / F^{\rm UV} = 4/(3 \sqrt{3})$ and for $p=3$ one obtains $F^{\rm IR}/F^{\rm UV} = 32 \sqrt{2} / (27 \sqrt{3})$.

One obvious question to ask is:  what are the gravity duals to these RG flows?  For $p = 2$, we believe this holographic RG flow was constructed in \cite{Corrado:2001nv} (for $p=3$, we are not aware of a similar holographic construction).  Let's examine the holographic RG flow of \cite{Corrado:2001nv}  in more detail.  This flow was originally found in 4-d ${\cal N} = 8$ gauged supergravity as a flow between two extrema of the gauged supergravity potential---the maximally supersymmetric one and the $U(1)_R \times SU(3)$-symmetric one found in \cite{Warner:1983vz}.  An uplift of this flow to 11-d supergravity was constructed in \cite{Corrado:2001nv} where in the UV the geometry aysmptotes to $AdS_4 \times S^7$, and in the IR it asymptotes to a warped product between $AdS_4$ and a stretched and squashed seven-sphere.  It was noticed in \cite{Corrado:2001nv} that the uplift of the 4-d flow to eleven dimensions was not unique in the sense that an $S^5 \subset S^7$ in the UV geometry could be replaced by the base of any $\text{CY}_3$ cone which is a regular Sasaki-Einstein manifold.  Such a generalization of the holographic RG flow \cite{Corrado:2001nv} should be dual to the flow induced by the superpotential perturbation $\tr (X_3)^2$ in all the gauge theories dual to $\text{CY}_3\times \C$.

 We can compare the field theory prediction \eqref{IRUVFree} with the gravity computation.  From a four-dimensional perspective, the free energy on $S^3$ is given by eq.~\eqref{FHolography} in terms of the radius $L$ of $AdS_4$ and the effective 4-d Newton constant $G_4$.  In the holographic RG-flow of \cite{Corrado:2001nv}, the 4-d Newton constant is kept fixed, so the ratio of free energies is
 \es{FreeRatio}{
  \frac{F_{\rm IR}}{F_{\rm UV}} = \left (\frac{L_{\rm IR}}{L_{\rm UV}} \right)^2 = \frac{4}{3\sqrt{3}} \,,
 }
where in the last equation we used $L_{\rm UV} / L_{\rm IR} = 3^{3/4} / 2$ \cite{Corrado:2001nv}.  Indeed, this expression is in agreement with eq.~\eqref{IRUVFree}.

Two comments are in order.  First, when $\text{CY}_3 = \C^3$ supergravity predicts that the IR theory has emergent $U(1)_R \times SU(3)$ symmetry.  We now explain why this is a consistent possibility in the field theory.  In the field theory, at the IR fixed point one can just integrate out $X_3 \sim [X_1, X_2]$ and obtain the effective superpotential $\tr \left( [X_1, X_2]^2 + q X_1 \tilde q \right)$.  The monopole operators have the OPE $T \tilde T \sim X_1$, which implies  $R[X_1] = R[T] + R[\tilde T]$.  From eq.~\eqref{IRUVRelation} we see that $\Delta_1^{\rm UV} = 1$ and $\Delta_2^{\rm UV} = 1/2$ in the UV (see section~\ref{BMODEL}) implies $\Delta_1^{\rm IR} = 2/3$ and $\Delta_2^{\rm IR} = 1/3$ in IR.   The fact that $\Delta_m = 0$ tells us $R[T] - R[ \tilde T] = 0$, and combining the above observations we conclude $R[T] = R[\tilde T] = 1/3$.  This leads us to conjecture that $T$, $\tilde T$, and $X_2$ form a triplet of $SU(3)$, making the expected symmetry enhancement to $U(1)_R \times SU(3)$ a consistent possibility.  We thus propose that this gauge theory is dual to Warner's $U(1)_R\times SU(3)$ invariant fixed point of gauged supergravity \cite{Warner:1983vz}. Another proposed gauge theory dual is a certain mass-deformed version of ABJM theory \cite{Benna:2008zy}; we will solve the corresponding matrix model in section~\ref{ABJMNOFLAVORS}.

The second comment starts with the observation that the IR free energy of the mass-deformed $\C \times {\cal C}$ theory, ${\cal C}$ being the conifold, is the same as that of the undeformed $\C^4$ theory.  There is a field theory argument that explains this match:  The $\C^4$ theory (whose superpotential is $W \sim \tr (X_1[X_2, X_3] + q X_1 \tilde q)$) has a marginal direction where one adds $X_1^2$ to the superpotential.  Integrating out $X_1$ one obtains $W \sim \tr ([X_2, X_3]^2 + q [X_2, X_3] \tilde q)$.  This theory is related by another marginal deformation to $W \sim \tr ([X_2, X_3]^2 + q X_2 X_3 \tilde q)$, which in turn can be obtained by integrating out $X_1$ from $W \sim \tr (X_1[X_2, X_3] + q X_2 X_3 \tilde q + mX_1^2 )$.  The theory with the latter superpotential has the same free energy as the mass-deformed $\C \times {\cal C}$ theory.

\subsection{A non-toric example:  The cone over $V_{5, 2}/\Z_n$}

It was proposed in \cite{Jafferis:2009th} that the theory dual to the $AdS_4 \times V_{5, 2}/\Z_n$ M-theory background is a Yang-Mills $U(N)$ gauge theory with three adjoint fields $X_i$ and $2n$ fields $q_j$ and $\tilde q_j$ transforming in $\bf N$ and $\bf \overline N$ of $U(N)$, respectively, and superpotential
 \es{YMSuperpot}{
  W \sim \tr \left[ X_1 [X_2, X_3] + \sum_{j=1}^n q_j (X_1^2 + X_2^2 + X_3^2) \tilde q_j\right] \,.
 }
The fact that the superpotential has R-charge $R[W] = 2$ implies that $X_i$ has R-charge $2/3$ and $q_j$ and $\tilde q_j$ have R-charge $1/3$.

The free energy functional is in this case
 \es{FreeV52}{
  F[\rho] = \frac{16 \pi^2}{27} N^{3/2} \int dx\, \rho^2 + \frac{2n}{3} N^{3/2} \int dx\, \rho \abs{x}
   + N^{3/2} \int dx\, \rho x \Delta_m \,.
 }
Extremizing with respect to $\rho$ under the constraint that $\rho$ is a density, one obtains
 \es{FreeV52Ext}{
  F = \frac{8 \pi \sqrt{n} N^{3/2}}{27} \sqrt{4 - \frac{9\Delta_m^2}{n^2}} \,.
 }
Maximizing this expression with respect to $\Delta_m$ gives $\Delta_m = 0$ and
 \es{FreeV52Again}{
  F = \frac{16 \pi \sqrt{n} N^{3/2}}{27} \,.
 }
Combining this expression with the M-theory expectation \eqref{MtheoryExpectation}, one obtains
 \es{VolV52Orbifold}{
  \Vol(Y) = \frac{27 \pi^4}{128\, n} \,,
 }
in agreement with the expectation that the space $Y$ is a $\Z_n$ orbifold of $V_{5, 2}$.

\section{Deforming the ABJM theory}
\label{ABJMNOFLAVORS}

\begin{figure}[htb]
\begin{center}
\leavevmode
\newcommand{\svgwidth}{.5\textwidth}
\definecolor{orange}{RGB}{255,165,0}
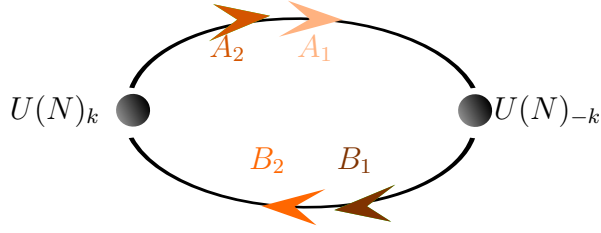
\end{center}
\caption{The quiver diagram for the ABJM theory at CS level $k$.   }
\label{ABJMQUIV}
\end{figure}

In this section we will study some deformations of the ABJM theory that lead to RG flow.  Before we do that though, we look at the ABJM theory and assign arbitrary R-charges to the bifundamental fields $A_i$ and $B_i$ that are consistent with the fact that the superpotential
 \es{W0ABJM}{
  W_0 \sim \tr \left[ \epsilon^{ij} \epsilon^{kl} A_i B_k A_j B_l \right]
 }
has R-charge two.  In other words, denoting $R[A_i] = \Delta_{A_i}$ and $R[B_i] = \Delta_{B_i}$, the constraint the R-charges satisfy is
 \es{ABJMRChargeConstraint}{
  \Delta_{A_1} + \Delta_{A_2} + \Delta_{B_1} + \Delta_{B_2} = 2 \,.
 }
Of course, assigning arbitrary R-charges $\Delta_{A_i}$ and $\Delta_{B_i}$ breaks SUSY from ${\cal N} = 6$ to ${\cal N} = 2$.  The quiver diagram for this theory is shown in figure~\ref{ABJMQUIV}

Using the general rules from section~\ref{RULES}, the matrix model free energy functional is
 \es{FreeABJM}{
  F[\rho, \delta y] &= \frac{k}{2 \pi} N^{3/2} \int dx\, \rho x \delta y - N^{3/2} \int dx\, \rho^2 \Biggl[(\delta y)^2
    + 2 \pi \delta y (\Delta_{A_1} \Delta_{A_2} - \Delta_{B_1} \Delta_{B_2}) \\
    &{}- 2 \pi^2 \biggl( \Delta_{A_1} \Delta_{A_2} (\Delta_{B_1} + \Delta_{B_2})
    + \Delta_{B_1} \Delta_{B_2} (\Delta_{A_1} + \Delta_{A_2}) \biggr) \Biggr]
    + N^{3/2} \int dx\, \rho x \Delta_m \,,
 }
where $\delta y \equiv y_1 - y_2$ and $\Delta_m = \Delta_{m1} + \Delta_{m2}$ is the sum of the bare monopole R-charges $\Delta_{m}^{(1)}$ and $\Delta_{m}^{(2)}$ for the two gauge groups.  In order to find a saddle point of the path integral on $S^3$, this free energy functional should be extremized as usual under the constraint that $\rho$ is a density, namely $\int dx\, \rho = 1$ and $\rho \geq 0$ pointwise.  So we should introduce a Lagrange multiplier $\mu$ and extremize
 \es{Ftilde}{
  \tilde F[\rho, \delta y] = F - \mu \frac{N^{3/2}}{2 \pi} \left(\int dx\, \rho -1 \right)
 }
instead of \eqref{FreeABJM}.  Assuming without loss of generality that $\Delta_{A_1} > \Delta_{A_2}$ and $\Delta_{B_1}> \Delta_{B_2}$, one can write the eigenvalue distribution that extremizes \eqref{Ftilde} as a piecewise smooth function:
 \begin{subequations}
 \label{SaddleABJM}
 \es{SaddleABJM1}{
   &-\frac{\mu}{2 \pi (k \Delta_{A_2} - \Delta_m)} < x < -\frac{\mu}{2 \pi (k \Delta_{A_1} - \Delta_m)} \;: \\[10pt]
     &\qquad  \qquad \qquad \qquad
       \rho = \frac{\mu + 2 \pi x (k \Delta_{A_2} - \Delta_m)}
       {8 \pi^3 (\Delta_{A_2} + \Delta_{B_2})(\Delta_{A_2} + \Delta_{B_1}) (\Delta_{A_1} - \Delta_{A_2})   } \,,
       \qquad
       \delta y = - 2 \pi \Delta_{A_2} \,, 
}
 \es{SaddleABJM2}{
   &-\frac{\mu}{2 \pi (k \Delta_{A_1} - \Delta_m)} < x < \frac{\mu}{2 \pi (k \Delta_{B_1} - \Delta_m)} \;: \\[10pt]
     &\qquad  \qquad \qquad \qquad
       \rho = \frac{\mu + \pi x \left[k (\Delta_{A_1} \Delta_{A_2} - \Delta_{B_1} \Delta_{B_2} ) -2 \Delta_m \right]}
       {4 \pi^3 (\Delta_{A_1} + \Delta_{B_1}) (\Delta_{A_1} + \Delta_{B_2}) (\Delta_{A_2} + \Delta_{B_1})(\Delta_{A_2} + \Delta_{B_2})   } \,, \\[10pt]
   &\qquad\qquad\qquad\qquad
       \delta y =  \frac{2 k \pi^2 x \left[\Delta_{A_1} \Delta_{A_2} (\Delta_{B_1} + \Delta_{B_2})
       + \Delta_{B_1} \Delta_{B_2} (\Delta_{A_1} + \Delta_{A_2}) \right]}
       {\mu + \pi x \left[k(\Delta_{A_1} \Delta_{A_2} - \Delta_{B_1} \Delta_{B_2}) -2 \Delta_m\right]}\,, \\[10pt]
       &\qquad\qquad\qquad\qquad
       \qquad + \frac{\pi (\Delta_{A_1} \Delta_{A_2} - \Delta_{B_1} \Delta_{B_2}) (2 \pi x \Delta_m - \mu)}
       {\mu + \pi x \left[k(\Delta_{A_1} \Delta_{A_2} - \Delta_{B_1} \Delta_{B_2}) -2 \Delta_m\right]}\,, \\[10pt]
  }
  \es{SaddleABJM3}{     
  &\frac{\mu}{2 \pi (k \Delta_{B_1} + \Delta_m)} < x < \frac{\mu}{2 \pi (k \Delta_{B_2} + \Delta_m)} \;: \\[10pt]
     &\qquad  \qquad \qquad \qquad
       \rho = \frac{\mu - 2 \pi x (k \Delta_{B_2} + \Delta_m)}
       {8 \pi^3 (\Delta_{A_1} + \Delta_{B_2})(\Delta_{A_2} + \Delta_{B_2}) (\Delta_{B_1} - \Delta_{B_2})   } \,,
       \qquad
       \delta y = 2 \pi \Delta_{B_2} \,,
 }
\end{subequations} 
where
 \es{GotmuABJM}{
  \mu^2 = \frac{32 \pi^4}{k^3} (k \Delta_{A_1} - \Delta_m)  (k \Delta_{A_2} - \Delta_m)
    (k \Delta_{B_1} + \Delta_m)  (k \Delta_{B_2} + \Delta_m) \,.
 }
By plugging this solution into \eqref{FreeABJM} one obtains
 \es{FreeExtABJM}{
  F = \frac{N^{3/2} \mu}{3 \pi} =
   \frac{N^{3/2} 4 \sqrt{2} \pi}{3 k^{3/2}}  \sqrt{(k \Delta_{A_1} - \Delta_m)  (k \Delta_{A_2} - \Delta_m)
    (k \Delta_{B_1} + \Delta_m)  (k \Delta_{B_2} + \Delta_m)}\,.
 }

As expected from the discussion in section~\ref{SYMMETRY}, for the $U(N) \times U(N)$ gauge theory the free energy has one flat direction under which $\Delta_{A_i} \to \Delta_{A_i} + \hat \delta$, $\Delta_{B_i} \to \Delta_{B_i} - \hat \delta$, and $\Delta_m \to \Delta_m + k \hat \delta$, corresponding in the notation of section~\ref{SYMMETRY} to $\delta^{(1)} = -\delta^{(2)} = \hat \delta/2$.  This flat direction is due to the fact that the bifundamental fields as well as the diagonal monopole operators $T$ and $\tilde T$ are charged under the $U(1)$ gauge symmetry corresponding to the gauge field $\tr (A_{1\mu} - A_{2\mu})$, so it is not meaningful to assign them individual R-charges.  Under this gauge symmetry, the operators $A_1$ and $A_2$ have charge $1$, $B_1$ and $B_2$ have charge $-1$, and the monopole operators $T$ and $\tilde T$ have charges $k$ and $-k$, respectively.  The gauge-invariant operators include for example $\tr \tilde T(A_i)^k$ and $\tr T (B_i)^k$ with R-charges $k \Delta_{A_i} - \Delta_m$ and $k\Delta_{B_i} + \Delta_m$, and these are indeed the combinations that appear in the expression for $F$ in eq.~\eqref{FreeExtABJM}.

Regarding $\Delta_m = 0$ as a gauge choice, we can maximize \eqref{FreeExtABJM} under the constraint \eqref{ABJMRChargeConstraint} that the R-charges of the $A_i$ and $B_i$ fields sum up to two.  The maximum is at $\Delta_{A_1} = \Delta_{A_2} = \Delta_{B_1} = \Delta_{B_2} = 1/2$, which are the correct R-charges for the ${\cal N} = 6$ ABJM theory.  The value of $F$ at the maximum is
 \es{FABJM}{
  F =
   \frac{\sqrt{2k} \pi N^{3/2}}{3} \,,
 }
which, when combined with eq.~\eqref{MtheoryExpectation}, implies $\Vol(Y) = \pi^4 / (3k)$, in agreement with the fact that ABJM theory is dual to $AdS_4 \times S^7 / \Z_k$, the volume of $S^7$ being $\pi^4/3$.

A superpotential deformation of the schematic form $\tr (\tilde T A_1)^2$ when $k=1$ or $\tr \tilde T A_1^2$ when $k=2$ causes an RG flow to a new IR fixed point where the field $A_1$ can be integrated out.  It was proposed in \cite{Benna:2008zy, Klebanov:2008vq} (see also \cite{Ahn:2008ya}) that the holographic dual of this RG flow was constructed in \cite{Corrado:2001nv}, first as a flow in 4-d ${\cal N} = 8$ gauged supergravity from the maximally symmetric point to the $U(1)_R \times SU(3)$-invariant extremum \cite{Warner:1983vz} of the gauged supergravity potential, and then uplifted to M-theory as a flow from $AdS_4 \times S^7$ to a warped product between $AdS_4$ and a stretched and squashed seven-sphere.  (See also section~\ref{STRETCHEDSQUASHED} for another gauge theory realization of the same holographic RG flow.)

Working in the gauge $\Delta_m = 0$, the superpotential deformation mentioned above imposes in the IR the constraint $\Delta_{A_1} = 1$, so
 \es{FreeA1FixedABJM}{
  F = \frac{ 4 \sqrt{2k} \pi N^{3/2}}{3}  \sqrt{\Delta_{A_2} \Delta_{B_1} \Delta_{B_2}}\,.
 }
This expression should be maximized under the constraint \eqref{ABJMRChargeConstraint} that $\Delta_{A_2} + \Delta_{B_1} + \Delta_{B_2} = 1$.  By the standard inequality between the geometric and arithmetic mean, the product of three numbers whose sum is kept fixed is maximized when all the numbers are equal, so $F$ has a maximum when $\Delta_{A_2} = \Delta_{B_1} = \Delta_{B_2} = 1/3$.  In the IR we therefore have
 \es{FIR}{
  F_{\rm IR} = \frac{ 4 \sqrt{2k} \pi N^{3/2}}{9\sqrt{3}}
   = \frac{4}{3 \sqrt{3}} F_{\rm UV} \,,
 }
where $F_{\rm UV}$ is the free energy of the ABJM theory in eq.~\eqref{FABJM}.  As already discussed in section~\ref{STRETCHEDSQUASHED}, the ratio of $F_{\rm IR}$ to $F_{\rm UV}$ given above is what one expects from the dual holographic RG flow of \cite{Corrado:2001nv}.

\section{Flavoring the ABJM quiver}
\label{FLAVORABJM}

\begin{figure}[htb]
\begin{center}
\leavevmode
\newcommand{\svgwidth}{.5\textwidth}
\definecolor{orange}{RGB}{255,165,0}
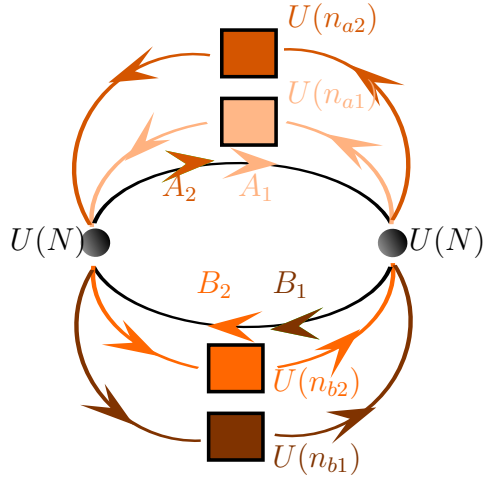
\end{center}
\caption{The quiver diagram for the flavored theories corresponding to the superpotential in equations~\eqref{W0ABJM} and~\eqref{deltaW}.   }
\label{ABJMGENQUIV}
\end{figure}

In this section we will analyze ${\cal N} = 2$ theories that come from adding flavors to the $U(N) \times U(N)$ ${\cal N} = 6$ ABJM theory \cite{Aharony:2008ug} at level $k$. In general, we could add four paris of bifundamental fields $(q_j^{(i)}, \tilde q_j^{(i)})$ with $i =1, 2$ and $j = 1, 2, \ldots, n_{ai}$ and $(Q_j^{(i)}, \tilde Q_j^{(i)})$ with $i =1, 2$ and $j = 1, 2, \ldots, n_{bi}$, and we could couple these fields to the ABJM theory \eqref{W0ABJM} through the superpotential coupling
 \es{deltaW}{
  \delta W \sim \tr \left[\sum_{j=1}^{n_{a1}} q_j^{(1)} A_1 \tilde q_j^{(1)} +
   \sum_{j=1}^{n_{a2}} q_j^{(2)} A_2 \tilde q_j^{(2)}
   + \sum_{j=1}^{n_{b1}} Q_j^{(1)} B_1 \tilde Q_j^{(1)} +
   \sum_{j=1}^{n_{b2}} Q_j^{(2)} B_2 \tilde Q_j^{(2)} \right] \,.
 }
As far as the matrix model goes, these extra fields corresponds to adding
 \es{ExtraFreeFunc}{
  \delta F[\rho, \delta y] = \frac{N^{3/2}}{2} \int dx\, \rho \abs{x} \left[
    \sum_{i = 1}^2 \left( n_{ai} \Delta_{A_i} + n_{bi} \Delta_{B_i} \right)
   + \frac{\delta y}{2 \pi} \left(n_{a1} + n_{a2} - n_{b1} - n_{b2} \right)
   \right]
 }
to the free energy functional for ABJM theory in eq.~\eqref{FreeABJM}.   The quiver diagram for this theory is given in figure~\ref{ABJMGENQUIV}.
It is straightforward to do the extremization of the free energy functional for arbitrary $n_{ai}$ and $n_{bi}$, but the resulting formulae are fairly long, so we will just examine a few particular cases.

\subsection{An infinite class of flavored theories}
\label{INFINITE}

The first particular case we examine is in some sense a generalization of the flavored quivers we studied in section~\ref{TORICONE}.  Like in the models in that section, there are no Chern-Simons terms and the number of arrows going out of any given node equals the number of arrows going in:
 \es{nRelation}{
  n_{a1} + n_{a2} = n_{b1} + n_{b2}  \,, \qquad k=0 \,.
 }
The $U(1)$ quantum corrected moduli space of these theories is given by the relation $T \tilde T = A_1^{n_{a1}} A_2^{n_{a2}} B_1^{n_{b1}} B_2^{n_{b2}}$ in $\C^6$ together with a K\"ahler quotient acting with charges $(0, 0, 1, 1, -1, -1)$ on $(T, \tilde T, A_1, A_2, B_1, B_2)$  \cite{Benini:2009qs, Jafferis:2009th}.  The free energy is
 \es{Freennnn}{
  F = \frac{2\pi N^{3/2}}{3}
   \sqrt{  \prod_{i, j = 1}^2 \left(\Delta_{A_i} + \Delta_{B_j} \right)
   \left[\sum_{i = 1}^2 \left( n_{ai} \Delta_{A_i} + n_{bi} \Delta_{B_i} \right)
   -\frac{4 \Delta_m^2}
   {\sum_{i = 1}^2 \left( n_{ai} \Delta_{A_i} + n_{bi} \Delta_{B_i} \right) } \right] } \,.
 }
In order to find the R-charges in the IR, this expression should be locally maximized under the constraint \eqref{ABJMRChargeConstraint}.  Clearly, the maximization over $\Delta_m$ yields simply $\Delta_m = 0$, so there is no asymmetry between the R-charges of the monopole operators $T$ and $\tilde T$, and the free energy as a function of $\Delta_{Ai}$ and $\Delta_{Bi}$ reduces to
 \es{FreennnnAgain}{
  F = \frac{2\pi N^{3/2}}{3}
   \sqrt{  \prod_{i, j = 1}^2 \left(\Delta_{A_i} + \Delta_{B_j} \right)
   \sum_{i = 1}^2 \left( n_{ai} \Delta_{A_i} + n_{bi} \Delta_{B_i} \right)
    } \,.
 }

In the $U(N) \times U(N)$ gauge theory, the free energy \eqref{FreennnnAgain} is invariant under $\Delta_{A_i} \to \Delta_{A_i} + \hat \delta$ and $\Delta_{B_i} \to \Delta_{B_i} - \hat \delta$, corresponding to $\delta^{(1)} = -\delta^{(2)} = \hat \delta/2$ in the notation of section~\ref{SYMMETRY}.  As discussed in section~\ref{SYMMETRY}, to remove this flat direction one can ungauge the gauge symmetry that rotates $A_i$ and $B_i$ by opposite phases and consider instead a gauge theory with $SU(N) \times SU(N) \times U(1)$ gauge group, where the remaining $U(1)$ comes from the diagonal $U(1)$ in $U(N) \times U(N)$.  The difference between the $SU(N) \times SU(N) \times U(1)$ gauge theory and the $U(N) \times U(N)$ one is that in the former there is an extra constraint
 \es{intConstraint}{
  \int dx\, \rho \delta y = 0 \,.
 }
Imposing this constraint removes the flat direction mentioned above.  An explicit calculation for the saddle point of the theories we are examining in this section gives that eq.~\eqref{intConstraint} is equivalent to
 \es{intrhodynnnn}{
  \Delta_{A_1} \Delta_{A_2} - \Delta_{B_1} \Delta_{B_2} = 0\,.
 }
In the $SU(N) \times SU(N) \times U(1)$ gauge theory one can therefore determine the R-charges of the bifundamental fields uniquely by maximizing \eqref{FreennnnAgain} under the constraints \eqref{ABJMRChargeConstraint} and \eqref{intrhodynnnn}.  In section~\ref{genflavABJM} we will show that this maximization problem is equivalent to the toric geometry $Z$-minimization, which means that the field theory free energy on $S^3$ as computed from maximizing \eqref{FreennnnAgain} agrees with the M-theory expectation \eqref{MtheoryExpectation}.  Moreover, the computation in section~\ref{genflavABJM} shows that the dimensions of baryonic operators computed from the volumes of five-cycles on the gravity side are consistent with $F$-maximization.

There are two particular cases where the quantum corrected moduli space can be expressed as a complete intersection and one can apply the methods of \cite{Bergman:2001qi} to compute the volume of the 7-d Sasaki-Einstein space $Y$ .   The first case is $n_{a1} = n_{b1}= 1$ and $n_{a2} = n_{b2} = 0$, where the cone over $Y$ can be described by the equation $z_1 z_2 = z_3 z_4 z_5$ in $\C^5$ \cite{Benini:2009qs}.  In fact, we encountered this space in section~\ref{D3ONE} where we found that the volume was $\Vol(Y) = 9 \pi^4/64$.  One can indeed reproduce this volume by minimizing \eqref{FreennnnAgain} explicitly and using eq.~\eqref{MtheoryExpectation}.

Another particular case is $n_{a1} = n_{a2} = n_{b1} = n_{b2} = 0$, where the Calabi-Yau cone over $Y$ is the ``cubic conifold'' described as a complete intersection by the equations $z_1 z_2 = z_3 z_4 = z_5 z_6$ in $\C^6$.  Eq.~(16) of\cite{Bergman:2001qi} with $n = 4$, $d = 4$, $\vec{w} = (1, 1, 1, 1, 1, 1)$ (so $w = 1$ and $\abs{w} = 6$) gives $\Vol(Y) = \pi^4/12$.  Indeed, extremizing \eqref{FreennnnAgain} and using \eqref{MtheoryExpectation} one can reproduce the volume of $Y$ in this case too.

\subsection{M2-branes probing $\C \times {\cal C}$}
\label{CCONABJM}

A quite non-trivial example where the bare monopole R-charge $\Delta_m$ plays a crucial role is the case $n_{a1} =1$ and $n_{a2} = n_{b1} = n_{b2} = 0$ at CS level $k = 1/2$.  The CS level is a half-integer because in the IR there is an extra $1/2$ shift in the CS level coming from integrating out the fermions in the chiral multiplets $q^{(1)}$ and $\tilde q^{(1)}$, which are massive at generic points on the moduli space.  The $U(1)$ quantum corrected moduli space is $\C$ times the conifold ${\cal C}$.

The $U(N) \times U(N)$ theory has a flat direction given by $\Delta_{A_i} \to \Delta_{A_i} + \hat \delta$, $\Delta_{B_i} \to \Delta_{B_i} + \hat \delta$, and $\Delta_m \to \Delta_m + \hat \delta /2$, so the free energy should only be a function of
 \es{hatRcharges}{
   \hat \Delta_{A_i} \equiv \Delta_{A_i} - 2\Delta_m \,, \qquad
    \hat \Delta_{B_i} \equiv \Delta_{B_i} + 2 \Delta_m \,.
 }
Indeed, an explicit extremization of the free energy functional gives
 \es{FreeCConTwo}{
  F = \frac{2 \sqrt{2} \pi N^{3/2}}{3} \sqrt{\frac
   {\hat \Delta_{A_1} \left(\hat \Delta_{A_2} + \hat \Delta_{B_1} \right)
   \left(\hat \Delta_{A_2} + \hat \Delta_{B_2} \right)
   \left(\hat \Delta_{A_1} + 2 \hat \Delta_{B_1} \right)
   \left(\hat \Delta_{A_2} + 2 \hat \Delta_{B_2} \right)}
   {4 - \hat \Delta_{A_1}}} \,,
 }
where $\hat \Delta_{A_i}$ and $\hat \Delta_{B_i}$ satisfy the constraint $\hat \Delta_{A_1} + \hat \Delta_{A_2} + \hat \Delta_{B_1} + \hat \Delta_{B_2} = 2$ coming from eq.~\eqref{ABJMRChargeConstraint}.  This expression is maximized for
 \es{CConMaximum}{
  \hat \Delta_{A_1} =1 \,,\qquad \hat \Delta_{A_2} = \frac12 \,, \qquad
   \hat \Delta_{B_1} = \hat \Delta_{B_2} = \frac 14 \,,
 }
yielding
 \es{CConFMax}{
  F = \frac{\sqrt{3} \pi N^{3/2}}{2\sqrt{2}} \,.
 }
From \eqref{MtheoryExpectation} one obtains that the Sasaki-Einstein base $Y$ of $\C \times {\cal C}$ has volume $\Vol(Y) = 16 \pi^4 / 81$, in agreement with the value we compute in section~\ref{CCONTORIC} using toric geometry techniques or in appendix~\ref{CCY5} using more elementary methods.

In the $SU(N) \times SU(N) \times U(1)$ theory the flat direction in $F$ is no longer there because one imposes as an additional constraint that $\int dx\, \rho \delta y = 0$.  From an explicit computation of the saddle point, one finds that this constraint reduces to
 \es{GotDeltamCCY}{
  \Delta_m = \frac{2 \hat \Delta_{B_1} \hat \Delta_{B_2} - \hat \Delta_{A_1} \hat \Delta_{A_2}}
   {2 \left(4 - \hat \Delta_{A_1} \right)} \,.
 }
Using \eqref{CConMaximum} one obtains
 \es{SUNCCY}{
  \Delta_m = -\frac{1}{16} \,,\qquad
   \Delta_{A_1} = \frac{7}{8} \,, \qquad \Delta_{A_2} = \Delta_{B_1} = \Delta_{B_2} = \frac 38 \,.
 }
As we will show in section~\ref{CCONTORIC} these values are in agreement with the dimensions of baryonic operators dual to M5-branes wrapping various cycles in $Y$ computed from the gravity side.

\subsection{Dual to $AdS_4 \times Q^{1, 1, 1}/\Z_n$}
\label{Q111SECTION}

Another example is the theory that was proposed in \cite{Jafferis:2009th, Benini:2009qs} as a dual of $AdS_4 \times Q^{1, 1, 1}/\Z_n$.  This theory has $n_{a1} = n_{a2} = n$, $n_{b1} = n_{b2} = 0$, and vanishing CS levels $k = 0$.  Obtaining an expression for the free energy as a function of arbitrary R-charges $\Delta_{A_i}$ and $\Delta_{B_i}$ is fairly involved, so using the symmetries of the quiver let's just focus on the subspace where
 \es{DeltaQ111}{
  \Delta_{A_i} = \Delta \,, \qquad \Delta_{B_i} = 1- \Delta \,,
 }
in agreement with the constraint \eqref{ABJMRChargeConstraint}, and allow an arbitrary bare monopole R-charge $\Delta_m$.  The extremization of the free energy functional gives
 \es{FQ111}{
 F = \frac{4\pi N^{3/2}}{3\sqrt{n}} \frac{\abs{n^2 - \Delta_m^2}}{\sqrt{3 n^2 - \Delta_m^2}}
 }
as well as
 \es{intrhodyQ111}{
  \int dx\, \rho \delta y = \pi \left[\frac{4 n^2}{3n^2 - \Delta_m^2} - 2 \Delta \right] \,.
 }

Notice that in this case the free energy $F$ is independent of $\Delta$, because the fact that $k = 0$ implies that the flat direction discussed in section~\ref{SYMMETRY} corresponds to $\Delta \to \Delta + \hat \delta$ (where $\hat \delta = \delta^{(1)} = -\delta^{(2)}$) leaving $\Delta_m$ invariant.  Maximizing \eqref{FQ111} with respect to $\Delta_m$ one obtains $\Delta_m = 0$ and
 \es{FQ111Max}{
  F = \frac{4 \pi \sqrt{n} N^{3/2}}{3 \sqrt{3}} \,,
 }
in agreement with the fact that the volume of $Y = Q^{1, 1, 1}/\Z_n$ is $\Vol(Y) = \pi^4 / (8 n)$.

As before, for the $U(N) \times U(N)$ theory it doesn't make sense to assign any meaning to $\Delta$ because one cannot construct a gauge-invariant operator just from the $A_i$ fields, for example.  In the $SU(N) \times SU(N) \times U(1)$ theory, on the other hand, the condition $\int dx\, \rho \delta y = 0$ combined with \eqref{intrhodyQ111} and $\Delta_m = 0$ implies $\Delta = 2/3$. It follows that the baryonic operators constructed out the $B_i$, such as ${\rm \mathcal{B}}(B_1)$, have dimensions $N/3$ in agreement with the dimension of wrapped M5-branes \cite{Fabbri:1999hw}.

\subsection{An infinite family of non-singular spaces}
\label{QKSECTION}

Except for $Q^{1, 1, 1}$, all the internal spaces $Y$ corresponding to the flavored conifold quiver that we examined so far have non-isolated singularities.  However, an infinite family of non-singular spaces $Q_k$ can be obtained by setting $n_{a1} = n_{a2} = 1$, $n_{b1} = n_{b2} = 0$, and arbitrary $k\geq 2$.  As in the previous section, obtaining $F$ as a function of arbitrary R-charges $\Delta_{A_i}$ and $\Delta_{B_i}$ is onerous, so we'll restrict to the case \eqref{DeltaQ111} with arbitrary bare monopole R-charge $\Delta_m$.  Since as discussed in section~\ref{SYMMETRY} there is a flat direction corresponding to $\Delta \to \Delta + \delta$ and $\Delta_m \to \Delta_m + k \delta$, the free energy only depends on the combination $\hat \Delta = \Delta_m - k \Delta$:
 \es{FQk}{
  F = \frac{4 \sqrt{2} \pi N^{3/2}}{3} \frac{\hat \Delta (\hat \Delta + k + 1)}
   {\sqrt{(k+1)^2 (k-1) - 4 (k+1) \hat \Delta - 2 \hat \Delta^2}} \,.
 }
but of course $\delta y$ depends both on $\Delta$ and $\hat \Delta$,
 \es{intrhodeltayQk}{
  \int dx\, \rho \delta y = - 2 \pi \Delta - \frac{2 \pi \hat \Delta (k+1)^2}
   {(k+1)^2 (k-1) - 4 (k+1) \hat \Delta - 2 \hat \Delta^2} \,.
 }
Because we choose to work in the $SU(N) \times SU(N) \times U(1)$ gauge theory, the right hand side of equation~\eqref{intrhodeltayQk} must vanish.  This constraint uniquely fixes $\Delta$ in terms of $\Delta_m$, or vice-versa, so that we are left with a prediction for the volume of the internal space $Y$ on the gravity side in terms of one free R-charge parameter.  In section~\ref{QKARBSEC} we reproduce this formula for the volume as a function of R-charge using toric geometry.  Thus, we again find that the field theory prediction for the volume of the internal space matches the gravitational one even for R-charges away from their critical values.

\section{$Z$-minimization from toric geometry}
\label{TORIC}

\subsection{General rules for finding volumes of base spaces}
\label{TORIC_RULES}

In this section we turn to the gravity side and describe how to find the volume of the seven-dimensional Sasaki-Einstein space $Y$ in the case where the four complex-dimensional Calabi-Yau cone over $Y$, which we call $X = C\left(Y\right)$, is toric.  This section relies heavily on the results of \cite{Martelli:2005tp}.

For the moment let's allow $X$ to be a toric cone of complex dimension $n$ that admits a Ricci flat metric.  That $X$ is toric means that the isometry group of $X$ has a $U(1)^n$ subgroup.  When we put a Ricci-flat metric on $X$, the base of the cone, which is the space $Y$ whose volume we want to find, has an induced Sasaki-Einstein metric.  The idea of \cite{Martelli:2005tp} was to allow more general K\"ahler metrics on $X$, which by definition induce metrics on $Y$ that are Sasakian, but not necessarily Einstein.  With each such Sasakian metric on $Y$ one can compute the volume of $Y$.   When varying over the space of K\"ahler metrics on $X$ (or equivalently Sasakian metrics $g$ on $Y$) one can consider the Einstein-Hilbert action $Z[g]$ on $Y$ with a fixed positive cosmological constant.  This action is extremized (in fact minimized) when the metric on $Y$ is also Einstein, and the extremum is in fact proportional to the volume $\Vol(Y)$ that we want to compute.  The strategy therefore is to parameterize the set of all Sasakian metrics on $Y$ and then extremize the Einstein-Hilbert action $Z$ over this set of metrics.

It turns out that Sasakian metrics are uniquely specified by the choice of a Reeb vector.  A Reeb vector $K$ is the vector field on $X$ that is paired up with the radial vector field in the complex structure.  The vector field $K$ generates isometries of the cone contained in the $U(1)^n$ isometry group, and various Sasakian metrics differ in how the orbits of $K$ sit within the $U(1)^n$ torus.

We may introduce symplectic coordinates $\left(y_i,\phi_i\right)$, with $i = 1, \ldots, n$, so that the K\"ahler form on $X$, which should also be viewed as a symplectic from, is given by
\es{kahlersym}{
  \omega = d y_i \wedge d \phi_i \,.
 }
The $\phi_i$ are the angular coordinates parameterizing the $U(1)^n$ group of isometries, and their ranges are taken to be between $0$ and $2 \pi$.  We assume that the toric cone $X$ is of Reeb type so that the coordinates $y_i$ live inside of a strictly convex polyhedral cone $\mathcal{C} \subset \mathbb{R}^n$.  Assume that the cone has $d$ facets.  Let the vectors $v_a$, with $a = 1, \ldots, d$, be the inward pointing normal vectors to these facets.  We only focus on cones with Gorenstein singularities, meaning that by an appropriate $SL(n; \Z)$ transformation we may fix the first components of the normal vectors so that $v_a = \left(1,w_a\right)$.

The Reeb vector $K = b_i \frac{\partial}{\partial \phi_i}$ can be thought of as the normal vector $b_i$ to a hyperplane (called the characteristic hyperplane) in $\mathbb{R}^n$.  When
\es{bcond}{
\left(b,y\right) = \frac{1}{2} \,
}
this hyperplane intersects the cone $\mathcal{C}$ to form a polytope $\Delta_b$ of finite volume.  Moreover, the volume of the internal space $Y$ is related to the Euclidean volume $\Vol\left(\Delta_b\right)$ of the finite polytope $\Delta_b$ by
\es{volP}{
\Vol\left(Y\right) = 2n \left(2 \pi \right)^n \Vol\left(\Delta_b\right) \, .
}
Additionally, each facet $\mathcal{F}_a$, with inward normal $v_a$, corresponds to a $\left(2n - 3\right)$-cycle $\Sigma_a$ in $Y$.  The volume of the cycle is related to the Euclidean volume $\Vol\left(\mathcal{F}_a\right)$ of the facet by
\es{volC}{
\Vol\left(\Sigma_a\right) = \left(2n - 2\right) \left(2 \pi \right)^{n-1} \frac{1}{|v_a|} \Vol\left(\mathcal{F}_a\right)  \, .
}
The volume of the internal manifold $Y$ can be neatly expressed in terms of the volumes of the facets:
\es{volY}{
\Vol\left(Y\right) = \frac{\left(2\pi\right)^n}{b_1} \sum_a \frac{1}{|v_a|} \Vol\left(\mathcal{F}_a\right) \, .
}
When $Y$ is Sasaki-Einstein there is a further condition that $b_1 = n$.

We now restrict to $n = 4$.  Suppose that the facet $\mathcal{F}_a$ is a tetrahedron.  The vertex of this tetrahedron is at the origin in $\mathcal{C}$ and the base is a triangle in the characteristic hyperplane.  There are three edges running from the characteristic hyperplane to the origin.  Each edge is the intersection between three hyperplanes, one of which is always the hyperplane normal to $v_a$.  Each of the other hyperplanes is involved in creating two edges.  Thus, in addition to the hyperplane normal to $v_a$ we have three other hyperplanes involved in creating the polytope.  We label these hyperplanes by their inward-pointing normal vectors $v_1$, $v_2$, and $v_3$.  It is then straightforward to derive a formula for the volume of the tetrahedron in terms of the normal vectors,
\es{volTet}{
\frac{1}{|v_a|} \Vol \left(\mathcal{F}_a\right) = \frac{1}{48} \frac{ \left(v_a,v_1,v_2,v_3\right)^2}{ | \left(b,v_a,v_1,v_2\right) \left(b,v_a,v_1,v_3\right) \left(b,v_a,v_2,v_3\right) |} \, .
}
If the projection of $\mathcal{F}_a$ onto the characteristic hyperplane is a polygon instead of a triangle, then the volume can still be computed this way by breaking the polygon up into triangles.

Each vertex of the toric diagram, which is the diagram of the vectors $w_a$, corresponds to a facet.  The sides of the diagram which contain this vertex at one of their corners coincide with edges of the polygon, running from the origin to the characteristic hyperplane.  If a vertex is at the edge of three planes its facet corresponds to a tetrahedron, if the vertex is at the edge of four planes the corresponding polygon is a pyramid, and so forth.  Given a Reeb vector and a toric diagram, this procedure gives a systematic way of finding the volume of the base space $Y$.

In most cases of interest to us the space $Y$ is Ricci flat.  There is a unique choice of Reeb vector $b$ that gives a Ricci flat metric over $Y$.  This is the $b$ that minimizes the function
\es{Zfcn}{
Z \left[b\right] = \frac{b_1 - \left(n - 1\right)}{\left(2 \pi \right)^n} \Vol\left(Y\right) \, .
}
We suspect that non-critical $b$ corresponds to more complicated supergravity solutions than the Freund-Rubin compactifications $AdS_4 \times Y$.  In all our examples, though,  there is an agreement between the volume of $Y$ for non-critical $b$ and the field theory prediction of this volume for non-critical R-charges.

\subsection{Toric gauge theories with one gauge group}

\subsubsection{Toric geometry computation with arbitrary flavoring}
\label{TORICARB}

\begin{figure}[htb]
\begin{center}
\leavevmode
\scalebox{0.7}{\includegraphics{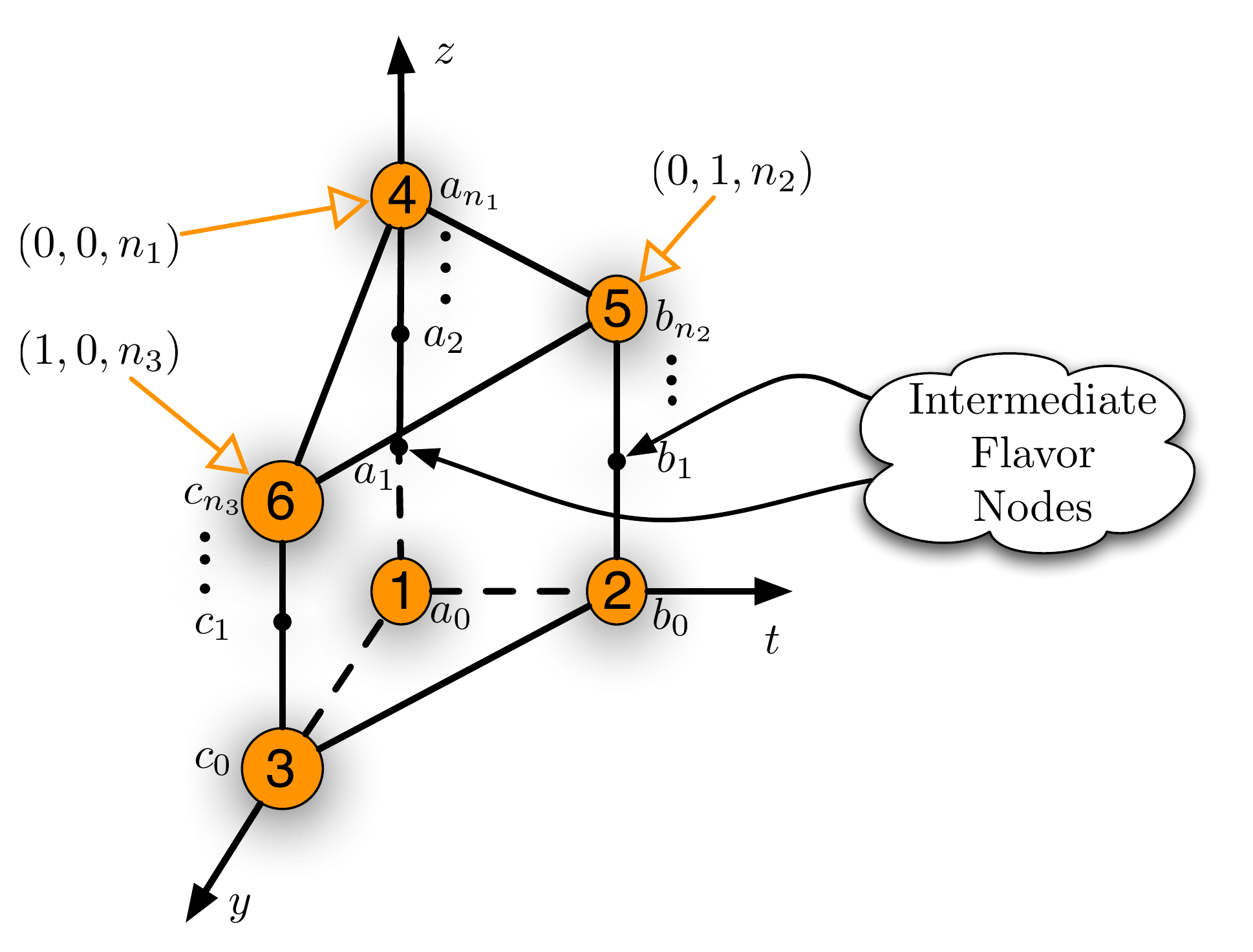}}
\end{center}
\caption{The toric diagram corresponding to an arbitrary flavoring of the ${\cal N}=8$ theory discussed in section~\ref{TORICONE}.   }
\label{C3GENPDF}
\end{figure}

In this section we consider the toric diagram \cite{Benini:2009qs} given in figure~\ref{C3GENPDF} corresponding to the moduli space \eqref{TTOPE} of an arbitrary flavoring of the ${\cal N} = 8$ theory that we discussed in section~\ref{TORICONE}.\footnote{There is a technical difference between the cases where we flavor one adjoint field, two adjoint fields, or all three adjoint fields.  The difference is that the toric diagrams have a different number of vertices in each of these cases.  However, the final expression for the volume of the internal space, given in equation \ref{vol3n3fR}, is consistent with all three toric diagram classes.  For that reason we only show the computation where all three adjoint fields are flavored.  That is, we require $n_1,n_2,n_3 >0$ in the following computation.}
We take the trial Reeb vector to be $b = \left(4,y,t,z\right)$.    The vertices labeled $1$ through $6$ correspond to tetrahedra.  The volumes of the five-cycles corresponding to these nodes in the toric diagram are
 \es{Vol3n3fCycles}{
\Vol \left( \Sigma_1 \right) &= \frac{\pi^3}{ytz} \,, \qquad
\Vol \left( \Sigma_2 \right) = \frac{\pi^3}{y\left(4 - y- t \right) z} \,, \qquad
\Vol \left( \Sigma_3 \right) = \frac{\pi^3}{\left(4- y-t\right)tz} \,, \\
\Vol \left( \Sigma_4 \right) &= \frac{\pi^3}{yt\left[n_1 \left(4-y-t\right) + n_2 t + n_3 y - z\right]} \,,\\
\Vol \left( \Sigma_5 \right) &= \frac{\pi^3}{y\left(4-t- y\right)\left[n_1 \left(4-y-t\right) + n_2 t + n_3 y - z\right]} \,, \\
\Vol \left( \Sigma_6 \right) &= \frac{\pi^3}{\left(4-t- y\right) t \left[n_1 \left(4-y-t\right) + n_2 t + n_3 y - z\right]} \,.
}
Using equations \eqref{volC} and \eqref{volY} we then calculate the volume of the base $Y$,
\es{Vol3n3fY}{
\Vol \left( Y_b \right) = \frac{\pi^4 \left[ n_3 y + n_2 t + n_1 \left(4 - y - t \right) \right]}{3 ytz \left(4 - y -t \right)  \left[ n_3 y + n_2 t + n_1 \left(4 - y - t \right) - z \right]  }
}
as a function of the trial Reeb vector $b = (4, y, t, z)$.

\subsubsection{Comparison with field theory}

We cannot construct supersymmetry preserving baryons in the field theory of section \ref{FLAVORC3}.  However, we can still match the field theory operators $\det \left(X_i \right)$ to giant gravitons on the gravity side, as discussed at the end of section~\ref{SYMMETRY}.  These are stabilized M5-branes wrapping topologically trivial cycles.  We propose that these cycles are simply the $\Sigma_i$ from the toric diagram.  More specifically, label the points $1$, $2$, and $3$ in the toric diagram by $a_0$, $b_0$, and $c_0$, respectively.  Let the towers of points above these three points be denoted by $a_i$, $b_i$, and $c_i$.  In this notation points $4$, $5$, and $6$ in the toric diagram correspond to $a_{n_1}$, $b_{n_2}$, and $c_{n_3}$.  We can solve the moduli space equations by writing \cite{Benini:2009qs}
\es{modspace}{
X_1 &= a_0 a_1 \cdots a_{n_1} \,, \qquad X_2 = b_0 b_1 \cdots b_{n_2} \,, \qquad X_3 = c_0 c_1 \cdots c_{n_3} \,, \\
T &= \left(a_0^{n_1} a_1^{n_1 - 1} \cdots a_{n_1 - 1}\right) \left(b_0^{n_2} b_1^{n_2 - 1} \cdots b_{n_2 - 1}\right) \left(c_0^{n_3} c_1^{n_3 - 1} \cdots c_{n_3 - 1}\right)\,, \\
\tilde T &= \left( a_1 a_2^2 \cdots a_{n_1}^{n_1} \right) \left( b_1 b_2^2 \cdots b_{n_2}^{n_2} \right) \left( c_1 c_2^2 \cdots c_{n_3}^{n_3} \right) \,.
}
This tells us, for example, that the field theory operator $\det \left(X_1\right)$ corresponds to an M5-brane wrapping the entire $a_i$ tower.  However, only the points $1$ and $4$ along that tower correspond to $5$-cycles.  Thus the operator $\det \left(X_1\right)$ is dual to an M5-brane wrapping $\Sigma_1$ and $\Sigma_4$.

On the field theory side we let the R-charges $R\left[X_1\right] = \Delta_1$, $R\left[X_1\right] = \Delta_2$, and $ R\left[T\right] - R[\tilde T] = 2 \Delta_m$ be free parameters.  We match these R-charges to the volumes of their corresponding cycles using eq.~\eqref{DeltaBaryon}.  For example, the equation for $\Delta_m$ is
\es{deltam}{
\frac{\pi}{6} \frac{ n_1 \left[ \Vol \left(\Sigma_4\right) - \Vol \left(\Sigma_1\right) \right] + n_2 \left[ \Vol \left(\Sigma_5\right) - \Vol \left(\Sigma_2\right) \right] + n_3 \left[ \Vol \left(\Sigma_6\right) - \Vol \left(\Sigma_3\right) \right]}{\Vol \left( Y_b \right)}  = 2 \Delta_m \, .
}
We can solve for the Reeb vector parameters $y$, $t$, and $z$ in terms of $\Delta_1$, $\Delta_2$, and $\Delta_m$.  The volume of the base space $Y$ can then be expressed as a function of the field theory R-charges, giving
\es{vol3n3fR}{
\Vol \left( Y_\Delta \right) = \frac{\pi^4}{12} \frac{\left(n_1 \Delta_1 + n_2 \Delta_2 + n_3 \Delta_3 \right)}{\Delta_1 \Delta_2 \Delta_3 \left[ \left(n_1 \Delta_1 + n_2 \Delta_2 + n_3 \Delta_3 \right)^2 - \left(2 \Delta_m \right)^2 \right]} \,.
}
Equation \eqref{vol3n3fR} is subject to the constraint $\Delta_1 + \Delta_2 + \Delta_3 = 2$.  Remarkably, equation \eqref{vol3n3fR} is in exact agreement with equation \eqref{FExtOneNode} for the free energy of the field theory combined with equation \eqref{MtheoryExpectation}.  $Z$-minimization requires that in order for the metric on $Y$ to be  Sasaki-Einstein one needs to minimize $\Vol(Y_\Delta)$ with respect to $\Delta_i$ and $\Delta_m$ under the constraint $\Delta_1 + \Delta_2 + \Delta_3 = 2$.  Since this minimization problem is equivalent to the maximization of $F$ in eq.~\eqref{FExtOneNode}, the dimensions of the determinant operators dual to giant gravitons are consistent with eq.~\eqref{DeltaBaryon}.

\subsection{Toric gauge theories with two gauge groups}

\subsubsection{ABJM at level $k$ with arbitrary R-charges}

\begin{figure}[htb]
\begin{center}
\leavevmode
\scalebox{0.7}{\includegraphics{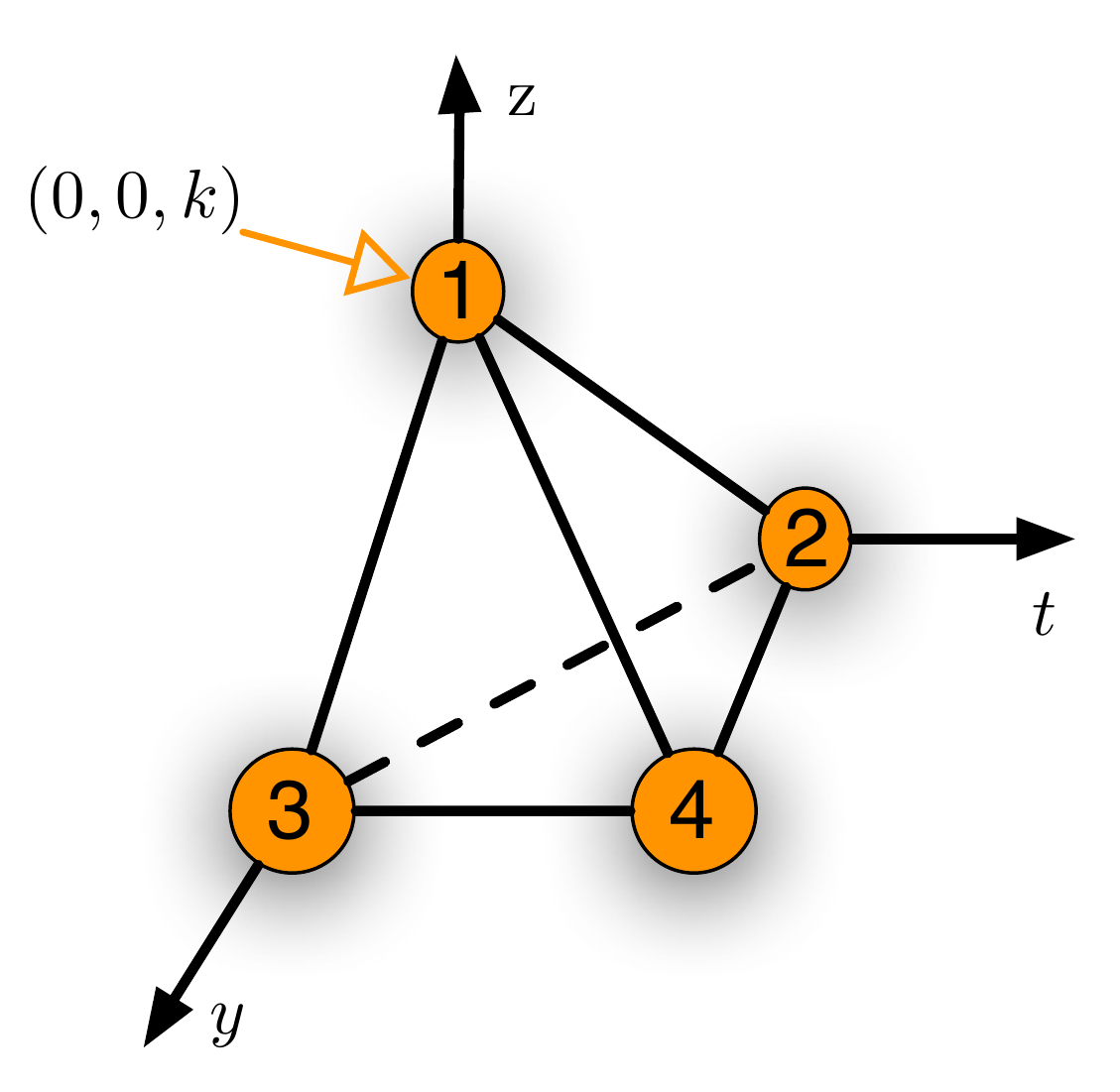}}
\end{center}
\caption{The toric diagram for the ABJM theory at CS level $k$.}
\label{ABJMPDF}
\end{figure}

The toric diagram corresponding to the $U(N) \times U(N)$ ABJM theory at CS level $k$ is given in figure~\ref{ABJMPDF} (see for example \cite{Benini:2009qs}).
  This field theory is discussed in section \ref{ABJMNOFLAVORS}.  The four vertices labeled 1, 2, 3, and 4 each correspond to facets that are tetrahedra.  In terms of a trial Reeb vector $b = \left(4,y,t,z\right)$, the volumes of these cycles are determined to be
\es{ABJMCycles}{
  \Vol \left( \Sigma_1 \right) &= \frac{k^2 \pi^3}{\left[ k \left(4 - y\right) - z \right] \left[ k \left(4 - t\right) - z \right] \left[ k \left(y + t - 4\right) + z \right]} \,,\\
  \Vol \left( \Sigma_2 \right) &=\frac{k^2 \pi^3}{\left[ k \left( y + t - 4\right) + z \right] \left[ k \left(4 - t\right) - z \right] z} \,,\\
  \Vol \left( \Sigma_3 \right) &= \frac{k^2 \pi^3}{\left[ k \left(4 - y\right) - z \right]\left[ k \left( y + t - 4\right) + z \right]  z} \,, \\
  \Vol \left( \Sigma_4 \right) &= \frac{k^2 \pi^3}{\left[ k \left(4 - y\right) - z \right] \left[ k \left(4 - t\right) - z \right]  z} \,.
}

At CS level $k$, the monopole operators $T$ and $\tilde T$ have gauge charge $k$ and $-k$ under the gauge field $A_{1\mu} - A_{2\mu}$, while the bifundamental fields $A_i$ and $B_i$ have gauge charges $1$ and $-1$, respectively.   We may therefore construct the four gauge invariant operators $\det (\tilde T A_i^k )$ and $\det ( T B_i^k)$.  The operator $\det (\tilde T A_1^k)$ is dual to a giant graviton wrapping the topologically-trivial cycle  $k \Sigma_1$.  Similarly,  $\det (\tilde T A_2^k)$ corresponds to $k \Sigma_4$, $\det (\tilde T B_1^k)$ to $k \Sigma_2$, and $\det (\tilde T B_2^k)$ to $k \Sigma_3$.  The R-charges of these operators in field theory are given by
\es{ABJMR}{
R[\tilde T A_i^k] = k \Delta_{A_i} - \Delta_m \,, \qquad
R[T B_i^k] =  k \Delta_{B_i} + \Delta_m   \,.
}
Matching these R-charges to the volumes of their corresponding cycles in field theory using eq.~\eqref{DeltaBaryon}, we may write the volume of the internal space $Y$ as a function of the R-charges,
\es{ABJMYD}{
\Vol \left(Y_\Delta \right) = \frac{\pi^4 k^3}{48 \left( k \Delta_{A_1} - \Delta_m \right) \left( k \Delta_{A_2} - \Delta_m \right) \left( k \Delta_{B_1} + \Delta_m \right) \left( k \Delta_{B_1} + \Delta_m \right)} \,.
}
Equation \eqref{ABJMYD} is subject to the constraint
\es{ABJMRC}{
\Delta_{A_1} + \Delta_{A_2} + \Delta_{B_1} +\Delta_{B_2} = 2 \,.
}
In this case too, the volume of the internal space as a function of the R-charges exactly matches equation \eqref{FreeExtABJM} for the free energy of the ABJM theory with arbitrary R-charges.

\subsubsection{A general flavoring of the conifold quiver at $k=0$}
\label{genflavABJM}

\begin{figure}[htb]
\begin{center}
\leavevmode
\scalebox{0.7}{\includegraphics{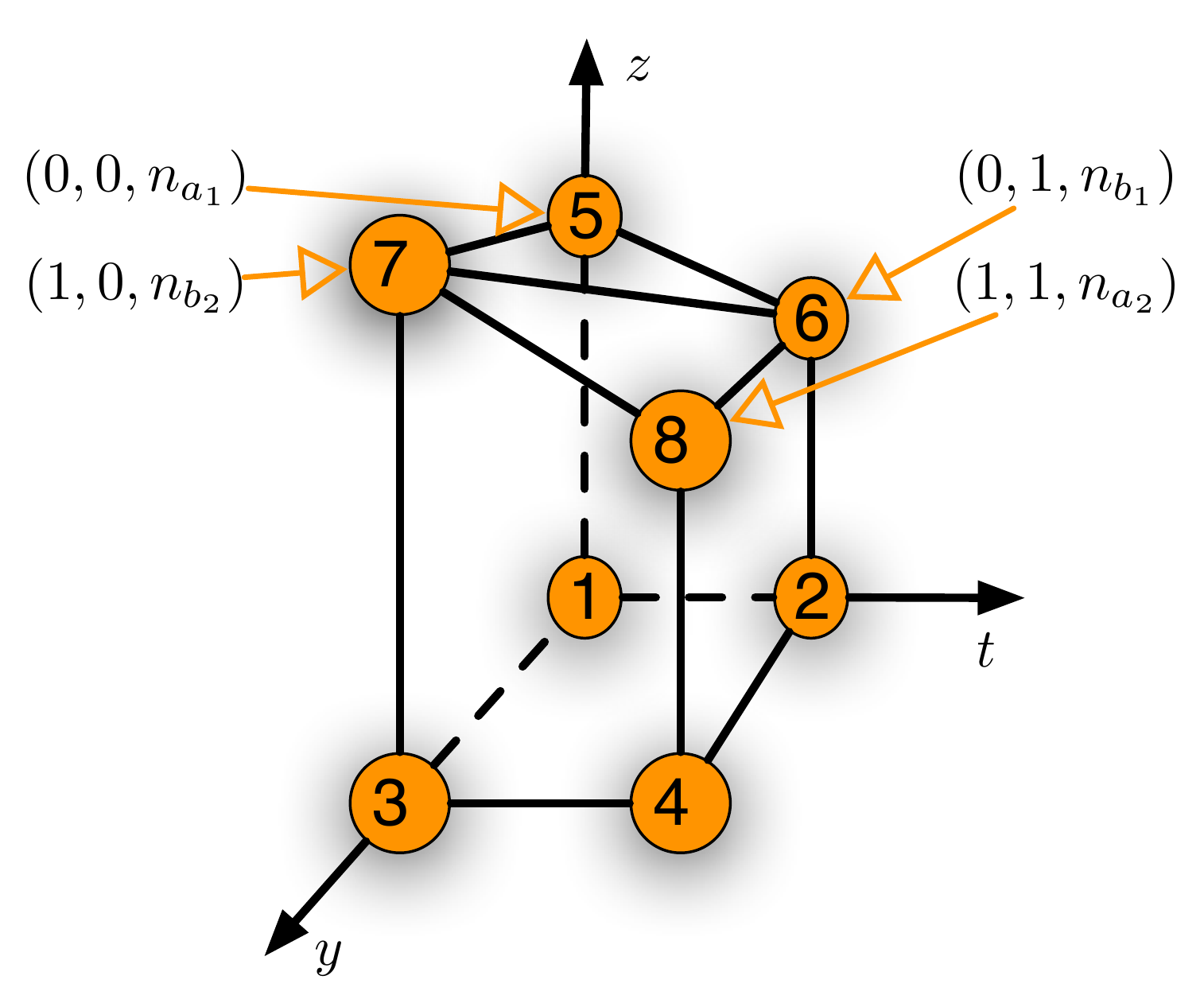}}
\end{center}
\caption{The toric diagram for the generally flavored conifold quiver of section~\ref{INFINITE}.  Intermediate flavor points are suppressed.   }
\label{ABJMGENPDF}
\end{figure}

We now consider flavoring all four bifundamental fields in ABJM as is discussed in section~\ref{INFINITE}.\footnote{For analogous reasons to those given in section \ref{TORICARB}, we only show the calculation for the cases where $\{n_{a1}\, ,n_{a2} \, ,n_{b1} \,, n_{b2} \}$ are greater than zero.  However, the final expression for the volume of the internal space will be valid more generally for any $\{n_{a1}\, ,n_{a2} \, ,n_{b1} \,, n_{b2} \}$ satisfying equation \eqref{nRelation}. }  The toric diagram for these theories is given in figure~\ref{ABJMGENPDF} \cite{Benini:2009qs}.
  In terms of the trial Reeb vector $b = \left(4,y,t,z\right)$, the volumes of the $5$-cycles corresponding to facets one through eight are
\es{VolCGenCycles}{
\Vol \left( \Sigma_1 \right) &= \frac{\pi^3}{ytz} \,, \qquad
   \Vol \left( \Sigma_2 \right) = \frac{\pi^3}{y\left(4 - t \right)z} \,, \\
   \Vol \left( \Sigma_3 \right) &= \frac{\pi^3}{\left(4-y\right)tz} \,, \qquad
\Vol \left( \Sigma_4 \right) = \frac{\pi^3}{\left(4-y\right)\left(4-t\right)z} \,, \\
   \Vol \left( \Sigma_5 \right) &= \frac{\pi^3}
   {yt\left[n_{a1} \left(4 - y - t\right) + n_{b1} t + n_{b2} y - z\right]} \,, \\
\Vol \left( \Sigma_6 \right) &= \frac{\pi^3}{y\left(4-t\right)
   \left[n_{a1} \left(4 - t\right) + n_{a2} y + n_{b1} \left(y - t\right) - z\right]} \,, \\
\Vol \left( \Sigma_7 \right) &= \frac{\pi^3}{\left(4-y\right)t\left[n_{a1} \left(4 - y\right)
   + n_{a2} t + n_{b2} \left( y - t \right) - z\right]} \,, \\
\Vol \left( \Sigma_8 \right) &= \frac{\pi^3}{\left(4-y\right)\left(4-t\right)
   \left(n_{a2} \left( y + t - 4\right) + n_{b1} \left(4 - y\right) + n_{b2} \left(4 - t\right) - z\right]} \,.
}

Consider the field theory with gauge group $SU\left(N\right) \times SU\left(N\right) \times U\left(1\right)$.  We can solve the moduli space equations in an analogous fashion to equations \eqref{modspace} and construct the baryon operators $\mathcal{B} \left(A_i\right)$, $\mathcal{B} \left(B_i\right)$, $\mathcal{B} \left(T \right)$, and $\mathcal{B} (\tilde T)$.  This tells us which cycles the baryon operators are dual to.  For example, the operator $\mathcal{B} (A_1)$ corresponds to an M5-brane wrapping  $\Sigma_1$ and $\Sigma_5$.  Following the by now familiar procedure of matching cycle volumes to baryonic R-charges, we then solve for the Reeb vector parameters $y$, $t$, and $z$ in terms of the field theory R-charges.  This allows us to express the volume of the internal space $Y$ as a function of the R-charges.  This function is exactly what one gets by combining equation \eqref{FreennnnAgain} for the free energy of the flavored ABJM theory with equation \eqref{MtheoryExpectation}, along with the $SU\left(N\right)$ condition in equation \eqref{intrhodynnnn}.

\subsubsection{$\C \times \mathcal{C}$: Flavoring one bifundamental in the conifold quiver at $k = 0$}
\label{CCONTORIC}

\begin{figure}[htb]
\begin{center}
\leavevmode
\scalebox{0.7}{\includegraphics{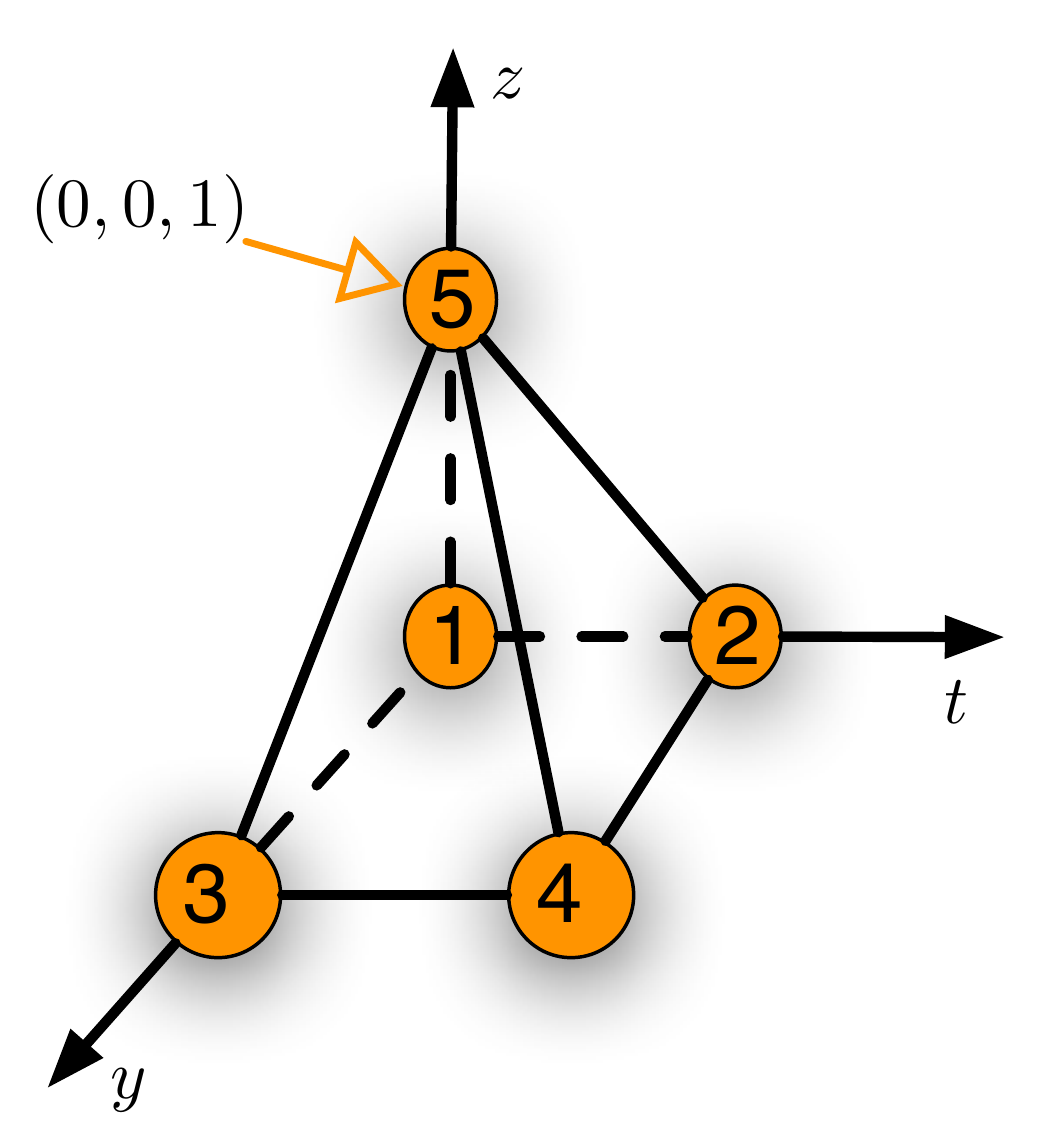}}
\end{center}
\caption{The toric diagram for $\C \times \mathcal{C}$.}
\label{CCONPDF}
\end{figure}

We now consider flavoring the bifundamental field $A_1$ in the ABJM model, as is discussed in section~\ref{CCONABJM}.  The toric diagram for this theory is given in figure~\ref{CCONPDF} \cite{Benini:2009qs}.
 Note that in this case the facets labeled one through four correspond to tetrahedra while the facet labeled number five corresponds to a pyramid in $\mathbb{R}^4$.  The volumes of the cycle corresponding to each node are
 \es{volTorCycles}{
  \Vol \left( \Sigma_1 \right) &= \frac{\pi^3}{ytz} \,, \qquad
    \Vol \left( \Sigma_2 \right) = \frac{\pi^3}{y\left(4 - t -z\right)z} \,, \\
   \Vol \left( \Sigma_3 \right) &= \frac{\pi^3}{\left(4- y - z\right)tz} \,, \qquad
   \Vol \left( \Sigma_4 \right) = \frac{\pi^3}{\left(4-y - z\right)\left(4-t-z\right)z} \,, \\
  \Vol \left( \Sigma_5 \right) &= \frac{\pi^3}{yt\left(4- y -z\right)} \,, \qquad
  \Vol \left( \Sigma_6 \right) = \frac{\pi^3 \left(4-z\right)}{yt\left(4-t-z\right)\left(4-y-z\right)} \,.
 }

Matching the volume of the base of the internal manifold of this theory to the free energy of the field theory proceeds analogously to the discussion in section \ref{genflavABJM}.  The most rigorous test of the theory is obtained from the $SU\left(N\right) \times SU\left(N\right) \times U\left(1\right)$ theory.  We again have the baryon operators $\mathcal{B} \left(A_i\right)$, $\mathcal{B} \left(B_i\right)$, $\mathcal{B} (T )$, and $\mathcal{B} (\tilde T )$.  Solving the moduli space equations tells us which baryons are dual to which wrapped cycles.  For example, an M5-brane wrapping the $5$-cycle $\Sigma_5$ is dual to the baryon $\mathcal{B} \left(T \right)$ while $\mathcal{B} (\tilde T )$ corresponds to the cycle $\Sigma_1$.  As $T\tilde T = A_1$, we see that the baryon $\mathcal{B} \left(A_1\right)$ is dual to an M5-brane wrapping both cycles $\Sigma_1$ and $\Sigma_5$.  Solving for the Reeb vector parameters in terms of the R-charges we reproduce eq.~\eqref{FreeCConTwo} for the free energy of the $\C \times \mathcal{C}$ field theory, as a function of the trial R-charges, given equation  \eqref{MtheoryExpectation}.

\subsubsection{$Q^{1,1,1}$: Flavoring $A_1$ and $A_2$ in the conifold quiver at $k = 0$}
\label{Qtoric}

In this section we present the toric geometry computation for the volume of the space dual to the $Q^{1,1,1}/\mathbb{Z}_n$ field theory of section \ref{Q111SECTION}.  With critical R-charges the internal space is simply $Q^{1,1,1}/\mathbb{Z}_n$.  However, we again find that the toric geometry computation for the volume matches the field theory free energy for non-critical R-charges.

The toric diagram for this theory is a special case of figure~\ref{ABJMGENPDF}, with $n_{a1} = n_{a2} = 1$ and $n_{b1} = n_{b2} = 0$.  For simplicity we only compare the field theory free energy to the volume of the internal manifold for arbitrary $\Delta_m$.  We know by symmetry that at the extremum of this theory
\es{deltaCond}{
\Delta_{A_1} = \Delta_{A_2} \, \quad \text{and} \, \quad \Delta_{B_1} = \Delta_{B_2} \,.
}
Again consider the $SU\left(N\right) \times SU\left(N\right) \times U\left(1\right)$ field theory.  The constraint in equation~\eqref{deltaCond}  in field theory is dual to the constraint in geometry
\es{deltaCondG}{
\Vol \left(\Sigma_1\right) + \Vol \left(\Sigma_6\right) = \Vol \left(\Sigma_4\right) + \Vol \left(\Sigma_5\right) \,,
 \qquad
  \Vol \left(\Sigma_2\right) = \Vol \left(\Sigma_3\right)  \, .
}
The constraint equations in \eqref{deltaCondG} are solved for $y = t = 2$.  After imposing this constraint, the volumes of the cycles corresponding to each node are
 \es{VolQ111Cycles}{
\Vol \left( \Sigma_1 \right) &= \Vol \left( \Sigma_5 \right) = \frac{\pi^3 \left( 4 n -z\right)}{4 \left( z - 2n \right)^2} \,, \\
 \Vol \left( \Sigma_2 \right) &= \Vol \left(\Sigma_3 \right) = \frac{n \pi^3}{4 n^2 - z^2} \,, \\
\Vol \left( \Sigma_4 \right) &= \Vol \left( \Sigma_6 \right) =  \frac{\pi^3 \left( 4 n + z\right)}{4 \left( z + 2n \right)^2} \,.
}
The monopole charge $\Delta_m$ is related to $z$ by the equation
\es{monChargeQ}{
2 \Delta_m = \frac{n \pi}{6} \frac{ \left[ \Vol\left(\Sigma_1\right) - \Vol\left(\Sigma_6\right) \right] + \left[ \Vol\left(\Sigma_5\right) - \Vol\left(\Sigma_4\right) \right] }{\Vol \left( Y_b \right)} \,.
}
Solving for $z$ we find that $z = 2 \Delta_m$, which gives the volume
\es{QVolume}{
\Vol \left( Y_\Delta \right) = \frac{n \pi^4}{24} \frac{3 n^2 - \Delta_m^2}{ \left(n^2 - \Delta_m^2 \right)^2}
}
of the internal manifold $ Y_\Delta$ as a function of the bare monopole R-charge $\Delta_m$.  This is in exact agreement with equation \eqref{FQ111} for the free energy of the field theory as a function of $\Delta_m$.

\subsubsection{$Q^k$: Flavoring $A_1$ and $A_2$ in the conifold quiver with $k > 0$}
\label{QKARBSEC}

This section generalizes the discussion in section \ref{Qtoric} to quivers with $k > 0$.  This is the field theory discussed in section~\ref{QKSECTION}.  For $k \neq 1$ the dual internal space $Q_k$ is nonsingular.  We compute the volume of the dual space as a function of monopole charge $\Delta_m$ and we state the volume at extremum for arbitrary $k$.

\begin{figure}[htb]
\begin{center}
\leavevmode
\scalebox{0.7}{\includegraphics{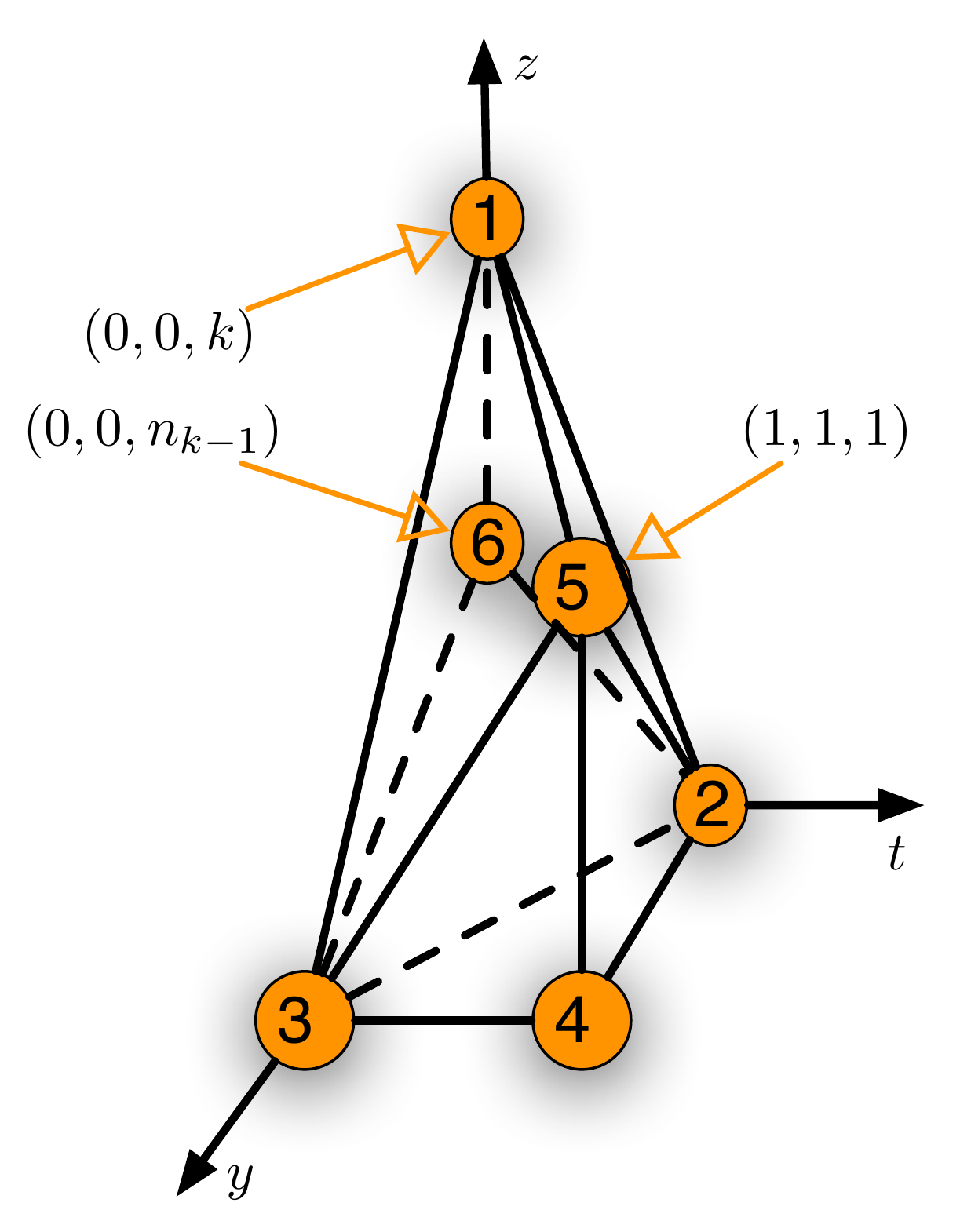}}
\end{center}
\caption{The toric diagram for the $Q_k$ theory at CS level $k$.   This theory is discussed in section~\ref{QKSECTION}.}
\label{ABJMQKPDF}
\end{figure}

The toric diagram for $Q_k$ is given in figure~\ref{ABJMQKPDF} \cite{Benini:2009qs}.
  We will use the same arguments presented in section \ref{Qtoric} and require the constraints \eqref{deltaCond} in field theory.  This again requires $y = t = 2$.  The volumes of the cycles corresponding to each node are given as functions of $z$ by
 \es{VolQk12Cycles}{
   \Vol \left( \Sigma_1 \right) &= \Vol \left( \Sigma_5 \right) = \frac{\pi^3 \left[ 4\left(1 + k\right)  -z\right]}{4 \left[2 \left(1 + k\right) - z \right]^2} \,, \\
    \Vol \left( \Sigma_2 \right) &= \Vol \left(\Sigma_3 \right) = \frac{ \pi^3 \left( k^2 - 1 + z\right)}{z^2\left[2 \left(1 + k\right) - z \right]} \,, \\
    \Vol \left( \Sigma_4 \right) &= \Vol \left( \Sigma_6 \right) =  \frac{\pi^3}{4z} \,.
}
Using equation \eqref{monChargeQ} we solve for $z$ in terms of $\Delta_m$.  The volume of the internal manifold $Y_\Delta$ is a nontrivial function of $\Delta_m$ when written explicitly.  However, it is not hard to show that the volume exactly matches the $U\left(N\right) \times U\left(N\right)$ free energy in equation \eqref{FQk} subject to the $SU\left(N\right)$ constraint in equation \eqref{intrhodeltayQk}.

It is useful to extremize the formula for the volume of the internal manifold $Y_\Delta$ in order to find the volume of the internal space $Q_k$.  An explicit computation gives
\es{volQ111k}{
\Vol{\left(Q_{k} \right)} = \frac{\pi^4 \left[2 \left(k-1\right)\left(1+k\right)^2 + 4 \left(1+k\right)z^*-{z^*}^2\right]}{6{z^*}^2\left[2\left(1+k\right)-z^*\right]^2} \,
}
with\footnote{We thank Stefano Cremonesi for pointing out a typo in the formula for $z_*$ in an earlier version of this paper.}
\es{zQ}{
z^* &= \frac{2}{3} \left(1+ k \right) \left[ 3 + \sqrt{1+k}\left(\sqrt{3} \cos \theta - 3 \sin \theta \right) \right] \,,\\
 \theta &\equiv \frac{1}{3} \arg\left( - 9 + i \sqrt{48 k -33} \right)  \, .
}

\section{Theories with $N^{5/3}$ scaling of the free energy}
\label{FIVETHIRDS}

\begin{figure}[htb]
\begin{center}
\leavevmode
\newcommand{\svgwidth}{.5\textwidth}
\definecolor{orange}{RGB}{255,165,0}
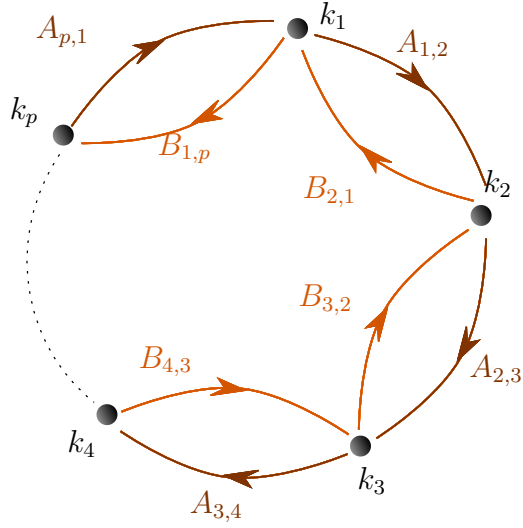
\end{center}
\caption{A ``necklace'' quiver diagram for a class of ${\cal N} = 3$ models with $U(N)$ gauge groups and CS levels $k_a$.  When the CS levels don't sum to zero, the free energy behaves as $N^{5/3}$ at large $N$.}
\label{NQUIV}
\end{figure}

In this section we consider a class of ${\cal N} = 3$ models where the field theory free energy scales as $N^{5/3}$ in the large $N$ limit.  These models are depicted in figure~\ref{NQUIV}. 
 They consist of $p$ $U\left(N\right)$ gauge groups at CS levels $k_a$, for $a = 1, \ldots, p$, with two non-chiral bifundamental pairs connecting adjacent nodes.  The matrix model for this theory is discussed in \cite{Herzog:2010hf} for the special case $k \equiv \sum_{a=1}^{p} k_a= 0$.  In the large $N$ limit the free energy scales as $N^{3/2}$ in this case.  However, we show below that when $k \neq 0$ the free energy has $N^{5/3}$ scaling.

Let $\lambda^{\left(a\right)} \left(x\right)$ denote the eigenvalues for the $a$th gauge group in the large $N$ continuum limit.  We make the ansatz that the eigenvalues corresponding to all the gauge groups are the same to leading order in $N$ and that they scale as $N^{\alpha}$:
\es{karbEig}{
\lambda^{\left(a\right)} \left(x\right) = N^{\alpha} \left(x + i y \left(x\right) \right)
}
for some $0 < \alpha < 1$.  The leading terms in the free energy for large $N$ may then be written as
\begin{eqnarray} \label{karbF}
F &=& \frac{N^{1+2\alpha}}{\pi} \frac{k}{p} \int dx \rho\left(x\right) x y\left(x\right) - i \frac{N^{1+2\alpha}}{2 \pi} \frac{k}{p} \int dx \rho\left(x\right) \left(x^2 - y^2\right)  \nonumber \\
&+& \pi^2 N^{2-\alpha} \int dx \frac{ \rho^2\left(x\right)}{1 + i y'} \,.
\end{eqnarray}
We assume that $k \neq 0$ in the following discussion.  Then we must take $\alpha = 1/3$ in order for the functions $\rho\left(x\right)$ and $y\left(x\right)$ to remain order one in the large $N$ limit.

The saddle point of the free energy in equation \eqref{karbF} may then be found using the same methods presented in the earlier sections.  After a small calculation one finds that the free energy is maximized when
\es{karbSoln}{
y \left(x\right) = \frac{1}{\sqrt{3}} x \,, \qquad \rho\left(x\right) = \frac{ 3^{1/6} k^{1/3} }{2 \pi p^{1/3}} - \frac{2 k x^2}{3 \sqrt{3} \pi^3 p} \,.
}
The free energy at the saddle point is given by
\es{karbFAgain}{
F =  \frac{3^{7/6} \left(3 - i \sqrt{3} \right) \pi}{20 p^{1/3}}    k^{1/3} N^{5/3}  \,.
}
According to \eqref{basicF}, only the real part of this quantity is a measure of the number of degrees of freedom.  The imaginary part is only defined modulo $2 \pi$ and contributes to the phase of the path integral. 

%The massive type IIA duals of theories with $k\neq 0$ were recently studied in \cite{Aharony:2010af}, and the $k^{1/3} N^{5/3}$ behavior of the free energy was observed there as well.

Theories with $k\neq 0$ have dual descriptions in massive type IIA theory \cite{Gaiotto:2009mv,Gaiotto:2009yz}.  These supergravity solutions were found in \cite{Petrini:2009ur,Lust:2009mb,Aharony:2010af}. In these backgrounds the free energy was found to scale as $k^{1/3} N^{5/3}$ \cite{Aharony:2010af} in agreement with the scaling we have found in the matrix models.

\section{Discussion}
\label{DISCUSSION}

In this paper we calculated the three-sphere free energy $F$ for a variety of ${\cal N}=2$ superconformal gauge theories with large numbers of colors. The localization of the free energy for such theories, which allows for varying the R-charges of the fields, was carried out in \cite{Jafferis:2010un,Hama:2010av}, and we used their results to write down and solve a variety of large $N$ matrix models with the method introduced in \cite{Herzog:2010hf}. The subsequent maximization of $F$ over the space of trial R-charges consistent with the marginality of the superpotential fixes them and the value of $F$.
The results we find are in complete agreement with the conjectured dual $AdS_4\times Y$ M-theory backgrounds.
We have also studied various RG flows and have found that $F$ decreases in all of them.  $F$ is also constant along exactly marginal directions.  Thus, $F$ seems to be a good candidate to serve as a 3-d analogue of the $a$-coefficient in the 4-d Weyl anomaly. This has led us to propose the $F$-theorem in three dimensions, analogous to the $a$-theorem in 4-d.

The reader will note that none of the models solved in this paper include chiral bi-fundamental fields. Instead, we have relied on
models with non-chiral bifundamentals, such as the ABJM model, which may be coupled to a rather general set of fundamental fields, either
chiral or non-chiral. Constructions of this kind were used in \cite{Jafferis:2009th, Benini:2009qs} to conjecture gauge theories dual to a variety of ${\cal N}=2$ M-theory backgrounds, including such well-known solutions as $AdS_4\times Q^{1,1,1}$ and $AdS_4\times V_{5,2}$.
These novel conjectures rely heavily on non-perturbative effects associated with monopole operators: in fact, in these theories the monopole operators play a geometrical role on an equal footing with the fields in the lagrangian. Our work, as well as the superconformal index calculation
for the flavored $AdS_4\times Q^{1,1,1}$ model \cite{Cheon:2011th}, provides rather intricate tests of these conjectures.

The earlier and perhaps better known conjectures for the gauge theories dual to $AdS_4\times M^{1,1,1}$, $AdS_4\times Q^{1,1,1}$ and $AdS_4\times Q^{2,2,2}$ \cite{Hanany:2008cd,Martelli:2008si,Franco:2008um,Franco:2009sp,Davey:2009sr,Amariti:2009rb} have instead involved chiral bifundamental fields. The rules derived in \cite{Jafferis:2010un,Hama:2010av} seem to apply to these models as well, and we have attempted to study these matrix models both numerically and analytically.  Unfortunately, the essential phenomenon in the matrix models exhibiting the $N^{3/2}$ scaling of the free energy, namely the cancellation of long range forces between the eigenvalues, cannot be achieved in the theories with chiral bifundamentals.  As a result, the range of the eigenvalues grows as $N$, rather than $\sqrt{N}$, and the free energy scales as $N^2$. The latter behavior is in obvious contradiction with the M-theory result \eqref{MtheoryExpectation}.  As $N$ increases, the eigenvalue distribution does not become dense; instead, the gaps do not shrink as $N$ is increased. This leads to an entirely different structure from what we have observed in the various matrix models that do produce the desired $N^{3/2}$ scaling of the free energy. The question whether the matrix models with chiral bifundamentals can be ``repaired'' is an interesting one and we hope it will be investigated further.\footnote{We thank F. Benini for useful discussions of this issue.}

More generally, we find it exciting that  the $F$-theorem for the three-sphere free energy might hold. Such a theorem should be applicable to all 3-d theories, either supersymmetric or not. Further tests of these ideas, as well as attempts at a general field theoretic proof, would be very useful at this stage.

\section*{Note Added}

After the calculations presented here were completed, and as the paper was being finalized, we received two papers \cite{Martelli:2011qj,Cheon:2011vi} that partly overlap with our results. We disagree, however, with the claim made in \cite{Cheon:2011vi} that the matrix model for the flavored $AdS_4\times Q^{1,1,1}$ theory is inconsistent with the $F$-maximization principle.
We demonstrate its consistency in section \ref{Q111SECTION}.

\section*{Acknowledgments}

We thank F.~Benini, C.~Herzog, M.~Kiermaier, T.~Klose, J.~Maldacena, T.~Tesileanu, and M.~Yamazaki for useful discussions.  The work of DLJ was supported in part by DOE grant DE-FG02-90ER40542.  IRK, SSP, and BRS were supported in part by the US NSF  under Grant No.~PHY-0756966.  The work of SSP was also supported in part by Princeton University through a Porter Ogden Jacobus Fellowship, and that of BRS by the NSF Graduate Research Fellowship Program.

\newpage

\appendix

\section{Derivation of general rules}
\label{DERIVATION}

In this section we provide a derivation of the rules we gave in section~\ref{RULES} for finding the continuum limit of the free energy in eq.~\eqref{PartitionFunction}.  We assume that at large $N$ the eigenvalues scale as
 \es{LargeNScaling}{
  \lambda_i^{(a)} = N^\alpha x_i + i y_{a, i} + o(1)
 }
for some number $\alpha \in (0, 1)$.  We will eventually be interested in setting $\alpha = 1/2$.  In writing eq.~\eqref{LargeNScaling} we implicitly assume that as we take $N$ to infinity, the $x_i$ and $y_{a, i}$ become dense, so in the continuum limit we can express $y_a$ as a continuous function $y_a(x)$.   It is convenient to define the density
 \es{rhoDef}{
  \rho(x) = \frac{1}{N} \sum_{i = 1}^N \delta(x - x_i) \,,
 }
which as we take $N \to \infty$ also becomes a continuous function of $x$ normalized so that $\int dx\, \rho(x) = 1$.

\subsection{First rule}

For each gauge group $a$ with CS level $k_a$ and bare monopole R-charge $\Delta_m^{(a)}$, the discrete contribution to $F$ is
 \es{F1Discrete}{
  F_1 &= \sum_{i = 1}^N \left[ -\frac{i k_a}{4 \pi} \left( \lambda_{a, i} \right)^2 + \Delta_m^{(a)} \lambda_{a, i} \right] \\
   &=\sum_{i = 1}^N \left[ -\frac{i k_a}{4 \pi} \left( N^\alpha x_i + i y_{a, i} \right)^2 + \Delta_m^{(a)} \left( N^\alpha x_i + i y_{a, i} \right) \right]  + o(N) \,,
 }
where in the second line we used \eqref{LargeNScaling}.  Expanding in $N$ we obtain
 \es{F1DiscreteApprox}{
  F_1 = -\frac{i k_a}{4 \pi} N^{2 \alpha} \sum_{i = 1}^N x_i^2 + 
  N^{\alpha} \sum_{i = 1}^N \left[ \frac{k_a}{2 \pi}  x_i y_{a, i} + \Delta_m^{(a)} x_i \right]  + o(N) \,.
 }
The first term in this sum vanishes when we sum over $a$ because we assume $\sum_a k_a = 0$.  In taking the continuum limit we therefore only need to keep the second term and replace $\sum_i\left( \cdots \right)$ by $N \int dx\, \rho(x) \left( \cdots \right)$.  We get
 \es{GotF1}{
  F_1 = N^{\alpha+1} \int dx\, \rho(x) \left[ \frac{k_a}{2 \pi}  x y_a(x) + \Delta_m^{(a)} x \right] + o(N) \,,
 }
reproducing eq.~\eqref{CSContribution} when $\alpha = 1/2$.

\subsection{Second and third rules}

The interaction terms between the eigenvalues contain two types of terms: one coming from the one-loop determinant of the fields in the vector multiplets
 \es{F2Vector}{
  F_{2, \text{vector}} = - \frac 12 \sum_{a = 1}^p \sum_{i, j = 1}^N \ln \left( 4 \sinh^2 \frac{\lambda_i^{(a)} - \lambda_j^{(a)}}{2} \right)
 }
for each gauge group $a$, and one coming from the one-loop determinants of the matter fields
 \es{F2Pair}{
  F_{2, \text{matter}} = -\sum_{\substack{\text{bifundamentals}\\(a, b)}} \sum_{i, j = 1}^N \ell\left(1 - \Delta_{(a, b)} + i \frac{\lambda_i^{(a)} - \lambda_j^{(b)}}{2 \pi} \right)  \,,
 }
where the function $\ell(z)$ was defined in eq.~\eqref{ellDef}.  Since for each gauge group we require the relation \eqref{DeltaSum} to hold, we can rewrite the contribution $F_{2, \text{vector}}$ as
 \es{F2VectorAgain}{
  F_{2, \text{vector}} = -\frac 14\sum_{\substack{\text{bifundamentals}\\(a, b)}}  \left(1-\Delta_{(a, b)}\right) \sum_{i, j=1}^N  \ln \left(16 \sinh^2 \frac{\lambda_i^{(a)} - \lambda_j^{(a)}}{2} \sinh^2 \frac{\lambda_i^{(b)} - \lambda_j^{(b)}}{2} \right) \,.
 }
Combining \eqref{F2VectorAgain} and \eqref{F2Pair} one can write the interaction term in the free energy as a sum over the bifundamental fields.  

In order to calculate $F_2 = F_{2, \text{matter}} + F_{2, \text{vector}}$ as $N \to \infty$ we find it easier to first calculate the derivatives of $F_2$ with respect to $y_{a}$.
We have
 \es{F2Discrete}{
  &\frac{\partial F_2}{\partial y_{a, i}} = \sum_{j=1}^N \bigg[ -\frac{i}{2} \left( 2 - \Delta_{(a, b)} - \Delta_{(b, a)} \right) \coth \frac{N^\alpha (x_i - x_j) + i (y_{a, i} - y_{a, j})}{2} \\
  {}&-\frac{1}{4 \pi} \cot \left[\pi \Delta_{(a, b)} + \frac{i N^\alpha (x_j - x_i) + y_{a, i} - y_{b, j}}{2} \right] 
  \left(2 \pi (\Delta_{(a, b)} -1) + i N^\alpha (x_j - x_i) + y_{a, i} - y_{b, j} \right) \\
  {}&+\frac{1}{4 \pi} \cot \left[\pi \Delta_{(b, a)} - \frac{i N^\alpha (x_j - x_i) + y_{a, i} - y_{b, j}}{2} \right] 
  \left(2 \pi (\Delta_{(b, a)} -1) - i N^\alpha (x_j - x_i) + y_{a, i} - y_{b, j} \right) \bigg] \,,
 }
where in the first term we should not let $j= i$.  In the continuum limit this expression becomes
 \es{F2Continuum}{
  &\frac{\delta F_2}{\delta y_a(x)} \approx N^2 \rho(x)^2 \sum_{\substack{\text{bifundamentals}\\ \text{$(a, b)$ and $(b, a)$}}} \text{ P.V.}\int dx'  \bigg[ \frac{i}{2} \left( 2 - \Delta_{(a, b)} - \Delta_{(b, a)} \right) \coth \frac{\lambda_a(x) - \lambda_a(x')}{2} \\
  {}&-\frac{1}{4 \pi} \cot \left[\pi \Delta_{(a, b)} + \frac{i (\lambda_b(x') -\lambda_a(x)  )}{2} \right] 
  \left(2 \pi (\Delta_{(a, b)} -1) + i  (\lambda_b(x') -\lambda_a(x)  ) \right) \\
  {}&+\frac{1}{4 \pi} \cot \left[\pi \Delta_{(b, a)} - \frac{i (\lambda_b(x') -\lambda_a(x)  )}{2} \right] 
  \left(2 \pi (\Delta_{(b, a)} -1) - i (\lambda_b(x') -\lambda_a(x)  ) \right) \bigg] \,,
 }
where P.V. denotes principal value integration and by $\lambda_a(x)$ we mean $N^\alpha x + i y_a(x)$.  In the sum over pairs of bifundamental fields, adjoint fields should be counted once and should come with an explicit factor of $1/2$.  Changing variables from $x'$ to $\xi = N^\alpha (x'-x)$ and taking $N \to \infty$, the integral in \eqref{F2Continuum} becomes  
 \es{F2ContinuumAgain}{
  \frac{\delta F_2}{\delta y_a(x)} &\approx N^{2 - \alpha} \rho(x)^2 \sum_{\substack{\text{bifundamentals}\\ \text{$(a, b)$ and $(b, a)$}}} \int_{-\infty}^{\infty} d \xi  \\
  {}&\Bigg[-\frac{1}{4 \pi} \cot \left[\pi \Delta_{(a, b)} + \frac{i \xi + y_a(x) - y_b(x)}{2} \right] 
  \left(2 \pi (\Delta_{(a, b)} -1) + i \xi + y_a(x) - y_b(x) \right) \\
  {}&+\frac{1}{4 \pi} \cot \left[\pi \Delta_{(b, a)} - \frac{i \xi + y_a(x) - y_b(x)}{2} \right] 
  \left(2 \pi (\Delta_{(b, a)} -1) - i \xi - y_a(x) + y_b(x) \right)
  \Biggr] \,.
 }
This integral converges and can be evaluated to
 \es{FinallyF2Deriv}{
   \frac{\delta F_2}{\delta y_a(x)} &\approx N^{2 - \alpha} \rho(x)^2 \sum_{\substack{\text{bifundamentals}\\ \text{$(a, b)$ and $(b, a)$}}} \bigg[ 
     -f_{ab}(x) \left( \frac{f_{ab}(x)}{4 \pi} + 1-\Delta_{(a, b)}  - \frac{y_a(x) - y_b(x)}{2 \pi} \right) \\
     {}&+ f_{ba}(x) \left( \frac{f_{ba}(x)}{4 \pi} + 1-\Delta_{(b, a)}  - \frac{y_b(x) - y_a(x)}{2 \pi} \right)
     \bigg] \,,
 }
where we have defined
    \es{fDef}{
     f_{ab}(x) \equiv -i \ln e^{i \left[ y_a(x) - y_b(x) + 2 \pi (\Delta_{(a, b)} - 1/2) \right]} \,.
    }
Integrating this expression with respect to $y_a(x)$ one obtains an expression for $F$ up to $y_a$-independent terms.  The $y_a$-independent terms can be found by approximating $F_2$ itself when $y_a = 0$ in the same way that we approximated $\delta F / \delta y_a(x) $ above.  The final answer is
   \es{BifundamentalContribution}{
    F_2 = - \frac{N^{2-\alpha}}{12 \pi} \sum_{\substack{\text{bifundamentals}\\ \text{$(a, b)$}}}  \int dx\, \rho(x)^2 \left[ \pi^2 -f_{ab}^2 \right]
      \left[ 2 f_{ab}
      + 3 \left(y_a - y_b + 2 \pi (\Delta_{(a, b)} - 1) \right) \right] \,.
   }
In the region where $ y_a(x) - y_b(x) + 2 \pi (\Delta_{(a, b)} - 1/2) \in (-\pi, 3 \pi)$, as will be the case most of the time, we have
    \es{fabSimple}{
     f_{ab}(x) = y_a(x) - y_b(x) + 2 \pi \left( \Delta_{(a, b)} - \frac 12\right) \,,
    }
and \eqref{BifundamentalContribution} becomes
 \es{PairAgain}{
    F_2 = -N^{2-\alpha} \sum_{\substack{\text{bifundamentals}\\ \text{$(a, b)$ and $(b, a)$}}} \frac{ 2- \Delta_{(a, b)}^+}{2} \int dx\, \rho(x)^2
      \left[ \left(y_a-y_b + \pi \Delta_{(a, b)}^- \right)^2
      - \frac 13 {\pi^2 \Delta_{(a, b)}^+  \left(4-\Delta_{(a, b)}^+\right)} \right] \,,
 }
where $\Delta_{(a, b)}^{\pm} \equiv \Delta_{(a, b)} \pm \Delta_{(b, a)}$, reproducing eq.~\eqref{Pair} for $\alpha = 1/2$.  Eq.~\eqref{PairAgain} is valid in the range 
 \es{Rangeappendix}{
   \abs{y_a - y_b + \pi \Delta_{(a, b)}^-} \leq \pi \Delta_{(a, b)}^+ \,.
 }

In order to reproduce eq.~\eqref{Adjoint} for a field transforming in the adjoint of the $a$th gauge group, we take $y_a = y_b$, $\Delta_{(a, a)}^+ = 2 \Delta_{(a, a)}$, and $\Delta_{(a, a)}^- = 0$ in one of the terms of \eqref{PairAgain}, which we then multiply by a factor of $1/2$ as explained above.

\subsection{Fourth rule}

The contribution from the fundamental and anti-fundamental fields is
 \es{FundContributionappendix}{
  F_{3} = -\sum_{\substack{\text{fundamental }\\a}} \sum_{i}^N \ell\left(1 - \Delta_{a} + i \frac{\lambda_i^{(a)}}{2 \pi} \right)  
   -\sum_{\substack{\text{anti-fundamental }\\a}} \sum_{i}^N \ell\left(1 - \tilde \Delta_{a} - i \frac{\lambda_i^{(a)}}{2 \pi} \right)  \,,
}
where we denoted the dimension of the fundamentals and anti-fundamentals by $\Delta$ and $\tilde \Delta$, respectively, to avoid confusion.  In the continuum limit, replacing $\sum_i$ by $N \int dx\, \rho(x)$ as usual and using the scaling ansatz \eqref{LargeNScaling} we get
 \es{FundContinuum}{
  F_3 =  \frac{i(n_f - n_a)}{8 \pi}  N^{1 + 2 \alpha}  \int dx \, \rho x^2
   + N^{1 + \alpha} \sum_{\substack{\text{fundamental }\\a}} 
   \int dx\, \rho(x) \abs{x} \left( \frac{1- \Delta_a}{2} -\frac{1}{4 \pi}  y_a(x) \right) \\
   + N^{1 + \alpha} \sum_{\substack{\text{anti-fundamental }\\a}} 
   \int dx\, \rho(x) \abs{x} \left( \frac{1- \tilde \Delta_a}{2} +\frac{1}{4 \pi}  y_a(x) \right) \,,
 }
where $n_f$ is the total number of fundamentals and $n_a$ is the total number of anti-fundamentals.  When $n_f = n_a$ and $\alpha = 1/2$ one reproduces eqs.~\eqref{FundContribution} and \eqref{AntifundContribution}.

\subsection{Why $\alpha = 1/2$?}

When the CS levels sum to zero and the number of fundamentals equals the number of anti-fundamentals, we find $F_1 + F_3 \sim N^{1 + \alpha}$ at large $N$ and $F_2 \sim N^{2 - \alpha}$.  In order to have a non-trivial saddle point we have to balance out these two terms, so $1 + \alpha = 2- \alpha$ implying $\alpha = 1/2$.  The free energy therefore scales as $N^{3/2}$.

Note that as is the case for the theories in section~\ref{FIVETHIRDS} where the Chern-Simons levels don't sum to zero, we have $F_1 + F_3 \sim N^{1 + 2 \alpha}$ and $F_2 \sim N^{2 - \alpha}$, which implies $\alpha = 1/3$ and therefore $F \sim N^{5/3}$.  Of course, the derivation presented above doesn't hold exactly for those theories because for the saddle points the imaginary parts of the eigenvalues also grow as $N^{1/3}$.

\section{Flavoring the Martelli-Sparks model}

In this section we deform the flavored ABJM construction of the theory dual to $AdS_4\times Q^{1,1,1}$ \cite{Benini:2009qs, Jafferis:2009th} by assuming general monomial superpotentials
for the two adjoint chiral superfields $\Phi_1$ and $\Phi_2$:
 \es{Q111SupVar}{
  W \sim \tr \left[ \Phi_1^{n+1} + \Phi_2^{n+1} + \Phi_2 (A_1 B_1 + A_2 B_2) - \Phi_1 (B_1 A_1 + B_2 A_2)
    + q_{2i} A_1 Q_{2i} + q_{2i+1} A_2 Q_{2i+1} \right] \,.
 }
 This is a flavored version of the Martelli-Sparks construction which led to the dual of $AdS_4\times V_{5,2}$ for $n=2$ \cite{Martelli:2009ga}.
The marginality of the superpotential implies that the dimension of $\Phi_a$ is $\delta = 2/(n+1)$, $\Delta_A + \Delta_B + \delta = 2$ and that the dimensions of the fundamental fields are $1 - \Delta_A/2$.  The free energy functional is
 \es{Q111AdjFree}{
  F_n[\rho, y_a] = -N^{3/2} \delta \int dx\, \rho^2 (\delta y - 2 \pi \Delta_B) (\delta y + 2 \pi \Delta_A)
    + N^{3/2} \frac{n_f}{2 \pi} \int dx\, \rho (\delta y + 2 \pi \Delta_A) \abs{x} \,,
 }
where $\delta y \equiv y_1 - y_2$.  One should really consider the contribution of $\Delta_m$ in equation~\eqref{Q111AdjFree}.  However, it is not hard to show that the free energy is locally maximized for vanishing bare monopole R-charge, so we do not include $\Delta_m$ in this discussion.

The free energy is maximized for
 \es{Q111AdjSoln}{
  \rho = \frac{\sqrt{n_f} (n+1)}{2\pi \sqrt{3 n}} - \frac{n_f (n+1)^2 \abs{x}}{16 \pi^2 n} \,,
   \qquad    \delta y = \frac{\sqrt{n_f n}}{\sqrt{3}\rho} - 2 \pi \Delta_A \,.
 }
At the extremum
 \es{FQ111Adj}{
  F_n = \frac{16 \pi n^{3/2} n_f^{1/2} N^{3/2}}{3^{3/2} (n+1)^2} = \frac{4 n^{3/2}}{(n+1)^2} F_1 \,,
 }
so that
 \es{volQ111Adj}{
  \Vol(Y_n) =  \frac{(n+1)^4}{16 n^3} \Vol(Y_1) \,,
 }
and in particular $\Vol(Y_2) = 81 \Vol(Y_1) / 128$.

If we were to solve the $SU(N)$ theory and impose the constraint $\int dx\, \rho\, \delta y = 0$, we would obtain
 \es{GotDeltaA}{
  \Delta_A = \frac{4 n}{3 (n+1)} \,.
 }
In particular, when $n=1$ we reproduce $\Delta_A = 2/3$, which matches the gravity prediction.  For $n=2$ we get the prediction that $\Delta_A = 8/9$.

\section{Comments on warped $\C \times \text{CY}_3$ geometries}
\label{CCY5}

Some of the examples we discussed involve theories where the seven-dimensional internal space $Y$ is the base of a Calabi-Yau cone $\text{CY}_4$ that can be written as the product $\C \times \text{CY}_3$ for some Calabi-Yau three-fold $\text{CY}_3$.  Such a $\text{CY}_3$ is a cone over a given five-dimensional Sasaki-Einstein manifold $X$ and there it can be used to construct a type IIB supergravity solution whose metric is $AdS_5 \times X$.   In this section we want to answer the following question:  what can we infer about M-theory compactified on $Y$ from our knowledge of type IIB string theory compactified on $X$?

\subsection{From type IIB string theory to M-theory}
\label{2btom}

Let's first try to understand the topology of the space $Y$.  Denoting the radial coordinate in $\text{CY}_3$ by $\rho$ and parameterizing $\C$ by the complex coordinate $z$, one can find $Y$ by intersecting $\C \times \text{CY}_3$ with the unit sphere $\abs{z}^2 + \rho^2 = 1$.  Since on each $\abs{z} \leq 1$ slice this intersection reduces to a copy of $X$ of radius $\rho = \sqrt{1 - \abs{z}^2}$, one obtains a description of $Y$ as an $X$ fibration over the unit disk $\abs{z} \leq 1$ where the fiber shrinks to a point on the boundary of the disk.  Topologically, one says that $Y$ is a double suspension of $X$, $Y = S^2 X$.  It is a standard consequence of the Mayer-Vietoris sequence that in performing a suspension the reduced homology just shifts by one unit, so the reduced homology of $Y$ can be given simply in terms of that of $X$:
 \es{HomologyY}{
  \tilde H_n(Y; \Z) = \tilde H_{n-2}(X; \Z) \,.
 }
In particular, the number of linearly independent $n$-cycles of $Y$ $(n>2)$ is precisely equal to the number of linearly independent $(n-2)$-cycles of $X$.  If such a cycle of $X$ is represented by an embedded closed surface $\Sigma$, the corresponding cycle of $Y$ is represented by the double suspension $S^2 \Sigma$, meaning that $\Sigma$ is fibered over the disk $D^2$ and the fiber shrinks to a point on $S^1 = \partial D^2$.

Given a Sasaki-Einstein metric $ds_X^2$ on $X$, one can easily write down a Sasaki-Einstein metric on $Y$:
  \es{dsY}{
  ds_Y^2 = d \theta^2 + \sin^2 \theta\, d \phi^2 + \cos^2\theta\, ds_X^2 \,,
  }
where the angels $\theta$ and $\phi$ are inside the ranges $\left(0,\pi/2\right)$ and $\left(0, 2\pi\right)$, respectively, parameterizing a disk of radius $\pi/2$ in $\R^2$.  Here, the metric $ds_X^2$ is normalized so that $R_{mn} = 4 g_{mn}$, just like on the unit five-sphere, and the metric on $ds_Y^2$ is normalized so that $R_{mn} = 6 g_{mn}$, just like on the unit seven-sphere.  In fact, if $\text{CY}_3 = \C^3$, then $X = S^5$ and $Y = S^7$ ,with the standard metrics on them both.  Using the metric \eqref{dsY}, one can find the relation between the volumes of $X$ and $Y$:
 \es{VolXY}{
  \Vol(Y) = \frac{\pi}{3} \Vol(X) \,.
 }
More generally, one finds
 \es{VolCycles}{
  \Vol\left(\text{$n$-cycle in $Y$}\right) = \frac{2 \pi}{n-1} \Vol\left(\text{$(n-2)$-cycle in $X$}\right) \,,
 }
where the $n$-cycle in $Y$ is obtained by the double suspension of the corresponding $(n-2)$-cycle in $X$.

The near-horizon limit of $N$ D3-branes at the tip of $\text{CY}_3$ is given by the type IIB solution
 \es{D3near}{
  ds_{10}^2 = ds_{AdS_5}^2 + \tilde L^2 ds_X^2 \,, \qquad
   F_5 = \frac {4}{\tilde L} (\vol_{AdS_5}  + \tilde L^5 \vol_X) \,,
 }
where $\tilde L$ is the radius of $AdS_5$ and $\vol_{AdS_5}$ and $\vol_X$ are the volume forms on $AdS_5$ and $X$, respectively.  The radius $\tilde L$ is quantized in string units and can be found from the requirement that there are $N$ units of D3-brane flux through $X$:
 \es{D3Quantization}{
  N = \frac{1}{g_s(2 \pi \ell_s)^4} \int_X F_5 = \frac{ \Vol(X)}{4 \pi^4 g_s} \frac{\tilde L^4}{\ell_s^4} \,,
 }
where $\ell_s \equiv \sqrt{\alpha'}$ is the string length.

Similarly, the near-horizon limit of $N$ M2-branes at the tip of $\text{CY}_4 = \C \times \text{CY}_3$ is the 11-d supergravity extremum
 \es{M2near}{
  ds_{11}^2 = ds_{AdS_4}^2 + 4 L^2 ds_Y^2\,, \qquad
   F_4 = \frac{3}{L} \vol_{AdS_4} \,, \qquad
   F_7 \equiv *_{11} F_4 = 384 L^6 \vol_Y \,,
 }
where $L$ is the radius of $AdS_4$, which is quantized in Planck units from the requirement that $F_7$ generates $N$ units of M2-brane flux through $Y$:
 \es{M2Quantization}{
  N = \frac{1}{(2 \pi \ell_p)^6} \int_Y F_7 = \frac{6 \Vol(Y)}{\pi^6} \frac{L^6}{\ell_p^6} \,.
 }

\subsection{Wrapped D3-branes and wrapped M5-branes}

Let's consider a 3-cycle $\Sigma_3$ in $X$ and the corresponding 5-cycle $\Sigma_5 = S^2 \Sigma_3$ in $Y$.  In the type IIB background, a D3-brane wrapped over $\Sigma_3$ looks from the $AdS_5$ perspective as a very massive particle with mass $m_\text{D3}$.  In the 11-d background, an M5-brane wrapped over $\Sigma_5$ looks like a very massive particle in $AdS_4$ with mass $m_\text{M5}$.  Let's find the relation between these masses.

The DBI action for a D3-brane with no worldvolume fluxes in type IIB theory is
 \es{D3Action}{
  S_\text{D3} = -\frac{2 \pi}{g_s (2 \pi \ell_s)^4} \int d^4x\, e^{-\phi} \sqrt{-g} \,.
 }
Integrating over the compact coordinates, the action reduces to $-m_\text{D3} \int ds$, where
 \es{D3Mass}{
  m_\text{D3} \tilde L = \frac{2 \pi}{g_s (2 \pi \ell_s)^4} \tilde L^4 \Vol(\Sigma_3)
   = \frac{\pi N}{2} \frac{\Vol(\Sigma_3)}{\Vol(X)} \,,
 }
where in the second equality we used the D3 charge quantization condition \eqref{D3Quantization}.  Similarly, the DBI action for an M5-brane with no worldvolume fluxes is
 \es{M5Action}{
  S_\text{M5} = -\frac{2 \pi}{(2 \pi \ell_p)^6} \int d^6x\,\sqrt{-g} \,.
 }
Integrating over the compact coordinates, we obtain the effective mass
 \es{M5Mass}{
  m_\text{M5} L = \frac{2 \pi}{(2 \pi \ell_p)^6} 2^5 L^6 \Vol(\Sigma_5)
   = \frac{\pi N}{6} \frac{\Vol(\Sigma_5)}{\Vol(Y)} \,,
 }
where in the second equality we used the M2 charge quantization condition \eqref{M2Quantization}.  Combining \eqref{D3Mass} and \eqref{M5Mass} with \eqref{VolXY} and \eqref{VolCycles}, one obtains
 \es{MassRatio}{
  m_\text{M5} L = \frac 12 m_\text{D3} \tilde L\,.
 }
In the dual field theories, $m_\text{M5} L$ and $m_\text{D3} \tilde L$ are usually interpreted as the conformal dimensions $\Delta_\text{baryon}^{(3)}$ and $\Delta_\text{baryon}^{(4)}$ of some baryonic operators.  Eq.~\eqref{MassRatio} shows
 \es{DeltaRelation}{
  \Delta_\text{baryon}^{(3)} = \frac 12 \Delta_\text{baryon}^{(4)} \,.
 }
Note that when making the comparison \eqref{DeltaRelation} we consider the 3-d and 4-d theories on the same number $N$ of M2- and D3-branes, respectively.

\subsection{The minimal scalar equation}

A related question one can try to answer is whether there is any relation between the spectra of the mesonic operators in the theories dual to $AdS_5 \times X$ and $AdS_4 \times Y$.  Mesonic operators are dual to fluctuations around these backgrounds.  We analyze the simplest such fluctuations, namely those described by a minimal scalar equation $\square \Phi = 0$ in ten and eleven dimensions.

In 10-d, a basis of solutions can be found by writing $\Phi = \phi(y) \psi(x)$, where $\phi(y)$ depends only on the $AdS_5$ coordinates and $\psi(x)$ is an eigenfunction of the laplacian on $X$ with eigenvalue $-\lambda$.  Then the equation of motion $\square_{10} \Phi = 0$ reduces to
 \es{10eom}{
  \square_{AdS_5} \phi - \frac \lambda {\tilde L^2} \phi = 0 \,,
 }
so $\phi$ is a massive field with mass $m_{AdS_5}$ given by $m_{AdS_5}^2 \tilde L^2 = \lambda$.  The AdS/CFT relation between the mass of this field and the conformal dimension $\Delta^{(4)}$ of the dual operator then implies
 \es{Delta4}{
  \Delta^{(4)} (\Delta^{(4)} - 4) = \lambda \,.
 }

In 11-d, let's consider the ansatz $\Phi = \phi(y) (\cos \theta)^{\Delta^{(4)}} \psi(x)$ where $\phi(y)$ depends only on the $AdS_4$ coordiantes, $\psi(x)$ is an eigenfunction of the laplacian on $X$ with eigenvalue $-\lambda$ as above.  Using the explicit metric \eqref{dsY} one can check that $(\cos \theta)^{\Delta^{(4)}} \psi(x)$ is an eigenfunction of the laplacian on $Y$ with eigenvalue $-(\Delta^{(4)} - 4)(\Delta^{(4)} + 2)$.  The minimal scalar equation $\square_{11} \Phi = 0$ reduces to
 \es{11eom}{
  \square_{AdS_4} \phi - \frac{(\Delta^{(4)} - 4)(\Delta^{(4)} + 2)}{4 L^2} \phi = 0 \,,
 }
so $\phi$ is a massive field with mass $m_{AdS_4}$ given by $m_{AdS_4}^2 L^2 = (\Delta^{(4)} - 4)(\Delta^{(4)} + 2)/4$.    The conformal dimension $\Delta^{(3)}$ of the dual CFT$_3$ operator then satisfies
 \es{Delta3Eq}{
  \Delta^{(3)} (\Delta^{(3)} - 3) = m_{AdS_4}^2 L^2 = \frac{(\Delta^{(4)} - 4)(\Delta^{(4)} + 2)}{4} \,,
 }
with the solution
 \es{GotDelta3}{
  \Delta^{(3)} = 1 + \frac 12 \Delta^{(4)} \,.
 }

The minimal scalar equation arises for example from fluctuations in the AdS part of the metric.  The dual operators are typically $\tr \left( T_{\mu\nu} \times (\text{chiral operator}) \right)$, where the chiral operators are constructed from the bifundamental fields in the quiver gauge theory.  It is then convenient to define $\tilde \Delta^{(3)} = \Delta^{(3)} - 3$ and $\tilde \Delta^{(4)} = \Delta^{(4)} - 4$, which are the dimensions of the chiral operators corresponding to the eigenfunction $\psi(x)$ of the Laplacian on $X$.  From \eqref{GotDelta3} we obtain
 \es{ChiralOps}{
  \tilde \Delta^{(3)} = \frac 12 \tilde \Delta^{(4)} \,.
 }
This relation is surprisingly similar to \eqref{DeltaRelation}.

\bibliographystyle{ssg}
\bibliography{matrixGeneral}

\end{document}

%% file: NAQuiv.pdf_tex
%% Creator: Inkscape inkscape 0.48.0, www.inkscape.org
%% Accompanies image file 'NAQuiv.pdf' (pdf, eps, ps)

\begingroup
  \makeatletter
  \providecommand\color[2][]{%
    \errmessage{(Inkscape) Color is used for the text in Inkscape, but the package 'color.sty' is not loaded}
    \renewcommand\color[2][]{}%
  }
  \providecommand\transparent[1]{%
    \errmessage{(Inkscape) Transparency is used (non-zero) for the text in Inkscape, but the package 'transparent.sty' is not loaded}
    \renewcommand\transparent[1]{}%
  }
  \providecommand\rotatebox[2]{#2}
  \ifx\svgwidth\undefined
    \setlength{\unitlength}{671.88540039pt}
  \else
    \setlength{\unitlength}{\svgwidth}
  \fi
  \global\let\svgwidth\undefined
  \makeatother
  \begin{picture}(1,0.79017658)%
    \put(0,0){\includegraphics[width=\unitlength]{NAQuiv.pdf}}%
    \put(0.49275182,0.6389895){\color[rgb]{0,0,0}\makebox(0,0)[lt]{\begin{minipage}{0.12417082\unitlength}\raggedright $k_1$\end{minipage}}}%
    \put(0.54393441,0.44097235){\color[rgb]{0,0,0}\makebox(0,0)[lt]{\begin{minipage}{0.06973979\unitlength}\raggedright $k_2$\end{minipage}}}%
    \put(0.54007302,0.21028454){\color[rgb]{0,0,0}\makebox(0,0)[lt]{\begin{minipage}{0.08845045\unitlength}\raggedright $k_3$\end{minipage}}}%
    \put(0.24656335,0.28419129){\color[rgb]{0,0,0}\makebox(0,0)[lt]{\begin{minipage}{0.09185243\unitlength}\raggedright $k_4$\end{minipage}}}%
    \put(0.22542476,0.47402041){\color[rgb]{0,0,0}\makebox(0,0)[lt]{\begin{minipage}{0.10205823\unitlength}\raggedright $k_p$\\ \end{minipage}}}%
    \put(0.20148692,0.59784267){\color[rgb]{0,0,0}\makebox(0,0)[lb]{\smash{$\color{O4}A_{p,1}$}}}%
    \put(0.55187209,0.58493949){\color[rgb]{0,0,0}\makebox(0,0)[lb]{\smash{$\color{O4}A_{1,2}$}}}%
    \put(0.62351924,0.26806079){\color[rgb]{0,0,0}\makebox(0,0)[lb]{\smash{$\color{O4}A_{2,3}$}}}%
    \put(0.35136397,0.13538508){\color[rgb]{0,0,0}\makebox(0,0)[lb]{\smash{$\color{O4}A_{3,4}$}}}%
    \put(0.44998489,0.44214348){\color[rgb]{0,0,0}\makebox(0,0)[lb]{\smash{$\color{O3}B_{2,1}$}}}%
    \put(0.45716603,0.33738784){\color[rgb]{0,0,0}\makebox(0,0)[lb]{\smash{$\color{O3}B_{3,2}$}}}%
    \put(0.30160245,0.27735577){\color[rgb]{0,0,0}\makebox(0,0)[lb]{\smash{$\color{O3}B_{4,3}$}}}%
    \put(0.31131299,0.46779657){\color[rgb]{0,0,0}\makebox(0,0)[lb]{\smash{$\color{O3}B_{1,p}$}}}%
    \put(0.4132577,0.77389311){\color[rgb]{0,0,0}\makebox(0,0)[lb]{\smash{$\color{O1}\Phi_1$}}}%
    \put(0.15981315,0.06288751){\color[rgb]{0,0,0}\makebox(0,0)[lb]{\smash{$\color{O1}\Phi_4$}}}%
    \put(0.50851208,0.00505454){\color[rgb]{0,0,0}\makebox(0,0)[lb]{\smash{$\color{O1}\Phi_3$}}}%
    \put(0.79257411,0.43369903){\color[rgb]{0,0,0}\makebox(0,0)[lb]{\smash{$\color{O1}\Phi_2$}}}%
    \put(-0.00177904,0.49833589){\color[rgb]{0,0,0}\makebox(0,0)[lb]{\smash{$\color{O1}\Phi_p$}}}%
  \end{picture}%
\endgroup

%% file: C3QUIV.pdf_tex
%% Creator: Inkscape inkscape 0.48.0, www.inkscape.org
%% Accompanies image file 'C3QUIV.pdf' (pdf, eps, ps)

\begingroup
  \makeatletter
  \providecommand\color[2][]{%
    \errmessage{(Inkscape) Color is used for the text in Inkscape, but the package 'color.sty' is not loaded}
    \renewcommand\color[2][]{}%
  }
  \providecommand\transparent[1]{%
    \errmessage{(Inkscape) Transparency is used (non-zero) for the text in Inkscape, but the package 'transparent.sty' is not loaded}
    \renewcommand\transparent[1]{}%
  }
  \providecommand\rotatebox[2]{#2}
  \ifx\svgwidth\undefined
    \setlength{\unitlength}{321.63693848pt}
  \else
    \setlength{\unitlength}{\svgwidth}
  \fi
  \global\let\svgwidth\undefined
  \makeatother
  \begin{picture}(1,1.44422062)%
    \put(0,0){\includegraphics[width=\unitlength]{C3QUIV.pdf}}%
    \put(0.24872775,1.35512331){\color[rgb]{0,0,0}\makebox(0,0)[lb]{\smash{${\color{O1}X_1}$}}}%
    \put(0.38552787,1.40486881){\color[rgb]{0,0,0}\makebox(0,0)[lb]{\smash{${\color{O3}X_2}$}}}%
    \put(0.54720084,1.37999602){\color[rgb]{0,0,0}\makebox(0,0)[lb]{\smash{${\color{O4}X_3}$}}}%
    \put(0.55963723,0.21141852){\color[rgb]{0,0,0}\makebox(0,0)[lb]{\smash{$\color{O3}U(n_2)$}}}%
    \put(0.39796426,0.8330165){\color[rgb]{0,0,0}\makebox(0,0)[lb]{\smash{$U(N)$}}}%
    \put(0.37309154,0.53454333){\color[rgb]{0,0,0}\makebox(0,0)[lb]{\smash{$\color{O1}U(n_1)$}}}%
    \put(0.54720088,0.0122152){\color[rgb]{0,0,0}\makebox(0,0)[lb]{\smash{$\color{O4}U(n_3)$}}}%
  \end{picture}%
\endgroup

%% file: ABJMQUIV.pdf_tex
%% Creator: Inkscape inkscape 0.48.0, www.inkscape.org
%% Accompanies image file 'ABJMQUIV.pdf' (pdf, eps, ps)

\begingroup
  \makeatletter
  \providecommand\color[2][]{%
    \errmessage{(Inkscape) Color is used for the text in Inkscape, but the package 'color.sty' is not loaded}
    \renewcommand\color[2][]{}%
  }
  \providecommand\transparent[1]{%
    \errmessage{(Inkscape) Transparency is used (non-zero) for the text in Inkscape, but the package 'transparent.sty' is not loaded}
    \renewcommand\transparent[1]{}%
  }
  \providecommand\rotatebox[2]{#2}
  \ifx\svgwidth\undefined
    \setlength{\unitlength}{397.22473145pt}
  \else
    \setlength{\unitlength}{\svgwidth}
  \fi
  \global\let\svgwidth\undefined
  \makeatother
  \begin{picture}(1,0.36226345)%
    \put(0,0){\includegraphics[width=\unitlength]{ABJMQUIV.pdf}}%
    \put(0.46006602,0.27176059){\color[rgb]{0,0,0}\makebox(0,0)[lb]{\smash{${\color{O1}A_1}$}}}%
    \put(0.31908791,0.27176059){\color[rgb]{0,0,0}\makebox(0,0)[lb]{\smash{${\color{O3}A_2}$}}}%
    \put(-0.00296775,0.17205912){\color[rgb]{0,0,0}\makebox(0,0)[lb]{\smash{$U(N)_k$ }}}%
    \put(0.77654602,0.17188006){\color[rgb]{0,0,0}\makebox(0,0)[lb]{\smash{$U(N)_{-k}$ }}}%
    \put(0.52363274,0.09150024){\color[rgb]{0,0,0}\makebox(0,0)[lb]{\smash{${\color{O4}B_1}$}}}%
    \put(0.38265463,0.09150024){\color[rgb]{0,0,0}\makebox(0,0)[lb]{\smash{${\color{O2}B_2}$}}}%
  \end{picture}%
\endgroup

%% file: ABJMGENQUIV.pdf_tex
%% Creator: Inkscape inkscape 0.48.0, www.inkscape.org
%% Accompanies image file 'ABJMGENQUIV.pdf' (pdf, eps, ps)

\begingroup
  \makeatletter
  \providecommand\color[2][]{%
    \errmessage{(Inkscape) Color is used for the text in Inkscape, but the package 'color.sty' is not loaded}
    \renewcommand\color[2][]{}%
  }
  \providecommand\transparent[1]{%
    \errmessage{(Inkscape) Transparency is used (non-zero) for the text in Inkscape, but the package 'transparent.sty' is not loaded}
    \renewcommand\transparent[1]{}%
  }
  \providecommand\rotatebox[2]{#2}
  \ifx\svgwidth\undefined
    \setlength{\unitlength}{457.09628906pt}
  \else
    \setlength{\unitlength}{\svgwidth}
  \fi
  \global\let\svgwidth\undefined
  \makeatother
  \begin{picture}(1,0.74495189)%
    \put(0,0){\includegraphics[width=\unitlength]{ABJMGENQUIV.pdf}}%
    \put(0.36480218,0.44527332){\color[rgb]{0,0,0}\makebox(0,0)[lb]{\smash{${\color{O1}A_1}$}}}%
    \put(0.24228972,0.44527332){\color[rgb]{0,0,0}\makebox(0,0)[lb]{\smash{${\color{O3}A_2}$}}}%
    \put(-0.00257903,0.35863099){\color[rgb]{0,0,0}\makebox(0,0)[lb]{\smash{$U(N)$ }}}%
    \put(0.63982888,0.35847538){\color[rgb]{0,0,0}\makebox(0,0)[lb]{\smash{$U(N)$ }}}%
    \put(0.44371645,0.59474941){\color[rgb]{0,0,0}\makebox(0,0)[lb]{\smash{${\color{O1}U(n_{a1})}$}}}%
    \put(0.42004278,0.2886239){\color[rgb]{0,0,0}\makebox(0,0)[lb]{\smash{${\color{O4}B_1}$}}}%
    \put(0.29753032,0.2886239){\color[rgb]{0,0,0}\makebox(0,0)[lb]{\smash{${\color{O2}B_2}$}}}%
    \put(0.42004278,0.1311078){\color[rgb]{0,0,0}\makebox(0,0)[lb]{\smash{${\color{O2}U(n_{b2})}$}}}%
    \put(0.42004278,0.0085953){\color[rgb]{0,0,0}\makebox(0,0)[lb]{\smash{${\color{O4}U(n_{b1})}$}}}%
    \put(0.44371645,0.7172619){\color[rgb]{0,0,0}\makebox(0,0)[lb]{\smash{${\color{O3}U(n_{a2})}$}}}%
  \end{picture}%
\endgroup

%% file: NQuiv.pdf_tex
%% Creator: Inkscape inkscape 0.48.0, www.inkscape.org
%% Accompanies image file 'NQuiv.pdf' (pdf, eps, ps)

\begingroup
  \makeatletter
  \providecommand\color[2][]{%
    \errmessage{(Inkscape) Color is used for the text in Inkscape, but the package 'color.sty' is not loaded}
    \renewcommand\color[2][]{}%
  }
  \providecommand\transparent[1]{%
    \errmessage{(Inkscape) Transparency is used (non-zero) for the text in Inkscape, but the package 'transparent.sty' is not loaded}
    \renewcommand\transparent[1]{}%
  }
  \providecommand\rotatebox[2]{#2}
  \ifx\svgwidth\undefined
    \setlength{\unitlength}{405.90004883pt}
  \else
    \setlength{\unitlength}{\svgwidth}
  \fi
  \global\let\svgwidth\undefined
  \makeatother
  \begin{picture}(1,0.82364749)%
    \put(0,0){\includegraphics[width=\unitlength]{NQuiv.pdf}}%
    \put(0.49534919,0.83202737){\color[rgb]{0,0,0}\makebox(0,0)[lt]{\begin{minipage}{0.20553968\unitlength}\raggedright $k_1$\end{minipage}}}%
    \put(0.75994636,0.56172855){\color[rgb]{0,0,0}\makebox(0,0)[lt]{\begin{minipage}{0.11544011\unitlength}\raggedright $k_2$\end{minipage}}}%
    \put(0.55732788,0.08307453){\color[rgb]{0,0,0}\makebox(0,0)[lt]{\begin{minipage}{0.14641182\unitlength}\raggedright $k_3$\end{minipage}}}%
    \put(0.08988353,0.14220241){\color[rgb]{0,0,0}\makebox(0,0)[lt]{\begin{minipage}{0.15204311\unitlength}\raggedright $k_4$\end{minipage}}}%
    \put(-0.00292728,0.67998431){\color[rgb]{0,0,0}\makebox(0,0)[lt]{\begin{minipage}{0.16893675\unitlength}\raggedright $k_p$\\ \end{minipage}}}%
    \put(0.03910129,0.77387181){\color[rgb]{0,0,0}\makebox(0,0)[lb]{\smash{$\color{O4}A_{p,1}$}}}%
    \put(0.61909303,0.75251321){\color[rgb]{0,0,0}\makebox(0,0)[lb]{\smash{$\color{O4}A_{1,2}$}}}%
    \put(0.73769038,0.22798463){\color[rgb]{0,0,0}\makebox(0,0)[lb]{\smash{$\color{O4}A_{2,3}$}}}%
    \put(0.2871924,0.00836684){\color[rgb]{0,0,0}\makebox(0,0)[lb]{\smash{$\color{O4}A_{3,4}$}}}%
    \put(0.46918246,0.51696794){\color[rgb]{0,0,0}\makebox(0,0)[lb]{\smash{$\color{O3}B_{2,1}$}}}%
    \put(0.46232632,0.34274154){\color[rgb]{0,0,0}\makebox(0,0)[lb]{\smash{$\color{O3}B_{3,2}$}}}%
    \put(0.20482229,0.2433706){\color[rgb]{0,0,0}\makebox(0,0)[lb]{\smash{$\color{O3}B_{4,3}$}}}%
    \put(0.23497416,0.58676292){\color[rgb]{0,0,0}\makebox(0,0)[lb]{\smash{$\color{O3}B_{1,p}$}}}%
  \end{picture}%
\endgroup